\documentclass{article}

\usepackage{PRIMEarxiv}

\usepackage[utf8]{inputenc} 
\usepackage[T1]{fontenc}    
\usepackage{hyperref}       
\usepackage{url}            
\usepackage{booktabs}       
\usepackage{amsfonts}       
\usepackage{nicefrac}       
\usepackage{microtype}      
\usepackage{lipsum}
\usepackage{fancyhdr}       
\usepackage{graphicx}       
\graphicspath{{media/}}     
\usepackage{graphicx}  
\usepackage{epstopdf}
\usepackage{ragged2e}
\usepackage{algorithm}
\usepackage{algpseudocode}
\usepackage{enumerate}
\usepackage{subfigure}
\usepackage{amsmath}
\usepackage{bm}
\usepackage{stfloats}
\usepackage{siunitx}
\usepackage{booktabs}
\usepackage{multirow}
\usepackage{pifont}
\usepackage[justification=centering]{caption}

\title{A multi-task learning for cavitation detection and cavitation intensity recognition of valve acoustic signals\thanks{\textit{Accepted by the Engineering Applications of Artificial Intelligence on 19 April 2022.}}}

\author{
  Yu Sha \\
   Xidian University\\
   FIAS \thanks{\textit{FIAS: Frankfurt Institute for Advanced Studies}}\\
   XF-IJRC\thanks{\textit{XF-IJRC: Xidian-FIAS international Joint Research Center}}\\
   \And
  Johannes Faber \\
  FIAS \\
    \And
  Shuiping Gou \\
  Xidian University \\
    \And
  Bo Liu \\
  Xidian University \\
    \And
  Wei Li \\
  FIAS \\
    \And
  Stefan Schramm \\
  FIAS \\ 
    \And
  Horst Stoecker \\
  FIAS \\  
  Goethe Universit{\"a}t \\
  GSI\thanks{\textit{GSI: GSI Helmholtzzentrum f{\"u}r Schwerionenforschung GmbH}}\\
    \And
  Thomas Steckenreiter \\
  SAMSOM AG \\  
    \And
  Domagoj Vnucec \\
  SAMSOM AG \\  
    \And
  Nadine Wetzstein \\
  SAMSOM AG \\      
  \And
  Andreas Widl \\
  SAMSOM AG \\  
  \And
  Kai Zhou \thanks{\textit{Kai Zhou is the corresponding author. Email: zhou@fias.uni-frankfurt.de}}\\
  FIAS \\  
}

\begin{document}
\maketitle

\begin{abstract}
With the rapid development of smart manufacturing, data-driven machinery health management has received a growing attention. As one of the most popular methods in machinery health management, deep learning (DL) has achieved remarkable successes. However, due to the issues of limited samples and poor separability of different cavitation states of acoustic signals, which greatly hinder the eventual performance of DL modes for cavitation intensity recognition and cavitation detection. Also different tasks were performed separately conventionally. In this work, a novel multi-task learning framework for simultaneous cavitation detection and cavitation intensity recognition framework using 1-D double hierarchical residual networks (1-D DHRN) is proposed for analyzing valves acoustic signals. Firstly, a data augmentation method based on sliding window with fast Fourier transform (Swin-FFT) is developed to alleviate the small-sample issue confronted in this study. Secondly, a 1-D double hierarchical residual block (1-D DHRB) is constructed to capture sensitive features from the frequency domain acoustic signals of valve. Then, a new structure of 1-D DHRN is proposed. Finally, the devised 1-D DHRN is evaluated on two datasets of valve acoustic signals without noise (\emph{\textbf{Dataset 1}} and \emph{\textbf{Dataset 2}}) and one dataset of valve acoustic signals with realistic surrounding noise (\emph{\textbf{Dataset 3}}) provided by SAMSON AG (Frankfurt). Our method has achieved state-of-the-art results. The prediction accurcies of 1-D DHRN for cavitation intensitys recognition are as high as \bm{$93.75\%$}, \bm{$94.31\%$} and \bm{$100\%$}, which indicates that 1-D DHRN outperforms other DL models and conventional methods. At the same time, the testing accuracies of 1-D DHRN for cavitation detection are as high as \bm{$97.02\%$}, \bm{$97.64\%$} and \bm{$100\%$}. In addition, 1-D DHRN has also been tested for different frequencies of samples and shows excellent results for frequency of samples that mobile phones can accommodate.
\end{abstract}



\section{Introduction}
Cavitation is a dynamical phenomenon occurs in fluid dynamics, during which bubbles, also called cavities, form (and later collapse) inside the liquid when the local pressure (e.g. at the contact point of a liquid and some solid surface) is lower than the liquid's vapor pressure \cite{jazi2009detecting,zhao2007experimental}. When the bubble flows to places where the liquid pressure exceeds the vapor pressure, the bubble collapses and the implosion instantaneously produces a great impact (shock waves) and high temperature \cite{brujan2002final,bonnier2002experimental}. Furthermore, severe pitting and wear can be induced on the solid surface \cite{song2014corrosion}.

Generally, cavitation can potentially bring dangers to process plants, especially for valves, pumps, pipes or propellers. These potential dangers might quickly cause damage to components of the plants and loss of efficiency \cite{mckee2011review}. On one hand, corrosion and destruction to valves, pumps or pipes can be induced by the cavitation occurrence \cite{schleihs20143d}, vibration of the internal structure and noise loudness levels might also increase \cite{yin2016numerical} due to the bubble collapse and rupture \cite{brennen2014cavitation}. On the other hand, the fluid bulk properties and flow rate inside the plants can be altered by the cavitation \cite{gholizadeh2012fluid,liu2005numerical} which further can lead to standstill of the process. Consequently, plants such as valves, pipes, pumps and related components are at the largest risk confronting cavitation. Many industries are fighting against cavitation, such as SAMSON AG for valves and other manufacturing plants \cite{clarke2011evaluating}. In the worst case, cavitation can lead to the closure of the factory, e.g. for situations with test rack system including control valve. Therefore, it is vital to detect the cavitation of process plants (e.g. valves, pumps or pipes) at its early stages for intervention so as to ensure security and reduce any possible economic loss.
Conventionally cavitation is detected by comparing the fault and healthy conditions of specific devices under the monitored signals. According to the type of monitoring sensors, it can be divided into vibration signals based cavitation detection and acoustic signals based detection. Recently, some researchers have explored machine learning application for cavitation detection, based on features handcrafted from the vibration or acoustic signals \cite{precup2015overview}.

Sakthivel et al. \cite{sakthivel2010vibration} extracted 11 statistical features from the vibration time domain signals. These features are then fed into the C4.5 decision tree \cite{quinlan2014c4} to classify the bearing fault, seal fault, impeller fault, bearing and impeller fault together with cavitation. Muralidharan et al.\cite{muralidharan2013feature} used the Continuous Wavelet Transform (CWT) \cite{rao2002wavelet} to replace the statistical feature extraction from vibration signals. Then the CWT achieved features are taken as input of the decision tree algorithm for similar classification task as in \cite{sakthivel2010vibration}. In addition, Muralidharan et al. \cite{muralidharan2014fault} also studied the influence of different families and levels of the CWT on fault diagnosis of single-piece centrifugal oils using Support Vector Machine (SVM). Samanta et al. \cite{samanta2003artificial} extracted features from the original and pre-processed signals as the input of two different classifiers, the SVM and the artificial neural network (ANN), to identify normal and defective bearings. The parameters of the SVM and ANN are optimized by genetic algorithms \cite{holland1992genetic}, and the results explained the importance of feature selection to the classifier. Yang et al. \cite{yang2005cavitation} extracted 4 statistical features from the vibration time domain signals as the input of the SVM to detect cavitation and no cavitation of the butterfly valve. Bordoloi et al. \cite{bordoloi2017identification} proposed a SVM method using directly the vibration signal data of bearing block and pump casing to diagnose blockage level and cavitation intensity. Panda et al. \cite{panda2018prediction} extracted statistical features from the vibration time domain signals of the pump as the input of SVM to distinguish cavitation and flow block. Rapur et al. \cite{rapur2018automation} proposed an intelligent detection method based on SVM to identify mechanical fault and flow rate using combination of motor line current and pump vibration signal as input. Shervani-abar. \cite{shervani2018cavitation} proposed a multi-class cavitation detection method based on the vibration signal of the axial flow pump using SVM. 

Zouari et al. \cite{zouari2004fault} proposed a vibration signal fault detection method for centrifugal pumps using neural network and neuro-fuzzy technology. Rajakarunakaran et al. \cite{rajakarunakaran2008artificial} proposed a centrifugal pump fault detection using a feedforward neural network and a binary adaptive resonance network (ART1). Siano et al. \cite{siano2018diagnostic} proposed a method combining ANN and nonlinear regression to diagnose cavitation of time domain vibration signals. Nasiri et al. \cite{nasiri2011vibration} extracted features from the vibration signal of the centrifugal pump as the input of the neural network to identify cavitation. Jia et al. \cite{jia2016deep} proposed a deep neural network (DNN) to directly extract features from the original rolling element bearings and planetary gearboxes data set for fault diagnosis. Zhao et al. \cite{zhao2016fault} proposed a deep learning method to extract features from non-stationary vibration signals and diagnose centrifugal pump faults. Tiwari et al. \cite{tiwari2021blockage} extracted 6 statistical features from the time domain pressure data, and then fed these features as input into the neural network to classify blockage and cavitation. Potocnik et al. \cite{potovcnik2021condition} extracted spectral and psychoacoustic features from the valve acoustic data and then took these features to be input of a variety of machine learning algorithms to classify the cavitation, flow noise, whistling and rattling. Muralidharan et al. \cite{muralidharan2012comparative} presented the use of Naïve Bayes algorithm and Bayes net algorithm for fault diagnosis through discrete wavelet features extracted from vibration signals of good and faulty conditions of the components of centrifugal pump. Liu et al. \cite{liu2015fault} proposed a fault diagnosis method for gear pumps based on the ensemble empirical mode decomposition (EEMD) and Bayesian networks. It decomposes the vibration signal by the EEMD method and calculates the energy of the intrinsic mode functions (IMFs) as fault features and adds these features to the Bayesian network. Tang et al. \cite{tang2022intelligent} studied intelligent fault diagnosis of hydraulic piston pumps based on deep learning and Bayesian optimization, where the Bayesian Optimization (BO) algorithm helps the neural network to automatically select hyperparameters.

It can be found from the literature review that many traditional machine learning methods (like, decision tree, SVM or ANN with shallow architectures) have been explored for cavitation or fault detection. However, the above methods suffer from two main weaknesses. First, the shallow network architecture can not capture sensitive features, especially when the acoustic signals are contaminated by background noise. Second, manual feature extraction and selection are time-consuming and require domain specific knowledge \cite{qin2020novel} which also might introduce bias. Fortunately, state-of-the-art deep learning methods have strong ability in automatically extracting relevant features from original data. This advantage enable the application of deep learning to be transferred and inspired from speech recognition and computer vision to cavitation detection and diagnosis. 

Deep convolutional neural networks (DCNN) have been successfully applied in image classification and also natural language processing (NLP) \cite{gu2018recent}. DCNN integrate the extraction of low, medium and high level features and classifiers in an end-to-end hierarchical fashion. The ``level" of features can be achieved by the number of layers (depth) in the stack. The depth of convolutional neural networks (CNN) has gradually increased with architecture evolved from AlexNet \cite{krizhevsky2012imagenet}, VGG \cite{simonyan2014very}, Inception \cite{szegedy2015going} to Residual networks (ResNet) \cite{he2016deep}. Recently, CNNs with 1-D convolutional layers have been widely applied on time series data, such as, electrocardiogram detection \cite{kiranyaz2015real}, electroencephalogram diagnosis \cite{sharan2020epileptic}, bearing fault diagnosis \cite{ince2016real,zhang2018deep,jia2018deep} and so on. The used 1-D convolution can acquire sensitive features from local segments of the entire time series data which are uncorrelated with each other \cite{ince2016real,zhang2018deep,jia2018deep,ronao2016human,kiranyaz2015real,sharan2020epileptic}. Therefore, time series data analysis can take 1-D CNN for efficient feature representation.

\begin{figure*}
    \centering
     \includegraphics[width=1\textwidth,height=50mm]{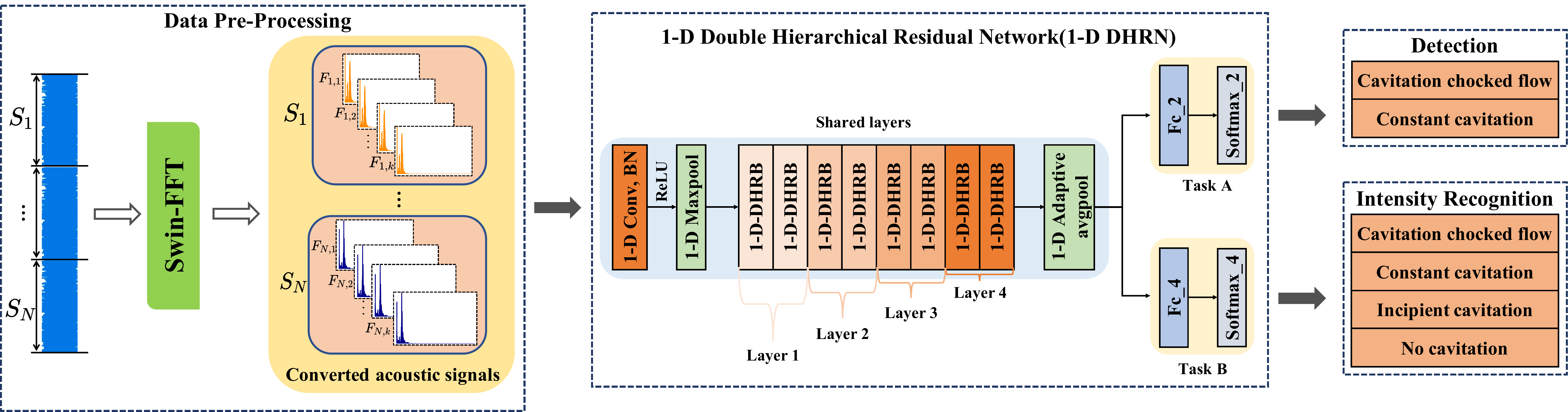}
    \caption{The framework of multi-task learning for cavitation detection and cavitation intensity recognition of valves acoustic signals using 1-D double hierarchical residual network. Task A is cavitation detection ("cavitation" and "non cavitation"). Task B is cavitation intensity recognition ("cavitation choked flow", "constant cavitation", "incipient cavitation" and "non cavitation").}
    \label{fig:1-D-DHRN}
\end{figure*}

In the literature, 1-D convolution has been successfully applied to do fault diagnosis. However, 
it's not yet clear how it works in cavitation detection and cavitation intensity recognition of valves, and under a combined task-demanding, which in practice is more required. Also, the modern residual structure was not discussed in before within this area.
In this study, targeted at a simultaneous multi-task learning, we introduce a special 1-D CNN based model that exploits 1-D double hierarchical residual blocks (1-D DHRB) as feature extractors. This novel method is termed as 1-D double hierarchical residual network (1-D DHRN). The architecture of our multi-task learning for cavitation detection and cavitation intensity recognition using 1-D DHRN is shown in Figure \ref{fig:1-D-DHRN}. Our 1-D DHRN is inspired by the ideas that, respective filed \cite{luo2016understanding} and residual structure \cite{he2016deep} with large enough perspective areas enable acquisition on more detailed information, and, the skip connection can increase the network non-linearity capability. Compared to our previous research \cite{SHA2022110897}, 1-D DHRN is a multi-task and end-to-end approach for cavitation detection and cavitation intensity recognition based upon the modern deep learning. In contrast to manual feature extraction that has to be performed in our previous research, the proposed 1-D DHRN here is able to capture sensitive features by itself and does not require specific knowledge, it thus allows the simultaneous cavitation detection and cavitation intensity recognition. Moreover, our proposed approach is more compatible with the requirements of the real-world industry. The main contributions of this work are summarized in follows:
\begin{itemize}
    \item In order to tackle the small-sample problem, the sliding window (Swin-FFT) data augmentation method is introduced in the study.
    \item The 1-D Double Hierarchical Residual Blocks (1-D DHRB) with large kernel are proposed as an automatic feature extractor to capture sensitive features of valve acoustic signals.
    \item The 1-D Double Hierarchical Residual Network (1-D DHRN) is developed for valve cavitation detection and cavitation intensity recognition using directly the emitted acoustic signal.
    \item Multi-task learning for simultaneous cavitation detection and cavitation intensity recognition is presented.
    \item The proposed method is tested on three different datasets of acoustic signals measured by SAMSON AG and compared with the state-of-the-art methods.
    \item The impact of different sampling rate of the acoustic signals on our proposed method for cavitation detection and cavitation intensity recognition is investigated.
\end{itemize}

The remainder of this paper is organized as follows. Section \ref{sec:2-Experimental setup} introduces the acquisition of datasets for valve cavitation acoustic signals. Section \ref{sec:3-methods} describes the details of the proposed 1-D DHRN model. Section \ref{sec:4-Experiments} introduces different case studies for each specific task. Section \ref{sec:5-Disscussions} discusses the effect of different experimental conditions on 1-D DHRN. The conclusion and future research outlook are given in Section \ref{sec:6-Conclusion}.

\section{Experimental Setup}
\label{sec:2-Experimental setup}
SAMSON AG devised a test rack with control valve (SAMSON AG type 3241, DN 80, PN40, Kvs 25 with positioner type 3730-6) and running water as the flowing medium inside to generate different flow status by gauging operation conditions accordingly: upstream pressure, downstream pressure and the valve stroke. The test bench is equipped with a set of sensors to measure the temperature of the test medium $T$, the pressures upstream ${p_1}$ and downstream ${p_2}$ of the test valve and the volumetric flow rate $Q$. Additionally, for the test valve mounted inside the bench, the absolute valve stroke $h$ and the sound intensity ${L_p}$ with a special sensor directly mounted on the valve body are measured. Furthermore, the test bench includes two additional control valves upstream and downstream of the test valve. A control system controls the pumps as well as these two valves to modify the total volumetric flow through the test bench. The two additional valves are used to vary the upstream and downstream pressure around the test valve. All sensors and the test valve are mounted between these two external control valves (more details see Table \ref{tab:sensors}). The pipes between the external valves and the test valve are long enough to ensure an undisturbed flow for proper measurements. Water, which can also be heated, was used as the test medium for all tests. The water temperature was maintained at 25 – 40$^{\circ}$C in order to hold the vapor pressure nearly constant and to eliminate the influence of temperature on the cavitation.

\begin{figure*}
    \centering
    \includegraphics[width=0.9\textwidth,height=50mm]{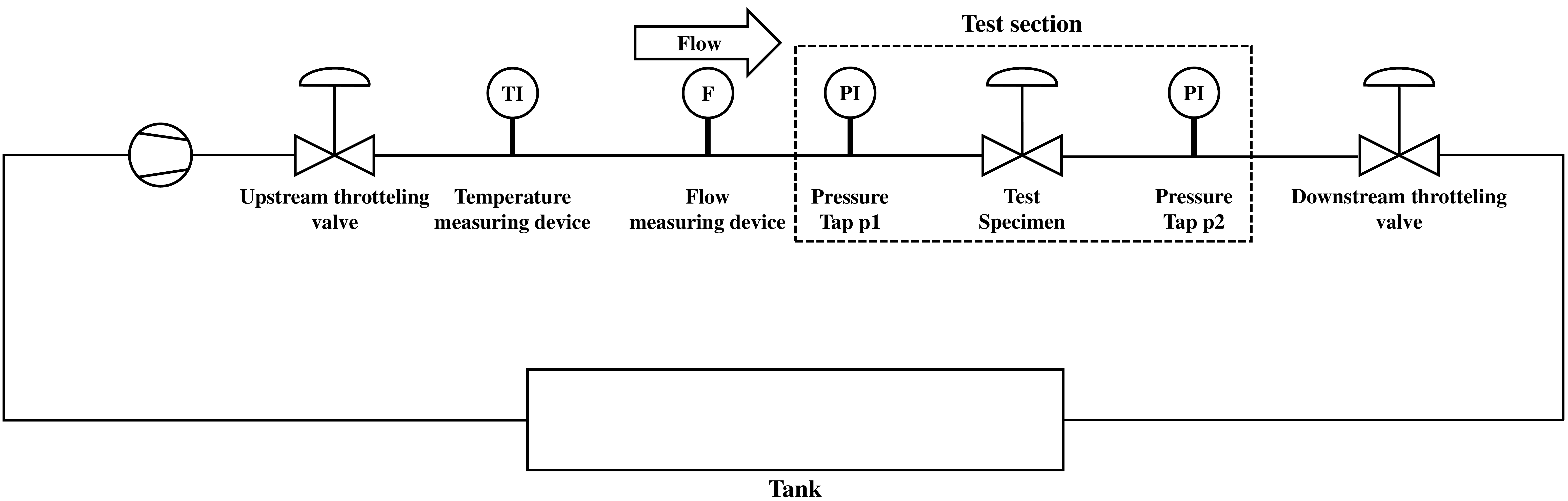}
    \caption{Schematic view of the test rack at SAMSON AG (Figure provided by SAMSON AG).}
    \label{fig:system}
\end{figure*}

Two piezo elements were introduced to record directly the structure-borne noise of the valve: one placed on the body of the control valve and the other to the NAMUR at the bonnet. Furthermore, two microphones were placed in a distance of 1m to the control valve to record the airborne noise in a frequency range between 40 Hz to 20 kHz: one high-end microphone as well as a low cost microphone which typically is used for smartphones. The test section in figure \ref{fig:system} of is in a separate room and all other noise sources from the plant like the pump and the upstream and downstream throttling valves are in another room (below the best section room, in the cellar). Therefore, there is a spatial separation of the test section to all possible noise sources. Also bellow expansion joints are integrated in the pipe system in order to the transport of vibrations into the test section. All these measures ensure that a surrounding noise of maximum 55 dB were present during the tests. The preparation of the different states is very simple. The principle procedure is described in the standard IEC 60534-8-2. The considered states depend on the differential pressure ratio ($({p}_{1}-{p}_{2})/({p}_{1}-{p}_{v})$) of the valve. Therefore, the valve opening will be held constant during the test. The temperature of the medium is during the test constant, which means the vapor pressure will be held constant. Furthermore, the upstream pressure of the valve will also be held constant (e.g. 10 bar (a)) with the process control system of the plant. Therefore, in the consequence the different states will be adjusted by varying the downstream pressure in steps of 0.1,…,0.4 bar starting at 1bar (a). For each step the downstream pressure will be adjusted with the process control system of the plant and based on the steady state of the flow the noise measurement concerning the adjusted step will be started. As a result, there is for a constant valve opening and a constant valve upstream pressure a noise emission characteristic (noise emission versus differential pressure ratio). And based on the noise emission characteristic the different states will be determined. This extensive procedure for determination of cavitation states cannot be realized in customers plants.
\begin{table*}[htbp]
\caption{The detailed information of various measurement sensors}
\label{tab:sensors}
\centering
\begin{tabular}{llcc}
\toprule
Physical parameters                                                                   & Sensor types       & Measuring ranges & Tolerances \\ 
 \midrule
Upstream pressure                                                                     & SAMSON Type 6054   & 0-16 bar          & ± 0.09 bar \\
Downstream pressure                                                                   & SAMSON Type 6054   & 0-16 bar         & ± 0.09 bar \\
Flow rate                                                                             & Krohne Type M950   & 0-180 m³/h       & ± 0.9 m³/h \\
Valve stroke                                                                          & Sylvac Type s229   & 0-50 mm          & ± 0.001 mm \\
Medium temperature                                                                    & SAMSON Type 5204   & -20-150 $^{\circ}$C       & ± 0.45 $^{\circ}$C  \\
\begin{tabular}[c]{@{}l@{}}Structure-borne noise \\ (at valve body)\end{tabular}      & Vallen Type VS45-H & 20-450 kHz       & \begin{tabular}[c]{@{}c@{}}Sensitivity -63 dB  re 1V/µbar\\ (Accuracy according to individual calibration)\end{tabular} \\
\begin{tabular}[c]{@{}l@{}}Structure-borne noise \\ (at valve body)\end{tabular}      & PCB Type M353 B17  & 0-30000 Hz       & Sensitivity 10 mV/g, Accuracy ± 3 dB    \\
\begin{tabular}[c]{@{}l@{}}Structure-borne noise \\ (at NAMUR at bonnet)\end{tabular} & PCB Type M353 B17  & 0-30000 Hz       & Sensitivity 10 mV/g, Accuracy ± 3 dB    \\ 
\bottomrule
\end{tabular}
\end{table*}

The cavitation we are identifying here takes place at control valves and the fluid inside is water. Cavitation takes place local at the narrowest throttling area, where the pressure falls to its minimum in the system, below the vapor pressure. Then, vapor bubbles are formed up in the liquid. These vapor bubbles implode when subjected to higher pressures with intensive sound pressure level. Depending on cavitation intensity erosion effects in the valve itself can occur additionally. Within SAMSON AG experimental setup in the laboratory, differential pressure at various constant upstream pressures of the control valve is varied, also different operation conditions (e.g. valve opening ratio, temperature) were adjusted, by which, five flow status are induced with corresponding acoustic signal measured for many events: cavitation choked flow, constant cavitation, incipient cavitation, turbulent flow and background no-flow. The turbulent flow is a flow through the control valve without any cavitation noise. Starting at a certain differential pressure ratio within a certain range only a few vapor bubbles will be generated and the implosion of these bubbles is causing an increase of the noise emission, which is defined as incipient cavitation. By increasing the concentration of vapor bubbles, the noise emission is increasing up to a maximum noise which is defined as constant cavitation. Achieving the noise maximum also the concentration of the vapor bubbles is maximum which leads to a choked flow condition with cavitation and a decreasing noise behavior.

\begin{figure*}
\centering
\subfigure[Cavitation Choked Flow]{
\begin{minipage}[t]{0.45\linewidth}
\centering
\includegraphics[width=\textwidth,height=35mm]{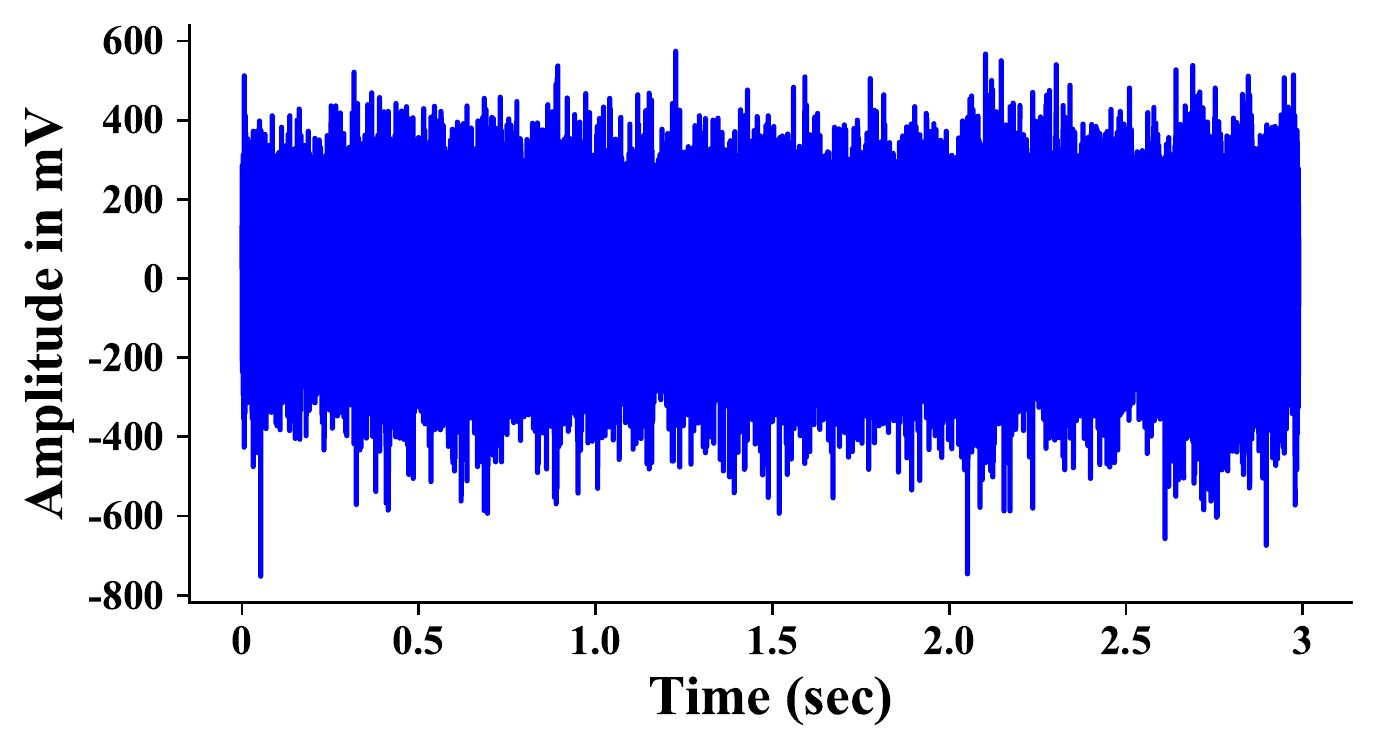}
\end{minipage}%
}%
\subfigure[Constant Cavitation]{
\begin{minipage}[t]{0.45\linewidth}
\centering
\includegraphics[width=\textwidth,height=35mm]{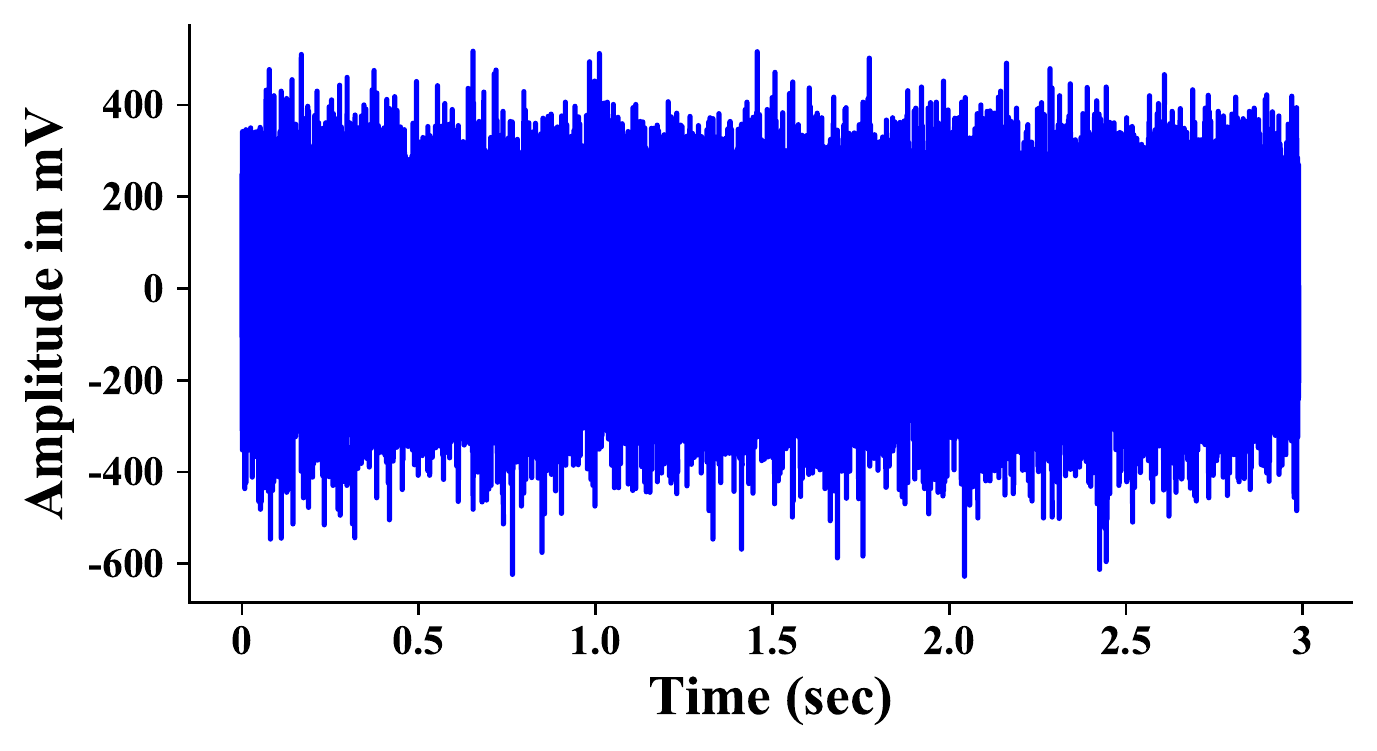}
\end{minipage}%
}%

\subfigure[Incipient Cavitation]{
\begin{minipage}[t]{0.45\linewidth}
\centering
\includegraphics[width=\textwidth,height=35mm]{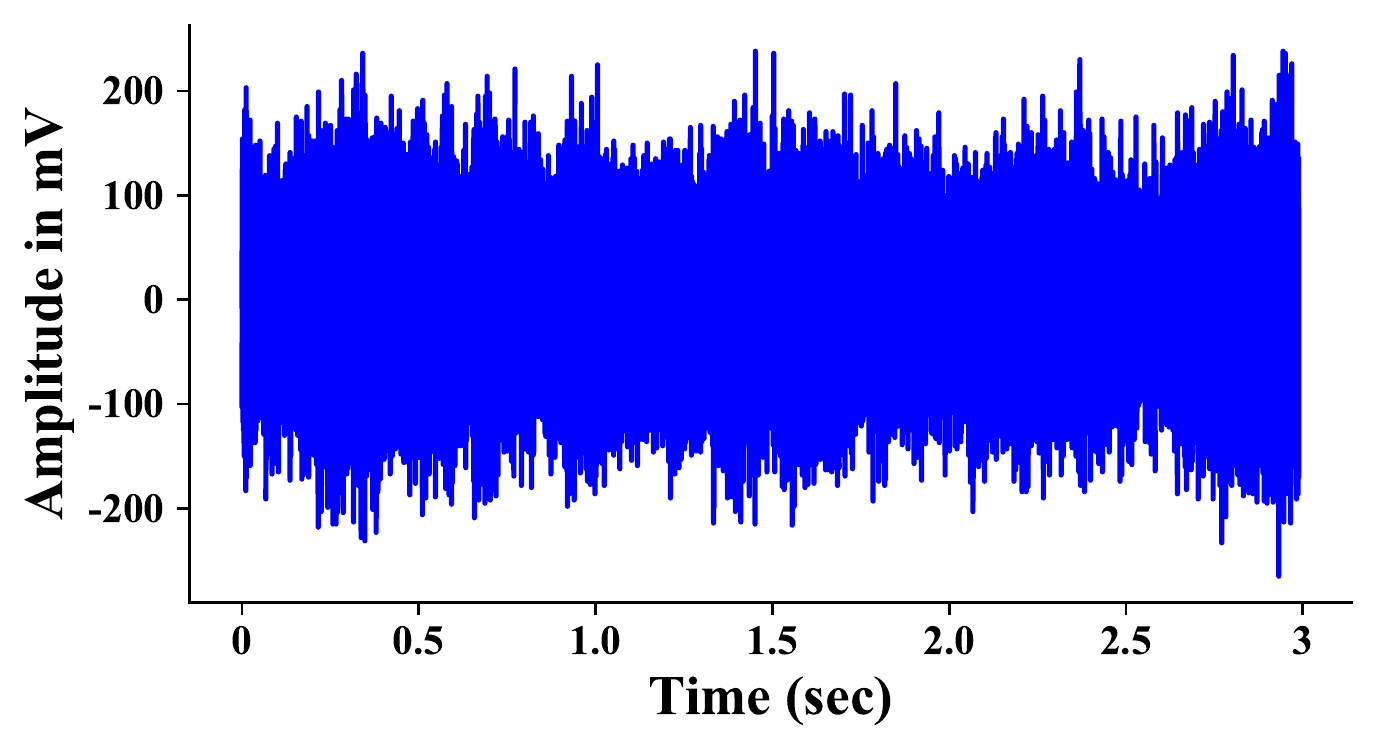}
\end{minipage}%
}%
\subfigure[Non cavitation]{
\begin{minipage}[t]{0.45\linewidth}
\centering
\includegraphics[width=\textwidth,height=35mm]{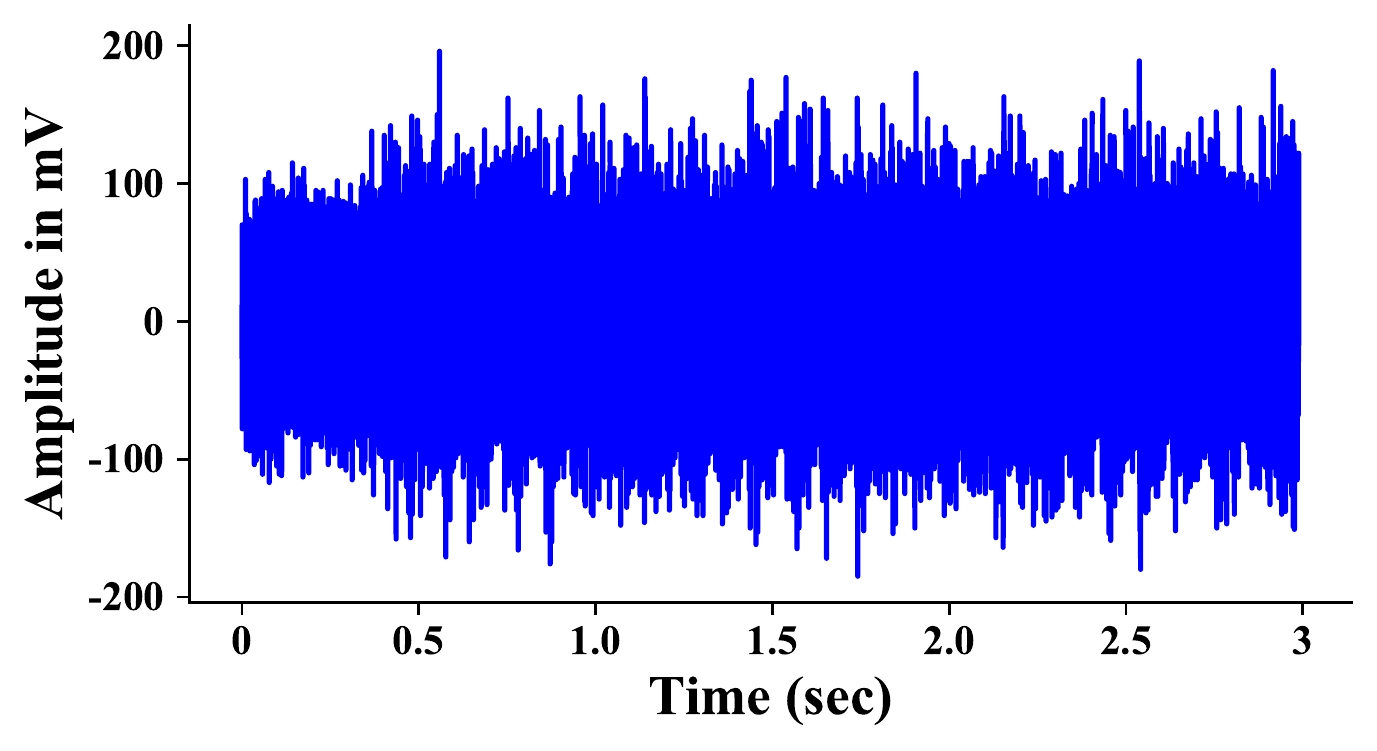}
\end{minipage}%
}%
\centering
\caption{Waveforms of acoustic signals for cavitation choked flow (a), constant cavitation (b), incipient cavitation (c) and non-cavitation (d), respectively.}
\label{fig:datashape}
\end{figure*}

Figure \ref{fig:datashape} shows an exemplary comparison of time waveformsw of the acoustic signals for cavitation choked flow, constant cavitation incipient and non cavitation. It is difficult to distinguish the sensitive features directly in Figure \ref{fig:datashape}. Therefore, in this study an intelligent 1-D DHRN method is used to extract the sensitive features from the raw acoustic signals and to recognize the levels of cavitation intensity.

\section{Methods}
\label{sec:3-methods}
This section presents the multi-task learning framework of cavitation detection and cavitation intensity recognition using 1-D DHRN, including data augmentation method Swin-FFT, 1-D DHRB construction, and the final 1-D DHRN structures.

\subsection{Architecture Overview}
The overview architecture of our multi-task learning framework of cavitation detection and cavitation intensity recognition using 1-D double hierarchical residual networks (1-D DHRN) is illustrated in Figure \ref{fig:1-D-DHRN}. First, the Swin-FFT (introduced in subsection \ref{data augmentation}) method is used for data augmentation to handle the small-sample issue in the data pre-processing module.  Before data augmentation, the total data is split into a training set and testing set (with a ratio of $80\%$ : $20\%$). Performing the train/test splitting in advance can ensure that a piece of signal data after any data augmentation would only exit in the training set or the testing set. Then, the pieces of the frequency domain acoustic signals after Swin-FFT are fed as the input of the 1-D DHRN (introduced in subsection \ref{1-D double hierarchical residual networks}). Finally, the results of cavitation detection and cavitation intensity recognition on the valve are achieved simultaneously. And the Swin-FFT, 1-D pooling layers (max pooling and adaptive average pooling) do not introduce additional parameters. Only our convolutional and fully concatenated layers introduce additional parameters. 

In our method, cavitation detection is taken as Task A and cavitation intensity recognition as Task B. Cavitation intensity recognition and cavitation detection correspond to different practical industrial applications. The cavitation detection and cavitation intensity recognition apply the mechanism of hard parameter sharing in multi-task learning \cite{ruder2017overview}. The two tasks share features extracted from the hidden layer. Then, the two tasks can be performed through different output layers. The essence of our proposed method is to minimize the cavitation detection error and the cavitation intensity recognition error, respectively. In other words, we transform data-driven multi-task supervised learning into an optimization problem, to achieve the minima of the two objective functions of cavitation detection $\mathcal{L'}$ and cavitation intensity recognition $\mathcal{L''}$ by choosing appropriate hyperparameters, as follows.
\begin{equation}
\mathcal{L'} = CE(f({X_{train}};{\theta'} ),{{Y'}_{train}})
\end{equation}

\begin{equation}
\mathcal{L''} = CE(f({X_{train}};{\theta''} ),{{Y''}_{train}})
\end{equation}
where, $CE(\cdot )$ is the cross-entropy function. The basic process of different optimization methods is similar, and we choose the Adam optimization method. First, the output of $f’=f({X_{train}};{\theta'} )$, $f’'=f({X_{train}};{\theta''} )$ and the optimization objectives $\mathcal{L'}$, $\mathcal{L''}$ of the model are computed using the initial parameters. The network parameters $\theta'$ and $\theta''$ are then tuned to decrease the objective functions from the final layer to the first layer. This process is repeated until the proper model and a small fit error, i.e. loss function value. And the proposed 1-D DHRN is summarized in Algorithm \ref{algo: 1D DHRN multi-task}.
\begin{algorithm}[htbp]
	\caption{Pseudocode of 1-D DHRN}
	\label{algo: 1D DHRN multi-task}
	\begin{algorithmic}
    \Require Sub-sequences $({X_{train}},({{Y'}_{train}},{{Y''}_{train}}))$ of the training set, Sub-sequences $({X_{val}},({{Y'}_{val}},{{Y''}_{val}}))$ of the validation set, learining rate: 0.0001, batch size: 4
    \Ensure results of cavitation detection and cavitation intensity recognition
    \Repeat
    \State Random sorting of sub-sequences in training set $({X_{train}},{{Y'}_{train}},{{Y''}_{train}})$.
    \State Random initialization of 1D DHRN weights $\theta$.
    \For{epoch $= 0,1,\ldots,N$}
    \State Select sub-sequences of specified batch size from training set $({X_{train}},({{Y'}_{train}},{{Y''}_{train}}))$.
    \State The forward propagation algorithm calculates the net input $z(l)$ and activation value $\sigma(l)$ for each layer of the 1D DHRN.
    \State Calculating the loss function: 
    \State $\mathcal{L'} = CE(f({X_{train}};{\theta'} ),{{Y'}_{train}})$ $\rhd$ cavitation detection loss  
    \State $\mathcal{L''} = CE(f({X_{train}};{\theta''} ),{{Y''}_{train}})$ $\rhd$ cavitation intensity recognition loss
    \State Update parameters of cavitation detection and cavitation intensity recognition:
    \State min $CE(f({X_{train}};\theta ),{{Y'}_{train}})$ $\Rightarrow$ ${{\partial L'} \mathord{\left/{\vphantom {{\partial L'} {\partial \theta '}}} \right.\kern-\nulldelimiterspace} {\partial \theta '}}$
    \State min $CE(f({X_{train}};\theta ),{{Y''}_{train}})$ $\Rightarrow$ ${{\partial L''} \mathord{\left/{\vphantom {{\partial L'} {\partial \theta '}}} \right.\kern-\nulldelimiterspace} {\partial \theta ''}}$
    \State Save parameters of cavitation detection and cavitation intensity recognition in current epoch.
    \EndFor
    \Until 
    \State 1D DHRN model on validation set for $\mathcal{L'}({X_{train}},{{Y'}_{train}})$ and $\mathcal{L''}({X_{train}},{{Y''}_{train}})$ not falling.
    \For{i $= 0,1,\ldots,len(({X_{test}},({{Y'}_{test}},{{Y''}_{test}})))$}
    \State The output of cavitation detection and cavitation intensity recognition:
    \State $f'({X_{test}})$ $\rhd$ cavitation detection
    \State $f''({X_{test}})$ $\rhd$ cavitation intensity recognition
    \State Calculate the accuracy of cavitation detection and cavitation intensity recognition:
    \State $Acc(f'({X_{test}}),{{Y'}_{test}})$ and $Acc(f''({X_{test}}),{{Y''}_{test}})$  
    \EndFor
	\end{algorithmic}
\end{algorithm}

\subsection{Data Augmentation}
\label{data augmentation}
In general, machine learning is driven by big data\cite{l2017machine}. However, our \emph{\textbf{Dataset 1}}, \emph{\textbf{Dataset 2}} and \emph{\textbf{Dataset 3}} only have 356, 806 and 160 measured acoustic signals, respectively. So data augmentation is essential in handling this kind of problem with small-sample issue. Data augmentation can improve the performance of the model and prevent over-fitting, it can also bring into the model the desired invariance for the task and robustness for the extracted features\cite{shorten2019survey, Pang:2016vdc}. 

\begin{figure}[htbp]
    \centering
     \includegraphics[width=0.45\textwidth,height=50mm]{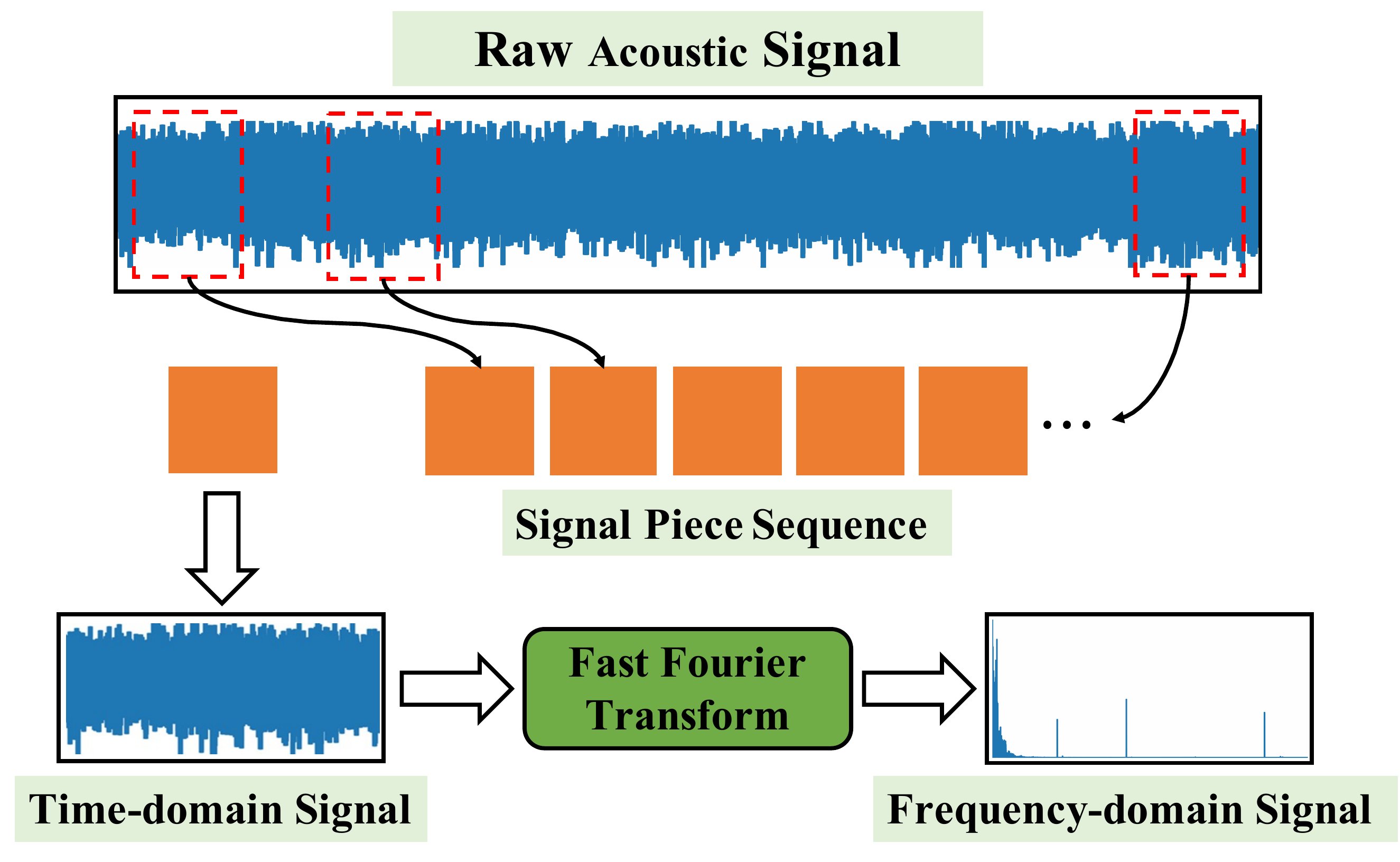}
    \caption{Data augmentation based on sliding window with fast fourier transform (SWin-FFT).}
    \label{fig:sliding-window}
\end{figure}

Considering the purposed maintaining for steady flow status (i.e. for each individual measurement it is the same fluid status class within 3 s or 25 s record duration) in each recorded data sample and the fine resolution for the sensor, one can actually split every sample into several pieces with each still holding enough essential information to decipher the flow status, but also with independent characteristics per piece due to the intrinsic randomness of the noise emission-given the piece is not so short. Therefore, we propose here a data augmentation method based on sliding window with fast Fourier transform (Swin-FFT), see Figure \ref{fig:sliding-window}. The method is divided into two steps:
\begin{enumerate}[Step 1:]
\item The signal data is split by a sliding window.
\item The time-domain data is transformed into frequency-domain by fast Fourier transform (FFT).
\end{enumerate}

The Swin-FFT contains only one parameter, which is the size of window ${W_{size}}$. In our following study, the value of ${W_{size}}$ will be analyzed and determined by experiments, which can reduce the effect of experts’ bias on cavitation intensity recognition and cavitation detection (see section \ref{sec:4-Experiments}) as much as possible.

\subsection{1-D Double Hierarchical Residual Networks}
\label{1-D double hierarchical residual networks}
\subsubsection{Comparison Between 1-D and 2-D Convolution}
\begin{figure*}
    \centering
     \includegraphics[width=0.95\textwidth,height=70mm]{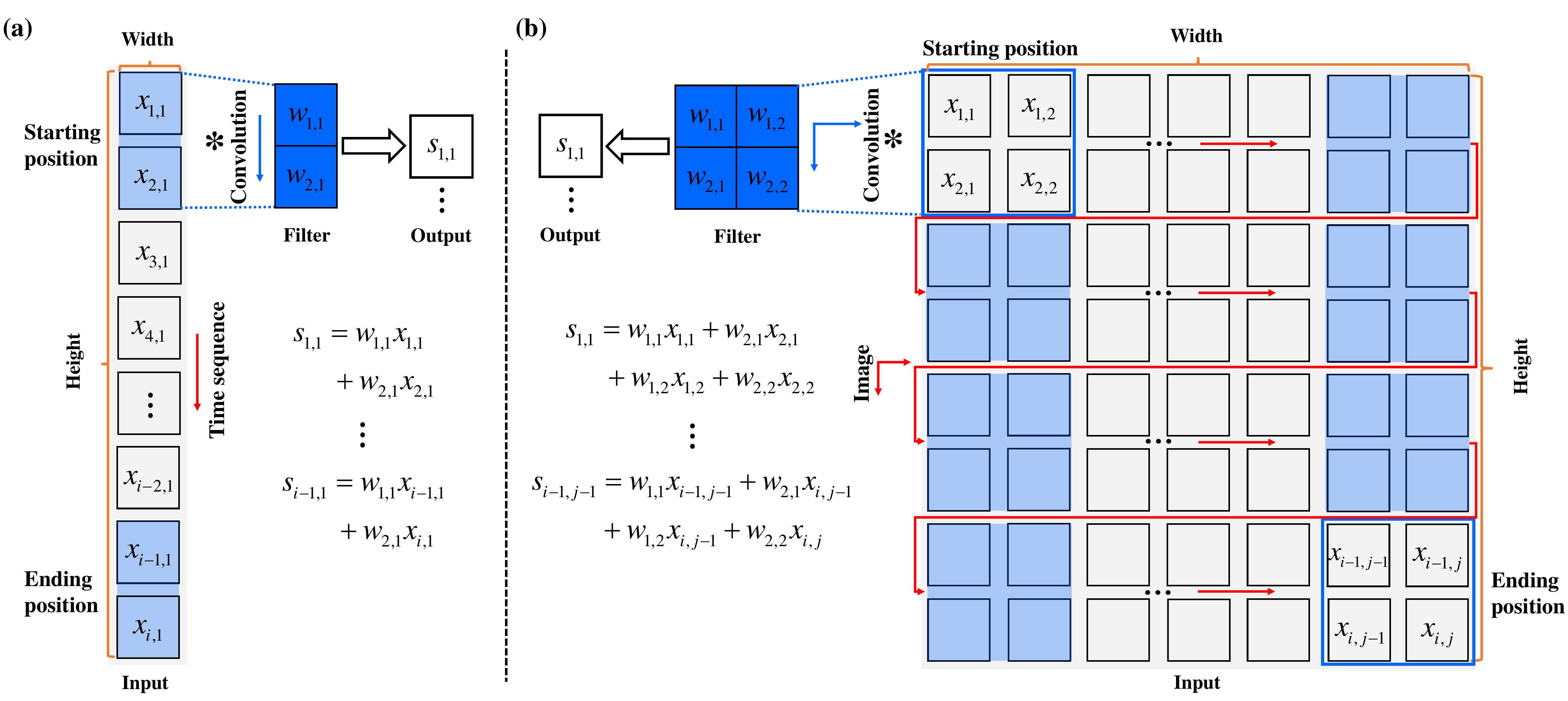}
    \caption{Comparison of working principle between 1-D and 2-D convolution. Figure (a) and Figure (b) show the working principle of 1-D and 2-D convolution, respectively.}
    \label{fig:1-D and 2-D conv working principle}
\end{figure*}

1-D convolution and 2-D convolution are very similar in structure, except that the filters are slid in a different way. For 1-D convolution operation, the convolution kernel slides towards only one direction, i.e. weighted summation in the width or height direction. In 2-D convolution, the convolution kernel slides into both horizontal and vertical directions, i.e. weighting and summing in both the width and height directions. Figure \ref{fig:1-D and 2-D conv working principle} compares the different sliding manners of the filters for 1-D and 2-D convolution operations.
\begin{figure}
    \centering
     \includegraphics[width=0.45\textwidth,height=40mm]{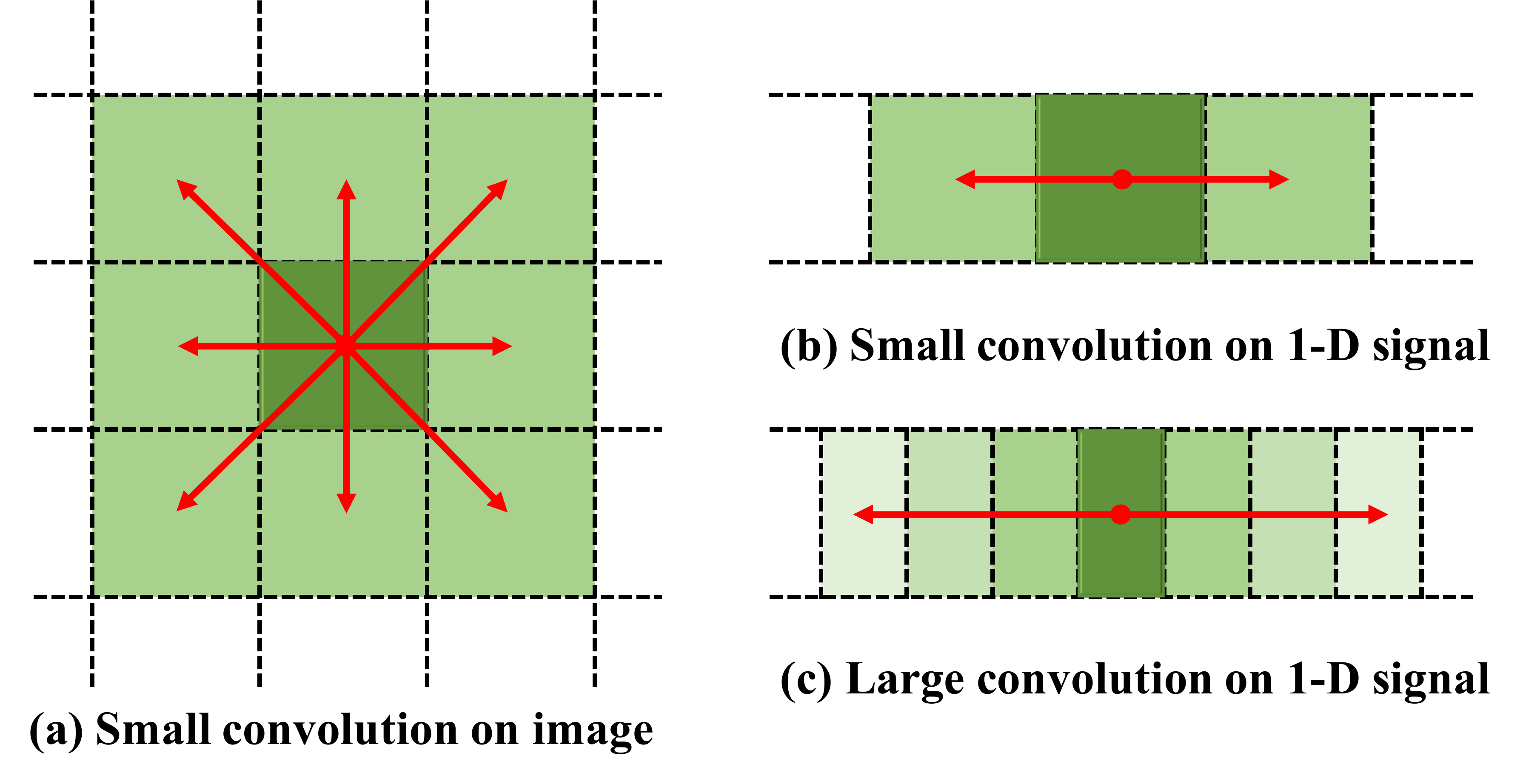}
    \caption{Comparison of small convolution kernel and large convolution kernel in catching information on signals and images. Figure (a) shows that a small convolution kernel can obtain neighbourhood information from eight directions for one pixel on images. Figures (b) and (c) show the ability of the small and large convolution kernels to capture neighbourhood information for one data point on the signal, respectively.}
    \label{fig:Small vs Large conv kernel}
\end{figure}
\begin{figure*}
    \centering
     \includegraphics[width=0.95\textwidth,height=45mm]{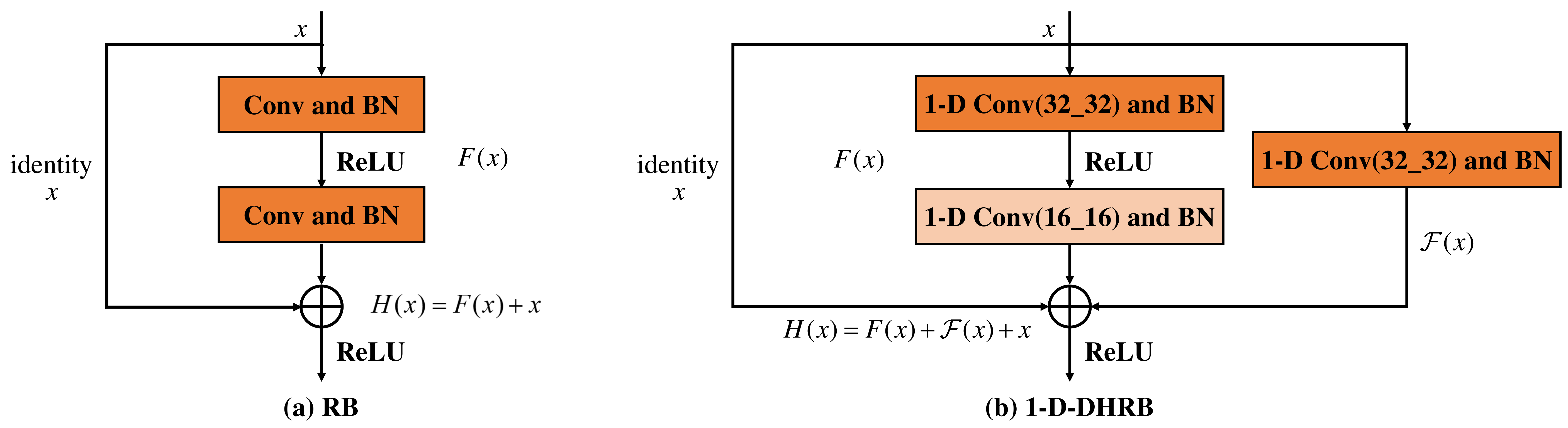}
    \caption{Comparison of structure the RB and 1-D DHRB. Figure (a) shows traditional residual block. Figure (b) is the proposed the 1-D double hierarchical residual block (1-D DHRB).}
    \label{fig:1-D-DHRB}
\end{figure*}

In the case of 1-D time series data as shown in Figure \ref{fig:1-D and 2-D conv working principle}(a), each element represents the original acoustic signal point measured by the sensor at each moment in time, and the column represents the original signal from one sensor or one channel with time directed downwards in the sketch. In general, the 1-D convolution kernel only slides vertically along the time axis (height) to extract features. The size of the 1-D convolution kernel determines the size of the receptive field for the filter, i.e. the number of samples (moments) to be calculated by the convolution operation. The stride of the 1-D convolution kernel is defined as the distance moved at each step on the time series signal, and it presents the accuracy of the extraction. In the case of 2-D array image-like data, the pixels of an image consist of a number of small blocks, the positions of which are presented by $x$ and $y$ direction (Figure \ref{fig:1-D and 2-D conv working principle}(b)). The width and height of the image is the sum of all x and y, respectively. The size of the 2-D convolution kernel, i.e. its width and height, determines the scope of the convolution operation at each step. Similarly, the stride of the 2-D convolution kernel is defined as the distance moved in each step over the width or height of the 2-D array (gray scale image). In general, the 2-D convolution is applied to the 2-D array or gray scale image in the order of left to right and top to bottom.

Regardless of whether it is a 1-D or 2-D convolution kernel, they all have the same and significant hyperparameters, such as the convolution kernel size and stride size, etc. They share weights when sliding to extract local features. However, the convolution with different kernel sizes have different roles on time series signals and images (see Figure \ref{fig:Small vs Large conv kernel}). In general, a small convolution kernel ($3\times 3$) can obtain information on eight neighbourhood directions for one pixel on an image, but it can obtain information on only two neighbourhood directions for one data point on a signal. To solve this problem, we use 1-D DHRB (introduced in \ref{1-D double hierarchical residual blocks}) with a large kernel on one-dimensional signals to obtain more neighbourhood information, although the information still comes from only the two neighbourhood directions.

\subsubsection{1-D Double Hierarchical Residual Block}
\label{1-D double hierarchical residual blocks}
Residual block (RB) is an important module of the residual networks (ResNet). RB is based on the idea of connecting blocks of convolutional layers by using skip connection, as shown in Figure \ref{fig:1-D-DHRB}(a). The output of the RB is $H(x) = F(x) + x$. RB changes the network learning objective from $H(x)$ to $H(x) - x$. This structure can help optimizing the trainable parameters in error backpropagation  to avoid the problem of vanishing and exploding gradients. Practically it can also enable the building of deeper network structures.

Motivated by above, we developed a one dimensional convolution-based double hierarchical residual block (1-D DHRB) to improve valve cavitation detection and intensity recognition. The 1-D DHRB consists of several convolutional layers of different filter size (Conv), batch normalization layers (BN), rectified linear unit (ReLU) activation function and two shortcuts, as shown in Figure \ref{fig:1-D-DHRB}(b). Both RB and 1-D DHRB have two Conv and BN layers. However, the size of two Conv layers of 1-D-DHRB are different and hierarchical compared to the traditional RB. The filter size of the first Conv layer is $32 \times 32$, and the second Conv layer has a filter size of $16 \times 16$ (The filter size has been determined through several experiments). 1-D DHRB uses large convolution kernels compared to RB, which can capture sensitive features of the acoustic signal with fewer layers of the network. The size of the first Conv layer is twice the size of the second Conv layer, which helps the network to focus more on the sensitive features of the feature map from the first Conv layer.

The 1-D DHRB adds one more shortcut compared to the traditional RB. One shortcut for 1-D DHRB is the identity $x$, the other shortcut for 1-D DHRB is a non-linear function ${\cal F}(x)$ of identity $x$ through one Conv and BN layer. The output of 1-D DHRB is given by concatenation as
\begin{equation}
\label{eq:1-D-DHRB}
H(x) = F(x) + {\cal F}(x) + x
\end{equation}

Several 1-D DHRBs of different kernel number are stacked after the first convolutional layer in 1-D DHRN. The 1-D DHRN has a depth of 18 and it consists of one 1-D convolutional layer, one maximum pooling layer, several 1-D DHRBs with different numbers of kernels, one global average pooling layer and fully connected layer. Table \ref{tab:1-D-HRN-parameters} shows the relevant parameters of the 1-D DHRN architecture. These layers enable 1-D DHRN to automatically extract features and classify cavitation and the levels of cavitation intensity.
\newcommand{\tabincell}[2]{\begin{tabular}{@{}#1@{}}#2\end{tabular}} 
\begin{table}
    \centering
    \caption{Parameters of the proposed 1-D DHRN architecture}
    \setlength{\tabcolsep}{1mm}{
     \begin{tabular}{llccc}
	\toprule 
	& No.& Layer type   & Kernel size/Stride & Kernel number   \\
	\midrule
	& 1  & 1-D Convolution& 32/1 & 64\\
	& 2  & Layer 1    & [32, 16]/1 & 64\\
	& 3  & Layer 2    & [32, 16]/2 & 128\\
	& 4  & Layer 3    & [32, 16]/2 & 256\\
	& 5  & Layer 4    & [32, 16]/2 & 512\\
	& 6  & Fully-connected    & 512 & 1\\
	& 7  & Softmax\_4  & 4 & 1\\
	& 8  & Softmax\_2  & 2 & 1\\
	\bottomrule
    \end{tabular}
    }
    \label{tab:1-D-HRN-parameters}
\end{table}
\begin{itemize}
    \item 1-D convolutional layer. The essence of the convolutional layer is a set of convolutional kernels. In this research, each convolutional layer has 64, 128, 256 and 512 convolutional kernels of sizes 32 and 16, respectively. In the one-dimensional forward propagation process, each filter convolves with the input vector to generate a corresponding output featured vector. For the $k$-th filter, the $j$-th of its output feature vector at layer $l$ is defined as follows.
    \begin{equation} 
    y_{j,k}^l = \sigma ({\bm{W}_{k}^l\ast \bm{X}_i^{l - 1} + b_{j,k}^l} )
   \end{equation}
\end{itemize}
where, $\sigma$ is the  activation function and the Rectified Linear Units (ReLU) is applied in this research, $\bm{X}_i^{l - 1}$ denotes the $i$-th local region at layer $l-1$, $\bm{W}_{k}^l$ and ${b_{j,k}^l}$ denote the weight vector and bias of the $k$-th filter kernel at layer $l$.
\begin{itemize}
    \item 1-D pooling layer. The pooling function uses the overall statistical features of the neighbouring outputs at a location to replace the network's output at the location. The pooling layer is usually added after the convolutional layer to perform the down-sampling operation. This operation can compress the feature vector and removes redundant information, and thus reducing the complexity of the network, decreasing computation and reducing memory consumption, etc. In this study, max pooling and adaptive average pooling are applied to different positions of the 1-D DHRN. The adaptive average pooling is a special pooling of average pooling and its output size is always $H$. The max pooling and average pooling are shown in Equation \eqref{eq:maxpooling} and Equation \eqref{eq:avgpooling}, respectively.
\end{itemize}
\begin{equation}
\label{eq:maxpooling}
    y_{i,k}^l = \mathop {\max }\limits_{(i - 1)H + 1 \le j \le iH} \{ y_{j,k}^{l - 1}\} 
\end{equation}
\begin{equation}
\label{eq:avgpooling}
    y_{i,k}^l = \frac{1}{S}\sum\limits_{m = 0}^{S - 1} {\{ y_{j,k}^{l - 1}\} } 
\end{equation}
where, ${y_{j,k}^{l - 1}}$ is the $j$-th element of the output feature vector of the $k$-th filter in layer $l-1$ and $H$ is its height, $y_{i,k}^l$ is the $i$-th element of the new feature vector after the down-sampling operation in layer $l$, $S$ is the filter size.
\begin{itemize}
    \item Fully connected layer. Each node of the fully connected layer is connected to all nodes of the previous layer and it can combine class-distinctive local information on convolutional or pooling layers. These information can be flattened into a vector and fed to the fully connected layer. The output ${\bm{y}^l}$ of the $l$-th fully connected layer is defined as follows.
    \begin{equation}
        {\bm{y}^l} = {({\bm{W}^l})^T}{y^{l - 1}} + {\bm{b}^l}
    \end{equation}
\end{itemize}
where, ${\bm{y}^{l - 1}}$, ${\bm{W}^l}$ and ${\bm{b}^l}$ are the input vector, weight matrix and bias vector of the fully connected layer at layer $l$, respectively.

\begin{itemize}
    \item Softmax layer. The softmax layer is also called the output layer. It is usually used to predict the multi-class classification. The softmax function is defined as follows.
    \begin{equation}
{p_{c}^{(n)}}=\frac{exp(\bm{\theta}_{c}^{T}{\bm{y}}^{(n),L-1})}{\sum_{c=1}^{C}exp(\bm{\theta}_{c}^{T}{\bm{y}}^{(n),L-1})}
    \end{equation}
\end{itemize}
where $L$ is the maximum of layer number, the superscript $n$ denotes that $n$-th sample, the subscript $c$ denotes the $c$-th class of total classification number $C$, ${\bm{y}}^{(n),L-1}$ is the output vector of the previous layer, and ${\bm{\theta}}_{c}$ is the weight vector for the $c$-th class.

For deep neural networks, the loss function is generally used to evaluate the degree of inconsistency between the predicted value of the model and the true value. It is a non-negative real-valued function, and the smaller the loss function, the better the robustness of the model. The cross-entropy loss function is the most popular loss function and is defined as follows.
\begin{equation}
{\cal L} =  - \frac{1}{N}\sum\limits_{n = 1}^N {\sum\limits_{c = 1}^C {sign({pred}^{(n)}=c)\log ({{p_{c}^{(n)}}})} } 
\end{equation}
where, $N$ is the total number of samples, $sign(\cdot )$ denotes the sign function returning 1 for real and 0 for fake. And ${pred}^{(n)}$ represents the prediction result for the $n$-th sample.

Neural networks are trained to determine the weights and biases of the network. In this research, Adam optimisation algorithm is used to update the weights and biases of the network during back propagation. The Adam optimisation algorithm \cite{kingma2014adam} is an extension of the stochastic gradient descent (SGD) \cite{bottou2012stochastic,bottou1998online} method and has been introduced in the literature \cite{kingma2014adam1,bock2018improvement,duchi2011adaptive}.

\section{Experiments}
\label{sec:4-Experiments}
\subsection{Overview}
In our experiments, first evaluation metrics are introduced. Then, \emph{\textbf{Dataset 1}}, \emph{\textbf{Dataset 2}} and \emph{\textbf{Dataset 3}} of the cavitation detection and cavitation intensity recognition are described. And we analyze the effect of the window size ${W_{size}}$ of Swin-FFT on the performance of 1-D DHRN. Finally, we compare the performance of 1-D DHRN and other state-of-the-art deep learning and machine learning models on \emph{\textbf{Dataset 1}}, \emph{\textbf{Dataset 2}} and \emph{\textbf{Dataset 3}}.

In all experiments, we use adaptive moment estimation (Adam) \cite{kingma2014adam} as our optimizer with learning rate set to be $1\times {10}^{-4}$. Both our model and the model of the comparison state-of-the-art method are trained on NVIDIA GPU and the number of training epochs is set to 100 on all datasets, upon which by monitoring the loss of validation in learning curve we observe saturation which gives convergence indication (see Appendix). These setting ensures that all trained models are more fairly analyzed for performance comparisons. In addition, both our method and the baseline methods are trained with the same data augmentation (Swin-FFT) in our implementation. Our comparison methods include five different approaches from both conventional machine learning and deep learning. Where, the SVM \cite{yang2005cavitation}, Decision Tree \cite{sakthivel2010vibration} and 1-D CNN \cite{shervani2018cavitation} are methods by other researchers for cavitation. The XGBoost + ASFE \cite{SHA2022110897} and 1-D ResNet-18 are our previous research and our benchmark, respectively. Our source code and data are released at \url{https://github.com/CavitationDetection/1-D-DHRN}.

\subsection{Evaluation Metric}
In order to evaluate the model after training, four metrics are chosen to comprehensively assess the model performance: Accuracy, Precision, Recall and F1-score. We first calculate the confusion matrix \cite{sokolova2009systematic} to more conveniently define the evaluation metrics and visualize model performance. In the confusion matrix shown in Figure \ref{fig:ConfusionMatrix}, each column represents the predicted class, and each row represents the actual class. TP (True Positive) is the fraction of positive samples those got correctly predicted by the model, and TN (True Negative) is for the correctly predicted negative samples. FP (False Positive) means the incorrectly classified positive samples those should be negative actually, and FN (False Negative) is for the incorrectly predicted negative samples.

Accuracy, Precision, Recall and F1-score are common evaluation metrics for classification problems, defined as:
\begin{figure}
    \centering
    \includegraphics[width=0.45\textwidth,height=45mm]{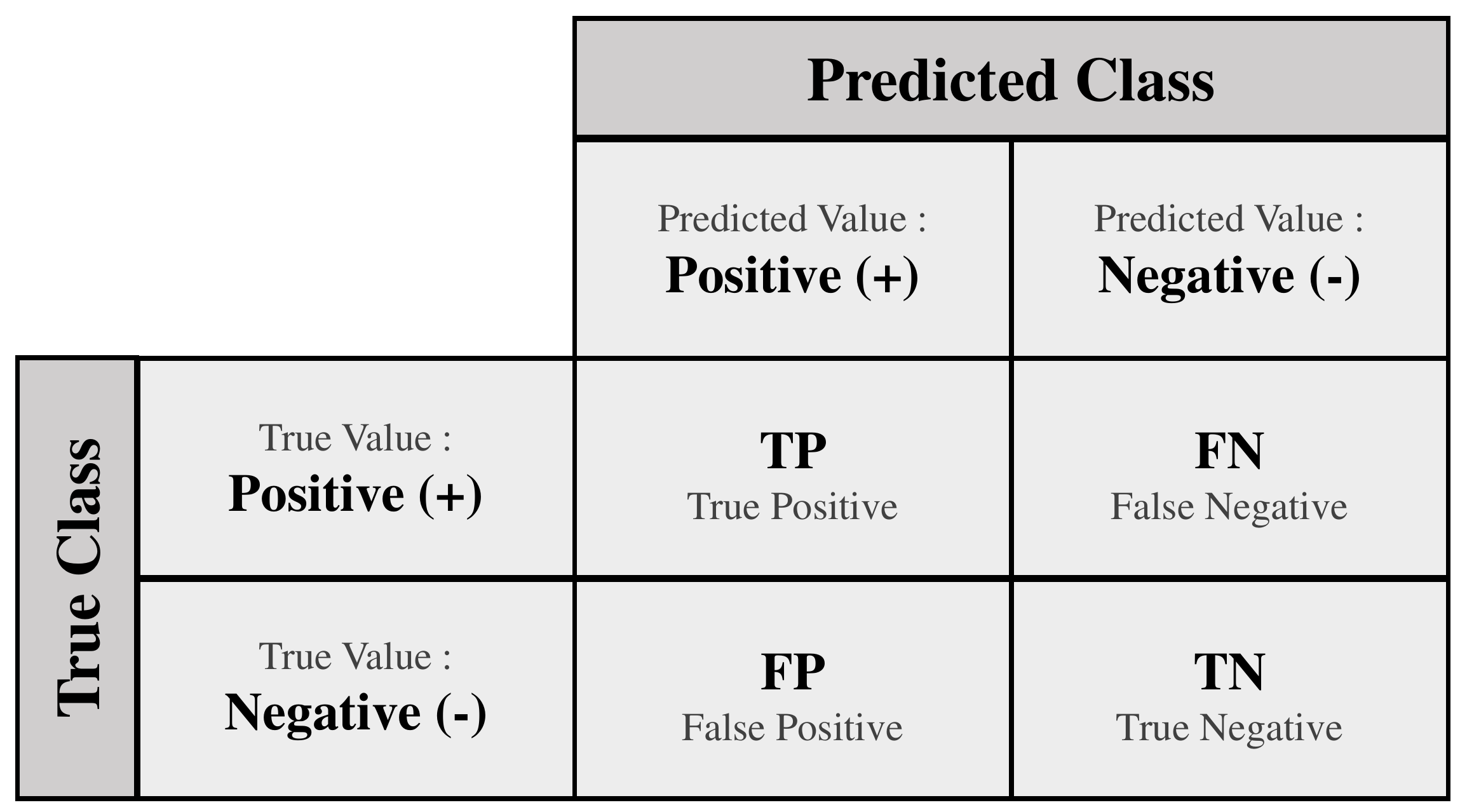}
    \caption{The Confusion Matrix.}
    \label{fig:ConfusionMatrix}
\end{figure}
\begin{equation}
Accuracy = \frac{{TP + TN}}{{TP + TN + FP + FN}}  
\end{equation}
\begin{equation}
Precision = \frac{{TP}}{{TP + FP}}
\end{equation}
\begin{equation}
{\mathop{\rm Re}\nolimits} call = \frac{{TP}}{{TP + FN}}
\end{equation}
\begin{equation}
F1-score = \frac{{2 \times Precision \times {\mathop{\rm Re}\nolimits} call}}{{Precision + {\mathop{\rm Re}\nolimits} call}}
\end{equation}

\subsection{Cavitation Intensity Recognition: Dataset 1}
\subsubsection{Data Description}
In this subsection, the proposed 1-D DHRN is evaluated on the condition monitoring of cavitation dataset (\emph{\textbf{Dataset 1}}) provided by SAMSON AG in Frankfurt. The hardware of this experiment is shown in Figure \ref{fig:system}, and five flow status are induced in the acoustic signal data by varying the differential pressure at various constant upstream pressures of the control valve different operation conditions: cavitation choked flow, constant cavitation, incipient cavitation, turbulent flow and background no-flow, as shown in Table \ref{tab:DataSet1-FlowStatus}. The turbulent flow and background no-flow are non-cavitation conditions. The experiments have been conducted using seven different valve strokes at four different upstream pressures and the operating parameters are shown in Table \ref{tab:DataSet1-OperationParameters}. \emph{\textbf{Dataset 1}} has a total of 356 acoustic signal samples and the frequency of samples is 1562.5 kHz within time duration of 3 s. It should be noted that the \emph{\textbf{Dataset 1}} has been measured by SAMSON AG in a professional environment. The details of the training set, validation set and test set for \emph{\textbf{Dataset 1}} without Swin-FFT is shown in Table \ref{tab: details of dataset1, dataset2 and dataset3}.
\begin{table*}[htbp]
\caption{The details of the training set, validation set and test set for \emph{\textbf{Dataset 1}}, \emph{\textbf{Dataset 2}} and \emph{\textbf{Dataset 3}} without Swin-FFT.}
\label{tab: details of dataset1, dataset2 and dataset3}
\footnotesize
\centering
	\setlength{\tabcolsep}{1mm}{
\begin{tabular}{lcccccccccccc}
\toprule
         & \multicolumn{4}{c}{Training set}         & \multicolumn{4}{c}{Validation set}              & \multicolumn{4}{c}{Test set}             \\ 
\midrule
         & choked flow & constant & incipient & non & choked flow & constant & incipient & non & choked flow & constant & incipient & non \\ 
\midrule
\emph{\textbf{Dataset 1}} & 53          & 68       & 29        & 109 & 5           & 7        & 3         & 12  & 14          & 18       & 8         & 30  \\
\emph{\textbf{Dataset 2}} & 107         & 286      & 47        & 143 & 11          & 31       & 5         & 15  & 30          & 79       & 12        & 40  \\
\emph{\textbf{Dataset 3}} & 29          & 29       & 29        & 29  & 3           & 3        & 3         & 3   & 8           & 8        & 8         & 8   \\ 
\bottomrule
\end{tabular}
}
\end{table*}
 \begin{table}
    \centering
    \caption{Details of the flow status condition of \emph{\textbf{Dataset 1}}.}
    \begin{tabular}{cccc}
	\toprule 
	\multicolumn{2}{c}{Flow status} &\tabincell{c}{ Number of \\samples}\\ 
	\midrule
	\multirow{3}{*}{Cavitation} 
	 & Cavitation choked flow & 72\\
	 & Constant cavitation  & 93 \\
	 & Incipient cavitation  & 40 \\
	 \midrule
	 \multirow{2}{*}{Non cavitation} 
	 & Turbulent flow  & 118 \\
	 & No flow  & 33 \\
	\bottomrule
    \end{tabular}
    \label{tab:DataSet1-FlowStatus}
\end{table}

\begin{table}
    \centering
    \caption{Operation parameters of \emph{\textbf{Dataset 1}}.}
    \begin{tabular}{llccc}
	\toprule 
	& No. & \tabincell{c}{Valve stroke\\($mm$)}&   \tabincell{c}{Upstream pressure \\($bar(a)$)} &  \tabincell{c}{Temperature\\($^{\circ}$C)}  \\
	\midrule
	& 1  & 15      & [10,9,6,4] & 25-50\\
	& 2  & 13.5    & [10,9,6,4] & 25-50\\
	& 3  & 11.25   & [10,9,6,4] & 25-50\\
	& 4  & 7.50    & [10,9,6,4] & 25-50\\
	& 5  & 3.75    & [10,9,6,4] & 25-50\\
	& 6  & 1.50    & [10,9,6,4] & 25-50\\
	& 7  & 0.75    & [10,9,6,4] & 25-50\\
	\bottomrule
    \end{tabular}
    \label{tab:DataSet1-OperationParameters}
\end{table}

\subsubsection{Results}
\begin{figure}
    \centering
    \includegraphics[width=0.5\textwidth,height=60mm]{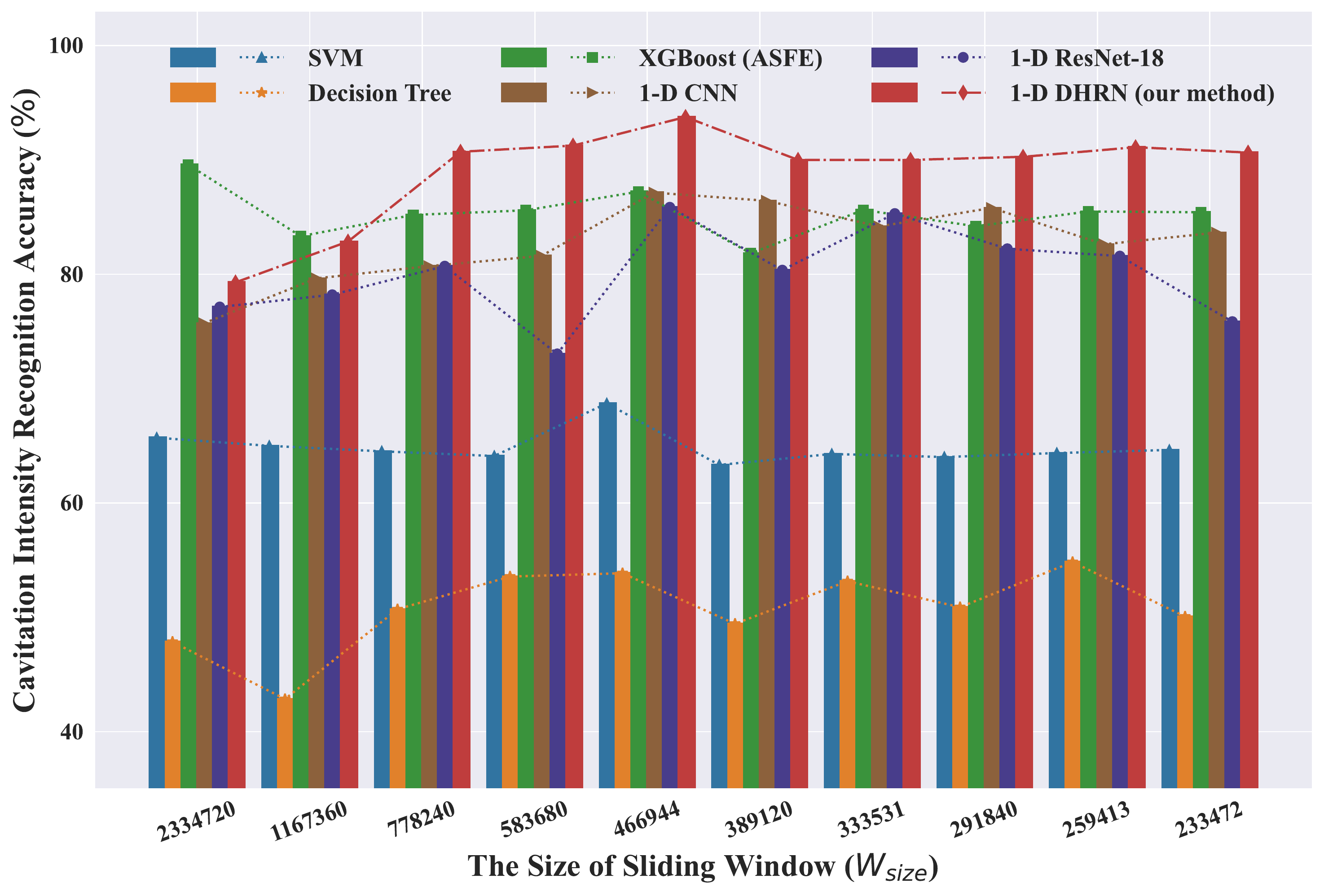}
    \caption{The effect of ${W_{size}}$ on cavitation intensity recognition accuracy for 1-D DHRN and comparison methods in \emph{\textbf{Dataset 1}}.}
    \label{fig:Accuracy intensity recognition Dataset1}
\end{figure}

The effect of ${W_{size}}$ on the accuracy of cavitation intensity recognition is studied (other metrics see Appendix). The ${W_{size}}$ is set to be 2234720, 1167360, 778240, 583680, 466944, 389120, 333531, 291840, 259413 and 233472. The results of cavitation intensity recognition of \emph{\textbf{Dataset 1}} are shown in Table \ref{tab: Accuracy intensity recognition dataset1} and Figure \ref{fig:Accuracy intensity recognition Dataset1}.
\begin{table*}
	\caption{Accuracy results of different ${W_{size}}$ of cavitation intensity recognition in \emph{\textbf{Dataset 1}} ($\%$).}
	\label{tab: Accuracy intensity recognition dataset1}
	\centering
	\setlength{\tabcolsep}{1.2mm}{
	\begin{tabular}{lcccccccccc}
	\toprule
	\multirow{2.5}{*}{Method} &\multicolumn{10}{c}{Window Size (${W_{size}}$)} \\
	\cmidrule(r){2-11}
	& 2334720 & 1167360 &778240 &583680 &466944 &389120 &333531 &291840 & 259413 &233472\\
	\midrule
	 SVM \cite{yang2005cavitation}                 & 65.71  & 65.00  & 65.42 & 64.11 & 68.71 & 63.33 & 64.29 & 64.02 & 64.37 & 64.64  \\
	 Decision Tree \cite{sakthivel2010vibration}   & 47.86  & 42.86  & 50.71 & 53.57 & 53.86 & 49.44 & 53.16 & 50.89 & 54.84 & 50.07  \\
	 1-D CNN \cite{shervani2018cavitation}         & 75.71  & 79.64  & 80.71 & 81.61 & 87.14 & 86.43 & 84.18 & 85.80 & 82.62 & 83.64   \\
	 XGBoost + ASFE \cite{SHA2022110897}            & 89.58  & 83.33  & 85.19 & 85.59 & 87.22 & 81.83 & 85.62 & 84.20 & 85.49 & 85.42   \\
	 1-D Resnet-18$^{*}$          & 77.14  & 78.21  & 80.71 & 73.04 & 85.86 & 80.36 & 85.31 & 82.23 & 81.59 & 75.86 \\
	 1-D DHRN (our method)         & 79.29  & 82.86  & 90.71 & 91.25 & \textbf{93.75} & 90.00 & 90.00 & 90.27 & 91.11 & 90.64 \\
	\bottomrule
	\end{tabular}
	}
	\begin{flushleft}
	\footnotesize{$^{*}$ our baseline: 1-D Resnet-18}
	\end{flushleft}
\end{table*}
\begin{figure}
    \centering
    \includegraphics[width=0.45\textwidth,height=60mm]{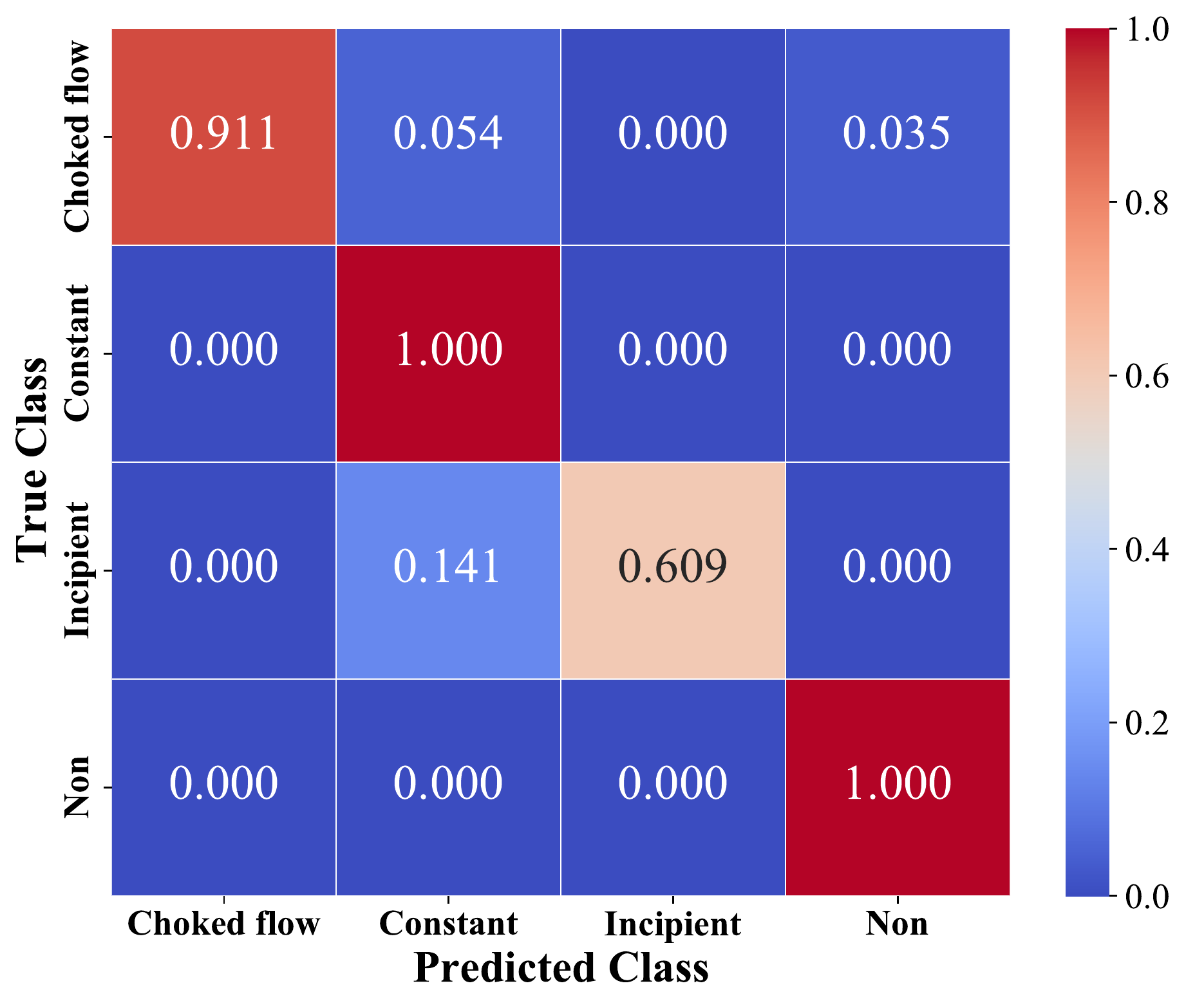}
    \caption{The confusion matrix for the best cavitation intensity recognition accuracy of the 1-D DHRN in \emph{\textbf{Dataset 1}}.}
    \label{fig:Confusion matrix FourDataset1}
\end{figure}

From Table \ref{tab: Accuracy intensity recognition dataset1} and Figure \ref{fig:Accuracy intensity recognition Dataset1}, it can be seen that the accuracy of 1-D DHRN gradually increases along with reducing window size when $2334720 \ge {W_{size}} \ge 466944$. And 1-D DHRN achieves the best cavitation intensity recognition accuracy of $93.75\%$ when ${W_{size}}$ is 466944. When $466944 \ge {W_{size}} \ge 233472$, the accuracy of 1-D DHRN progressively decreases but remains above $90\%$. Compared to other methods in the literature, the accuracy of 1-D DHRN is improved by $25.04\%$, $39.89\%$, $6.53\%$, $6.61\%$ and $7.89\%$ compared to SVM, Decision Tree, XGBoost (ASFE), 1-D CNN and 1-D ResNet-18, respectively. 

In general, the performances of 1-D DHRN are obviously better than SVM, Decision Tree, XGBoost (ASFE), 1-D CNN and also 1-D ResNet-18 at each particular value of ${W_{size}}$. The best cavitation intensity recogniton accuracies for the SVM, Decision Tree, XGBoost (ASFE), 1-D CNN and 1-D ResNet-18 methods are $68.71\%$, $54.84\%$, $89.58\%$, $87.14\%$ and $85.86\%$, respectively. 

When ${W_{size}}$ is large such as 2334720 and 1167360, the
1-D DHRN has poorer performance compared to XGBoost (ASFE), which is mainly because of the small amount of data obtained from data augmentation. However, even for this setup, the advantage of our approach compared to conventional XGBoost (ASFE) is the end-to-end implementation of cavitation intensity identification using only the valve acoustic signal, instead of feature extraction separately.

To demonstrate the cavitation intensity recognition results in more detail, the confusion matrix of the testing accuracy is evaluated in Figure \ref{fig:Confusion matrix FourDataset1}. It can be seen that the constant cavitation and the non cavitation are most easily to be distinguished with the highest accuracy compared to other levels of cavitation intensity. 
This can be explained by the fact that the constant cavitation and the non cavitation have special features that distinguish them from other states of cavitation. It can also be indicated that our 1-D DHRB structure can capture more sensitive cavitation features. However, the incipient cavitation is challenging to be recognized, which has the lowest identification accuracy compared to choked flow and constant cavitation. This is understandable because in physics it is the critical state between cavitation and non cavitation. Technically another reason is that the amount of incipient cavitation data is very small. In the study of Lehmann and Young \cite{lehman1964experimental} it also shows that the end stages of cavitation can be more easily detected than incipient cavitation. From a practical perspective, it's acceptable that the incipient cavitation is recognized as the constant cavitation to still give rise alarm to the process/plant. With this consideration, the accuracy of the incipient cavitation in our method reaches $75\%$ for \emph{\textbf{Dataset 1}}.

\subsection{Cavitation Intensity Recognition: Dataset 2}
\subsubsection{Data Description}
In this case study, the proposed 1-D DHRN is tested on condition monitoring of cavitation dataset (\emph{\textbf{Dataset 2}}) provided by SAMSON AG in Frankfurt. The principle of the experimental setup is shown in Figure \ref{fig:system}. \emph{\textbf{Dataset 2}} uses different valves and sensors compared to \emph{\textbf{Dataset 1}}. The five flow states are shown in Table \ref{tab:DataSet2-FlowStatus}. The experiments have been conducted using seven different valve strokes at three different upstream pressures and the operating parameters are shown in Table \ref{tab:DataSet2-OperationParameters}. The amount of \emph{\textbf{Dataset 2}} is 806 and the frequency of samples is 1562.5 kHz within time duration of 25 s. It should be noticed that the \emph{\textbf{Dataset 2}} has been collected by SAMSON AG in a professional environment. The details of the training set, validation set and test set for \emph{\textbf{Dataset 2}} without Swin-FFT is shown in Table \ref{tab: details of dataset1, dataset2 and dataset3}.
 \begin{table}
    \centering
    \caption{Details of the flow status condition of \emph{\textbf{Dataset 2}}.}
    \begin{tabular}{cccc}
	\toprule 
	\multicolumn{2}{c}{Flow status} &\tabincell{c}{ Number of \\samples}\\ 
	\midrule
	\multirow{3}{*}{Cavitation} 
	 & Cavitation choked flow & 148\\
	 & Constant cavitation  & 396 \\
	 & Incipient cavitation  & 64 \\
	 \midrule
	 \multirow{2}{*}{Non cavitation} 
	 & Turbulent flow  & 183 \\
	 & No flow  & 15 \\
	\bottomrule
    \end{tabular}
    \label{tab:DataSet2-FlowStatus}
\end{table}

\begin{table}
    \centering
    \caption{Operation parameters of \emph{\textbf{Dataset 2}}.}
    \begin{tabular}{llccc}
	\toprule 
	& No. & \tabincell{c}{Valve stroke\\($mm$)}&   \tabincell{c}{Upstream pressure \\($bar(a)$)} &  \tabincell{c}{Temperature\\($^{\circ}$C)}  \\
	\midrule
	& 1  & 60      & [10,6,4] & 23-52\\
	& 2  & 55    & [10,6,4] & 23-52\\
	& 3  & 45   & [10,6,4] & 23-52\\
	& 4  & 30    & [10,6,4] & 23-52\\
	& 5  & 25    & [10,6,4] & 23-52\\
	& 6  & 15    & [10,6,4] & 23-52\\
	& 7  & 6    & [10,6,4] & 23-52\\
	\bottomrule
    \end{tabular}
    \label{tab:DataSet2-OperationParameters}
\end{table}
\begin{table}
	\caption{Accuracy results of different ${W_{size}}$ of cavitation intensity recognition in \emph{\textbf{Dataset 2}} ($\%$).}
	\label{tab: Accuracy intensity recognition dataset2}
	\centering
	\footnotesize
	\setlength{\tabcolsep}{0.5mm}{
	\begin{tabular}{lccccc}
	\toprule
	\multirow{2.5}{*}{Method} &\multicolumn{5}{c}{Window Size (${W_{size}}$)} \\
	\cmidrule(r){2-6}
	& 2334720 & 1167360 &778240 &583680 &466944 \\
	\midrule
	 SVM \cite{yang2005cavitation}                 & 50.49  & 49.73  & 47.88 & 48.56 & 50.27   \\
	 Decision Tree \cite{sakthivel2010vibration}   & 65.07  & 62.86  & 62.27 & 62.39 & 60.29   \\
	 1-D CNN \cite{shervani2018cavitation}         & 85.05  & 85.09  & 86.13 & 85.12 & 85.11    \\
	 XGBoost + ASFE \cite{SHA2022110897}                            & 84.93  & 88.00  & 89.22 & 88.71 & 87.01  \\
	 1-D Resnet-18$^{*}$                          & 87.34  & 89.58  & 91.71 & 90.57 & 90.00  \\
	 1-D DHRN (our method)                         & 89.39  & 91.25  & \textbf{94.31} & 92.71 & 92.29  \\
	\bottomrule
	\end{tabular}
	}
\end{table}

\subsubsection{Results}
\begin{figure}
    \centering
    \includegraphics[width=0.5\textwidth,height=60mm]{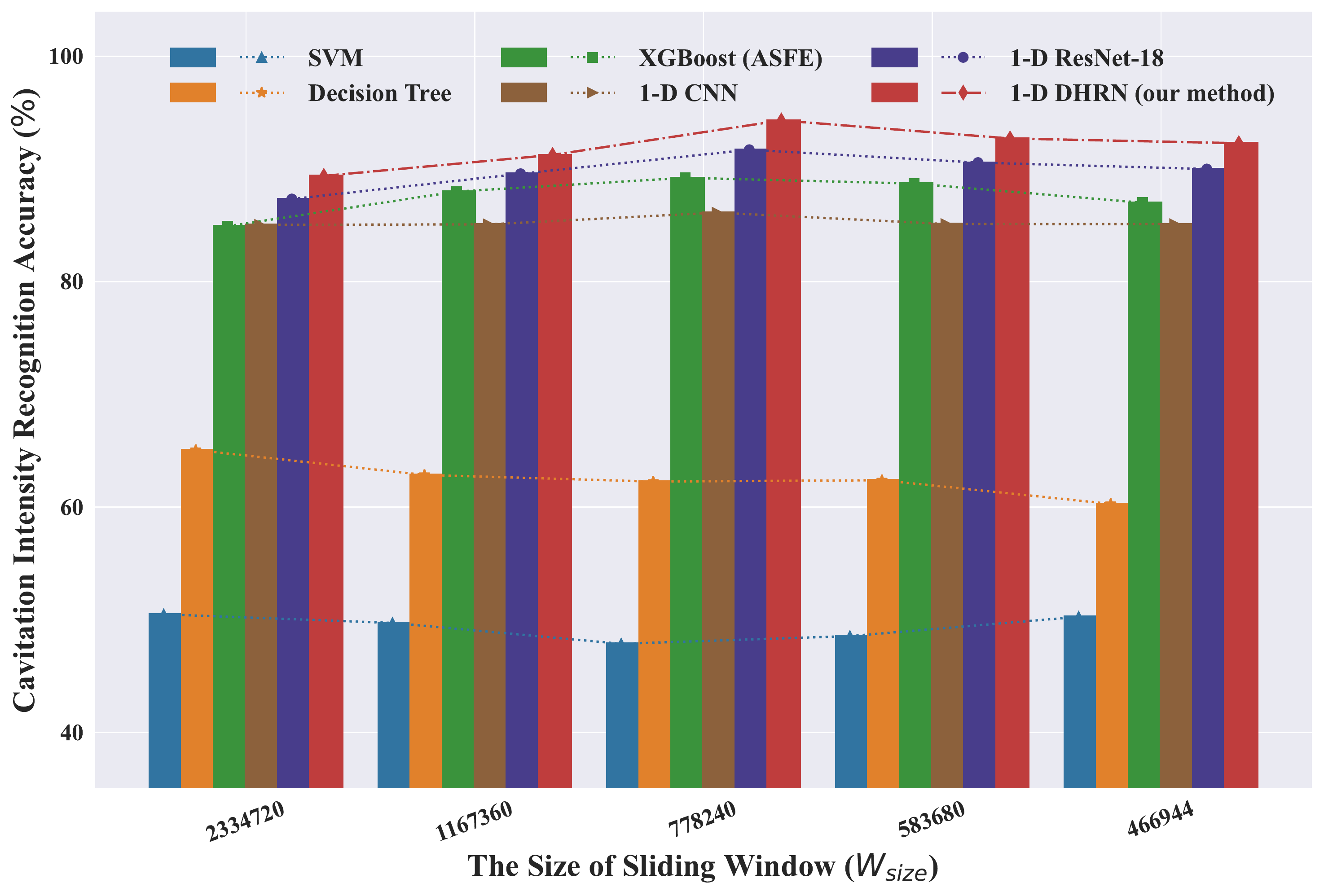}
    \caption{The effect of ${W_{size}}$ on cavitation intensity recognition accuracy for 1-D DHRN and comparison methods in \emph{\textbf{Dataset 2}}.}
    \label{fig:Accuracy intensity recognition Dataset2}
\end{figure}

The effect of ${W_{size}}$ on the accuracy of cavitation intensity recognition is studied (see Appendix of other metrics). The ${W_{size}}$ is set to be 2234720, 1167360, 778240, 583680 and 466944. The comparsion results of cavitation intensity recognition of \emph{\textbf{Dataset 2}} are shown in Table \ref{tab: Accuracy intensity recognition dataset2} and Figure \ref{fig:Accuracy intensity recognition Dataset2}.
\begin{figure}
    \centering
    \includegraphics[width=0.45\textwidth,height=60mm]{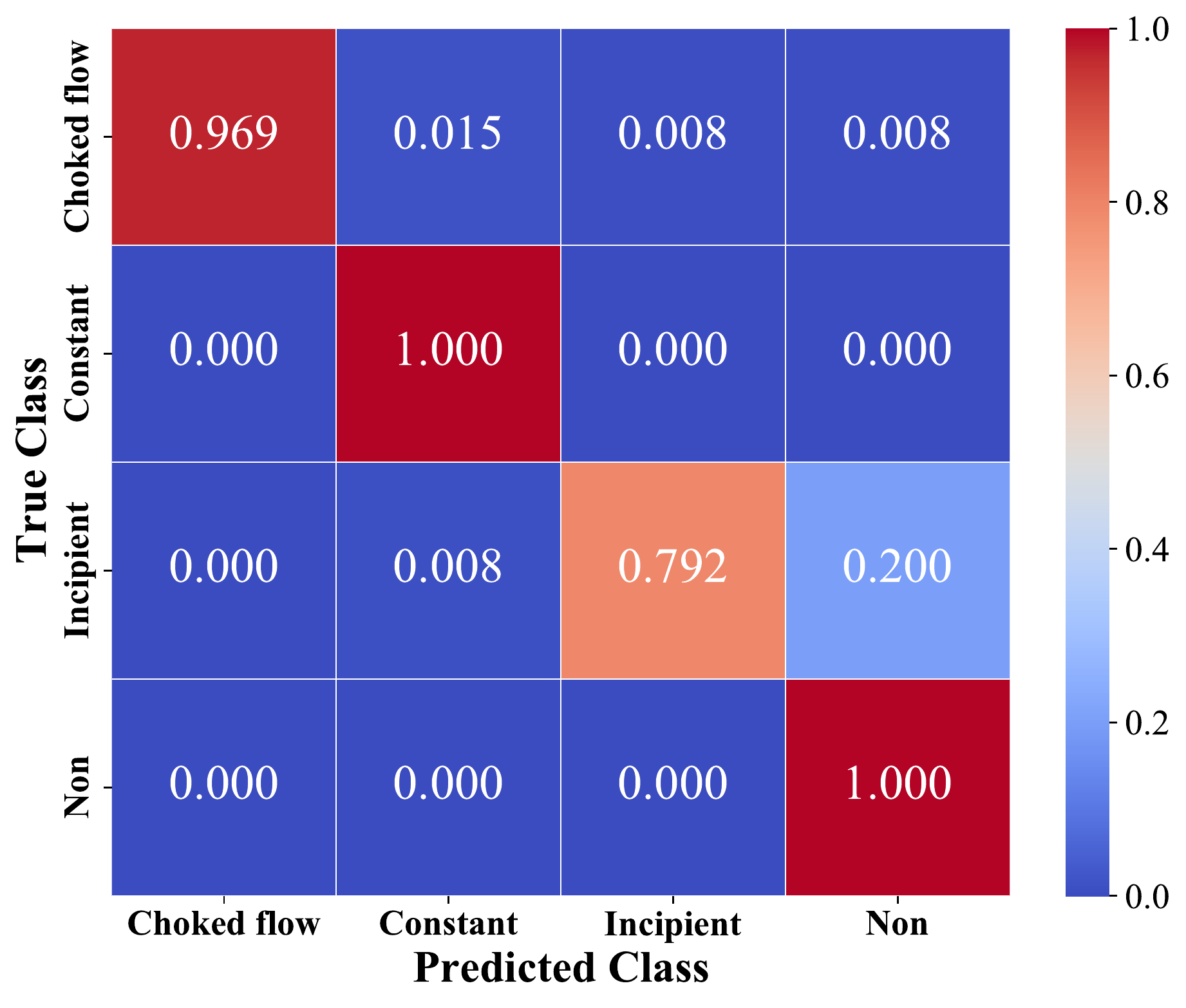}
    \caption{The confusion matrix for the best cavitation intensity recognition accuracy of the 1-D DHRN in \emph{\textbf{Dataset 2}}.}
    \label{fig:Confusion matrix FourDataset2}
\end{figure}

From Table \ref{tab: Accuracy intensity recognition dataset2}, it can be found that 1-D DHRN gets the best cavitation intensity recognition accuracy of $94.31\%$ when ${W_{size}}$ is 778240. The accuracy of 1-D DHRN gradually increases when $2334720 \ge {W_{size}} \ge 778240$. When $778240 > {W_{size}} \ge 466944$, the accuracy of 1-D DHRN progressively decreases and remains above $92\%$. The best cavitation intensity recognition of results from the SVM, Decision Tree, XGBoost (ASFE), 1-D CNN and 1-D Resnet-18 are $50.49\%$, $65.07\%$, $89.22\%$, $86.13\%$ and $91.71\%$, respectively, which were surpassed by $40.82\%$, $29.24\%$, $5.09\%$, $8.18\%$ and $2.6\%$ compared to 1-D DHRN. As can be seen from Figure \ref{fig:Accuracy intensity recognition Dataset2}, 1-D DHRN has achieved with higher accuracy than all other approaches at each value of ${W_{size}}$. 

The confusion matrix for testing accuracy is evaluated and shown in Figure \ref{fig:Confusion matrix FourDataset2}. It can be noted that choked flow, constant cavitation and non-cavitation are most easily identified with the highest accuracy compared to the incipient cavitation. And the recognition accuracy of the constant cavitation and the non cavitation reaches $100\%$. Take a practical point of view (i.e., can cause alarm) as mentioned earlier, the recognition accuracy of initial cavitation can reach $80\%$. It can also be seen that our method achieves excellent performance on datasets with different types of valves and sensors, which demonstrates the generalisation capability of the methodology.

\subsection{Cavitation Intensity Recognition: Dataset 3}
\subsubsection{Data Description}
As the third case study, the proposed 1-D DHRN is evaluated on the condition monitoring of cavitation dataset (\emph{\textbf{Dataset 3}}) provided by SAMSON AG in Frankfurt. \emph{\textbf{Dataset 3}} is different from \emph{\textbf{Dataset 1}} and \emph{\textbf{Dataset 2}} by carrying the noise of the real working environment. The five flows states about \emph{\textbf{Dataset 3}} are shown in Table \ref{tab:DataSet3-FlowStatus}. The experimental operational parameters about \emph{\textbf{Dataset 3}} are shown in Table \ref{tab:DataSet3-OperationParameters}. The quantity of \emph{\textbf{Dataset 3}} is 160 for each flow states and the frequency of samples is 1562.5 kHz within time duration of 25 s. It should be noted out that the \emph{\textbf{Dataset 3}} has been collected by SAMSON AG inside a professional environment. The details of the training set, validation set and test set for \emph{\textbf{Dataset 3}} without Swin-FFT is shown in Table \ref{tab: details of dataset1, dataset2 and dataset3}.
 \begin{table}
    \centering
    \caption{Details of the flow status condition of \emph{\textbf{Dataset 3}}.}
    \begin{tabular}{cccc}
	\toprule 
	\multicolumn{2}{c}{Flow status} &\tabincell{c}{ Number of \\samples}\\ 
	\midrule
	\multirow{3}{*}{Cavitation} 
	 & Cavitation choked flow & 40\\
	 & Constant cavitation  & 40 \\
	 & Incipient cavitation  & 40 \\
	 \midrule
	 \multirow{1}{*}{Non cavitation} 
	 & Turbulent flow  & 40 \\
	\bottomrule
    \end{tabular}
    \label{tab:DataSet3-FlowStatus}
\end{table}

\begin{table}
    \centering
    \caption{Operation parameters of \emph{\textbf{Dataset 3}}.}
    \begin{tabular}{llccc}
	\toprule 
	& No. & \tabincell{c}{Valve stroke\\($mm$)}&   \tabincell{c}{Upstream pressure \\($bar(a)$)} &  \tabincell{c}{Temperature\\($^{\circ}$C)}  \\
	\midrule
	& 1  & 15      & 10 & 32-39\\
	\bottomrule
    \end{tabular}
    \label{tab:DataSet3-OperationParameters}
\end{table}

\subsubsection{Results}
\begin{figure}
    \centering
    \includegraphics[width=0.5\textwidth,height=60mm]{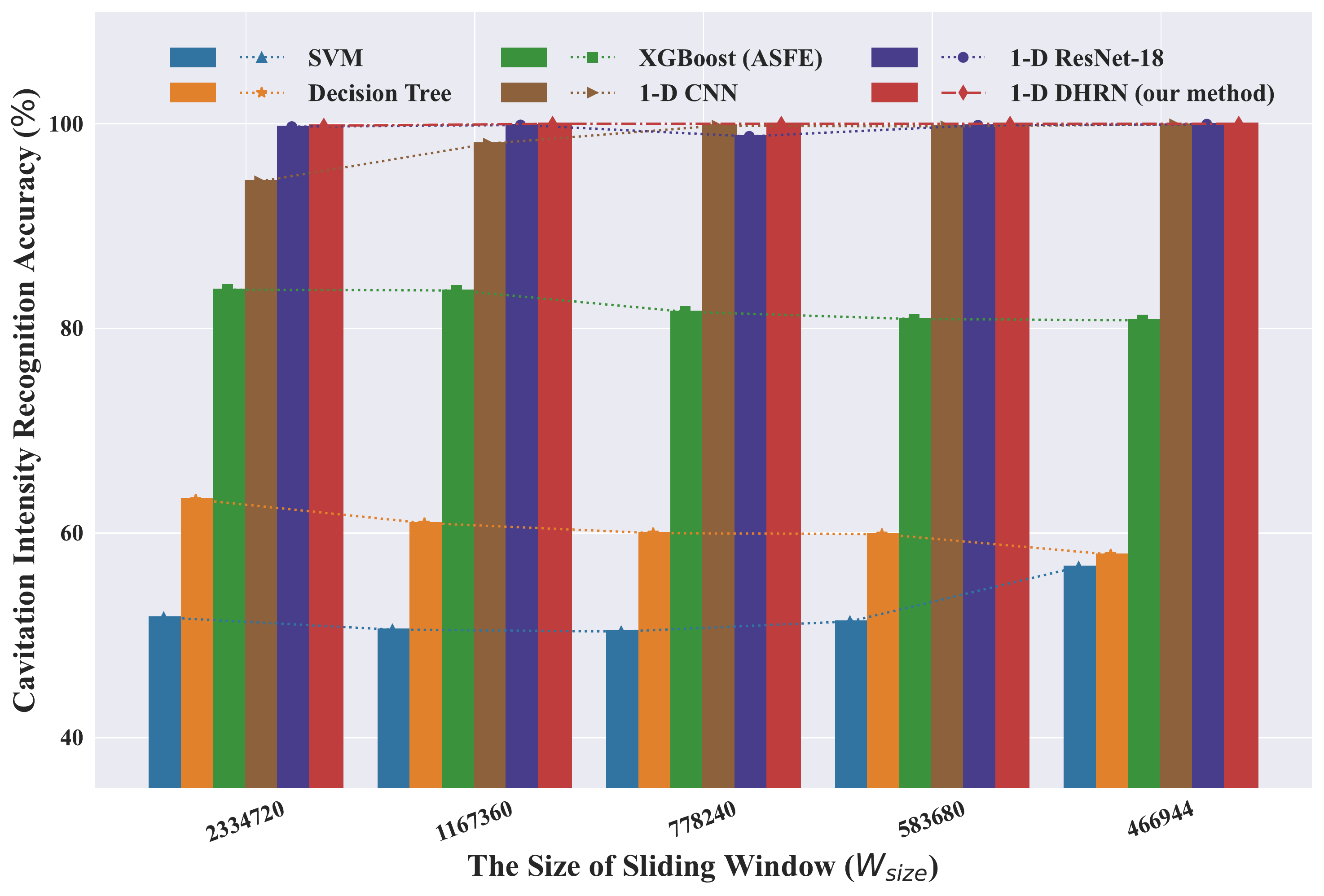}
    \caption{The effect of ${W_{size}}$ on cavitation intensity recognition accuracy for 1-D DHRN and comparison methods in \emph{\textbf{Dataset 3}}.}
    \label{fig:Accuracy intensity recognition Dataset3}
\end{figure}

The influence of ${W_{size}}$ on the accuracy of cavitation intensity recognition is investigated (see Appendix of other metrics). The ${W_{size}}$ is the same as the ${W_{size}}$ of \emph{\textbf{Dataset 2}}. The accuracy of cavitation intensity recognition of \emph{\textbf{Dataset 3}} is shown in Table \ref{tab: Accuracy intensity recognition dataset3} and Figure \ref{fig:Accuracy intensity recognition Dataset3}.
\begin{table}
	\caption{Accuracy results of different ${W_{size}}$ of cavitation intensity recognition in \emph{\textbf{Dataset 3}} ($\%$).}
	\label{tab: Accuracy intensity recognition dataset3}
	\centering
	\footnotesize
	\setlength{\tabcolsep}{0.5mm}{
	\begin{tabular}{lccccc}
	\toprule
	\multirow{2.5}{*}{Method} &\multicolumn{5}{c}{Window Size (${W_{size}}$)} \\
	\cmidrule(r){2-6}
	& 2334720 & 1167360 &778240 &583680 &466944 \\
	\midrule
	 SVM \cite{yang2005cavitation}                 & 51.76   & 50.57   & 50.38  & 51.37  & 56.70   \\
	 Decision Tree \cite{sakthivel2010vibration}   & 63.28   & 60.98   & 60.00  & 59.90  & 57.91   \\
	 1-D CNN \cite{shervani2018cavitation}         & 94.37   & 98.06   & 99.81  & 99.75  & 99.94   \\
	 XGBoost + ASFE \cite{SHA2022110897}                             & 83.79   & 83.71   & 81.63  & 80.92  & 80.80  \\
	 1-D Resnet-18$^{*}$                          & 99.72   & 99.88   & 98.77  & 99.85  & 99.96  \\
	 1-D DHRN (our method)                         & 99.81   & \textbf{100.00}     & \textbf{100.00}    & \textbf{100.00}    & \textbf{100.00}   \\
	\bottomrule
	\end{tabular}
	}
\end{table}
\begin{figure}
    \centering
    \includegraphics[width=0.45\textwidth,height=60mm]{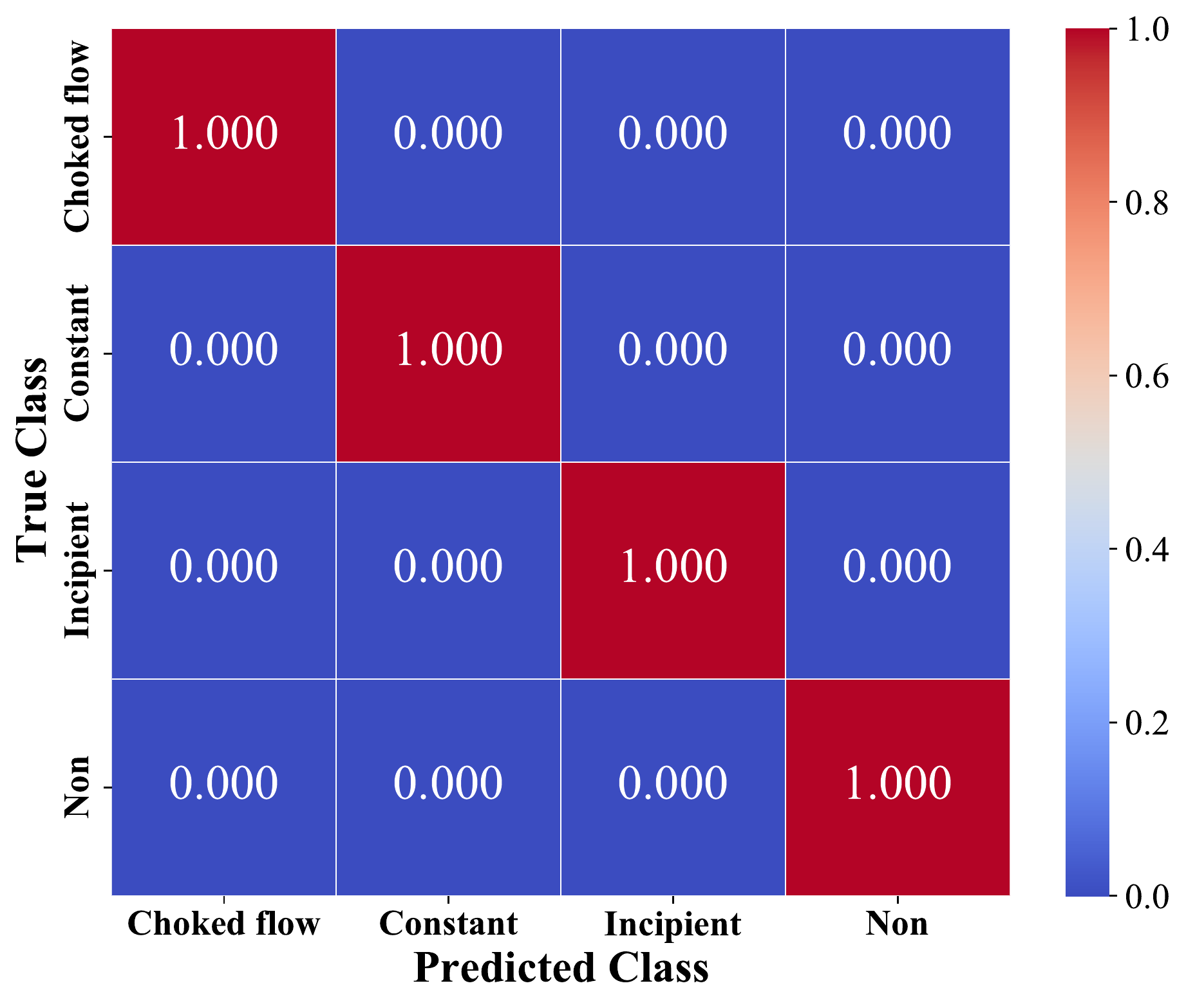}
    \caption{The confusion matrix for the best cavitation intensity recognition accuracy of the 1-D DHRN in \emph{\textbf{Dataset 3}}.}
    \label{fig:Confusion matrix FourDataset3}
\end{figure}

From Table \ref{tab: Accuracy intensity recognition dataset3} and Figure \ref{fig:Accuracy intensity recognition Dataset3}, it can be observed that the best accuracy of our method is $100\%$. And the accuracy of our method is above $99\%$ for every value of ${W_{size}}$. Our method has achieved $100\%$ accuracy of recognition of all cavitation states (see Figure \ref{fig:Confusion matrix FourDataset3}).

The reasons why our method and compared methods have achieved excellent results on \emph{\textbf{Dataset 3}} compared to \emph{\textbf{Dataset 1}} and \emph{\textbf{Dataset 2}} are as follows. First, although \emph{\textbf{Dataset 3}} are with real background noise compared to \emph{\textbf{Dataset 1}} and \emph{\textbf{Dataset 2}}, our data augmentation of the Swin-FFT operation can filter most of the noise (see subsection \ref{sec: analysis of the Swin-FFT}) in frequency domain. Second, \emph{\textbf{Dataset 3}} are obtained with only one setup for the valve 
stroke and upstream pressure (see Table \ref{tab:DataSet3-OperationParameters} and subsection \ref{sec analysis of data acquisition operations}). Third, \emph{\textbf{Dataset 3}} is balanced for each flow state especially for cavitation state (see Table \ref{tab:DataSet3-FlowStatus} and subsection \ref{sec analysis of data acquisition operations}). From the above, we can infer that different valve openings and upstream pressures can affect the accuracy of the cavitation intensity recognition.

\subsection{Cavitation Detection}
\begin{table*}
	\caption{Accuracy results of different ${W_{size}}$ of cavitation detection in \emph{\textbf{Dataset 1}} ($\%$).}
	\label{tab: Accuracy cavitation detection dataset1}
	\centering
	\setlength{\tabcolsep}{1.2mm}{
	\begin{tabular}{lcccccccccc}
	\toprule
	\multirow{2.5}{*}{Method} &\multicolumn{10}{c}{Window Size (${W_{size}}$)} \\
	\cmidrule(r){2-11}
	& 2334720 & 1167360 &778240 &583680 &466944 &389120 &333531 &291840 & 259413 &233472\\
	\midrule
	 SVM \cite{yang2005cavitation}                   & 81.43  & 82.86  & 80.24 & 79.82 & 83.29 & 79.29 & 79.29 & 80.00 & 80.56 & 79.14  \\
	 Decision Tree \cite{sakthivel2010vibration}     & 73.57  & 76.07  & 74.29 & 76.96 & 77.00 & 76.55 & 77.86 & 76.07 & 79.44 & 81.86  \\
	 1-D CNN \cite{shervani2018cavitation}           & 94.29  & 94.29  & 94.05 & 94.46 & 94.14 & 95.24 & 94.90 & 95.09 & 95.71 & 95.21   \\
	 XGBoost + ASFE \cite{SHA2022110897}                               & 93.56  & 92.36  & 92.36 & 91.32 & 91.25 & 91.08 & 88.99 & 89.93 & 89.51 & 90.00   \\
	 1-D Resnet-18$^{*}$          & 95.00  & 95.71  & 95.71 & 95.54 & 95.43 & 96.19 & 96.12 & 95.89 & 95.48 & 95.64 \\
	 1-D DHRN (our method)         & 96.43  & 96.71  & 96.67 & 96.43 & 96.43 & \textbf{97.02} & 96.82 & 96.07 & 96.03 & 96.23 \\
	\bottomrule
	\end{tabular}
	}
\end{table*}
\begin{table}
	\caption{Accuracy results of different ${W_{size}}$ of cavitation detection in \emph{\textbf{Dataset 2}} ($\%$).}
	\label{tab: Accuracy cavitation detection dataset2}
	\centering
	\footnotesize
	\setlength{\tabcolsep}{0.5mm}{
	\begin{tabular}{lccccc}
	\toprule
	\multirow{2.5}{*}{Method} &\multicolumn{5}{c}{Window Size (${W_{size}}$)} \\
	\cmidrule(r){2-6}
	& 2334720 & 1167360 &778240 &583680 &466944 \\
	\midrule
	 SVM \cite{yang2005cavitation}                 & 89.83  & 89.36  & 89.06 & 89.92 & 89.30   \\
	 Decision Tree \cite{sakthivel2010vibration}   & 89.58  & 88.59  & 89.96 & 89.84 & 88.97   \\
	 1-D CNN \cite{shervani2018cavitation}         & 94.29  & 95.24  & 94.05 & 94.46 & 94.14    \\
	 XGBoost + ASFE \cite{SHA2022110897}                            & 90.56  & 90.73  & 91.33 & 91.02 & 90.43  \\
	 1-D Resnet-18$^{*}$                          & 95.48  & 96.19  & 95.71 & 95.54 & 95.43  \\
	 1-D DHRN (our method)                         & 96.67  & 97.02  & \textbf{97.64} & 96.43 & 96.43  \\
	\bottomrule
	\end{tabular}
	}
\end{table}
\begin{table}
	\caption{Accuracy results of different ${W_{size}}$ of cavitation detection in \emph{\textbf{Dataset 3}} ($\%$).}
	\label{tab: Accuracy cavitation detection dataset3}
	\centering
	\footnotesize
	\setlength{\tabcolsep}{0.5mm}{
	\begin{tabular}{lccccc}
	\toprule
	\multirow{2.5}{*}{Method} &\multicolumn{5}{c}{Window Size (${W_{size}}$)} \\
	\cmidrule(r){2-6}
	& 2334720 & 1167360 &778240 &583680 &466944 \\
	\midrule
	 SVM \cite{yang2005cavitation}                 & 100.00  & 100.00  & 100.00 & 100.00 & 100.00   \\
	 Decision Tree \cite{sakthivel2010vibration}   & 100.00  & 100.00  & 100.00 & 100.00 & 100.00   \\
	 1-D CNN \cite{shervani2018cavitation}         & 100.00  & 100.00  & 100.00 & 100.00 & 100.00    \\
	 XGBoost + ASFE \cite{SHA2022110897}                             & 100.00  & 100.00  & 100.00 & 100.00 & 100.00  \\
	 1-D Resnet-18$^{*}$                          & 100.00  & 100.00  & 100.00 & 100.00 & 100.00  \\
	 1-D DHRN (our method)                         & 100.00  & 100.00  & 100.00 & 100.00 & 100.00  \\
	\bottomrule
	\end{tabular}
	}
\end{table}
In this subsection, the proposed 1-D DHRN is evaluated on condition monitoring of cavitation dataset (\emph{\textbf{Dataset 1}}, \emph{\textbf{Dataset 2}} and \emph{\textbf{Dataset 3}}) for the binary classification task, cavitation detection. The comparison of accuracies on \emph{\textbf{Dataset 1}}, \emph{\textbf{Dataset 2}} and \emph{\textbf{Dataset 3}} are shown in Tables \ref{tab: Accuracy cavitation detection dataset1}, \ref{tab: Accuracy cavitation detection dataset2} and \ref{tab: Accuracy cavitation detection dataset3} (see Appendix of other metrics).

\textbf{Dataset 1} From Table \ref{tab: Accuracy cavitation detection dataset1}, it can be found that 1-D DHRN achieves the best accuracy, $97.02\%$, when ${W_{size}}$ is 389120. 1-D DHRN is always better than other methods under the same value of ${W_{size}}$. And the accuracy of the 1-D DHRN is above $96\%$ at each value of ${W_{size}}$.

\textbf{Dataset 2} From Table \ref{tab: Accuracy cavitation detection dataset2}, it can be noted that 1-D DHRN achieves the best accuracy of $97.64\%$ when ${W_{size}}$ is 778240. 1-D DHRN has better accuracy compared to other methods under the same value of ${W_{size}}$. And the accuracy of the 1-D DHRN is above $96\%$ for each value of ${W_{size}}$.

\textbf{Dataset 3} As can be seen from Table \ref{tab: Accuracy cavitation detection dataset3}, both the 1-D DHRN and comparative methods have achieved $100\%$ accuracy for each ${W_{size}}$ values. It can be concluded that the cavitation and non-cavitation features of \emph{\textbf{Dataset 3}} are more easily distinguished compared to \emph{\textbf{Dataset 1}} and \emph{\textbf{Dataset 2}}.

\section{Discussions}
\label{sec:5-Disscussions}
\subsection{Analysis of Swin-FFT}
\label{sec: analysis of the Swin-FFT}
\begin{figure}
\centering
\subfigure[Cavitation Choked Flow]{
\begin{minipage}[t]{0.5\linewidth}
\centering
\includegraphics[width=0.7\textwidth,height=65mm]{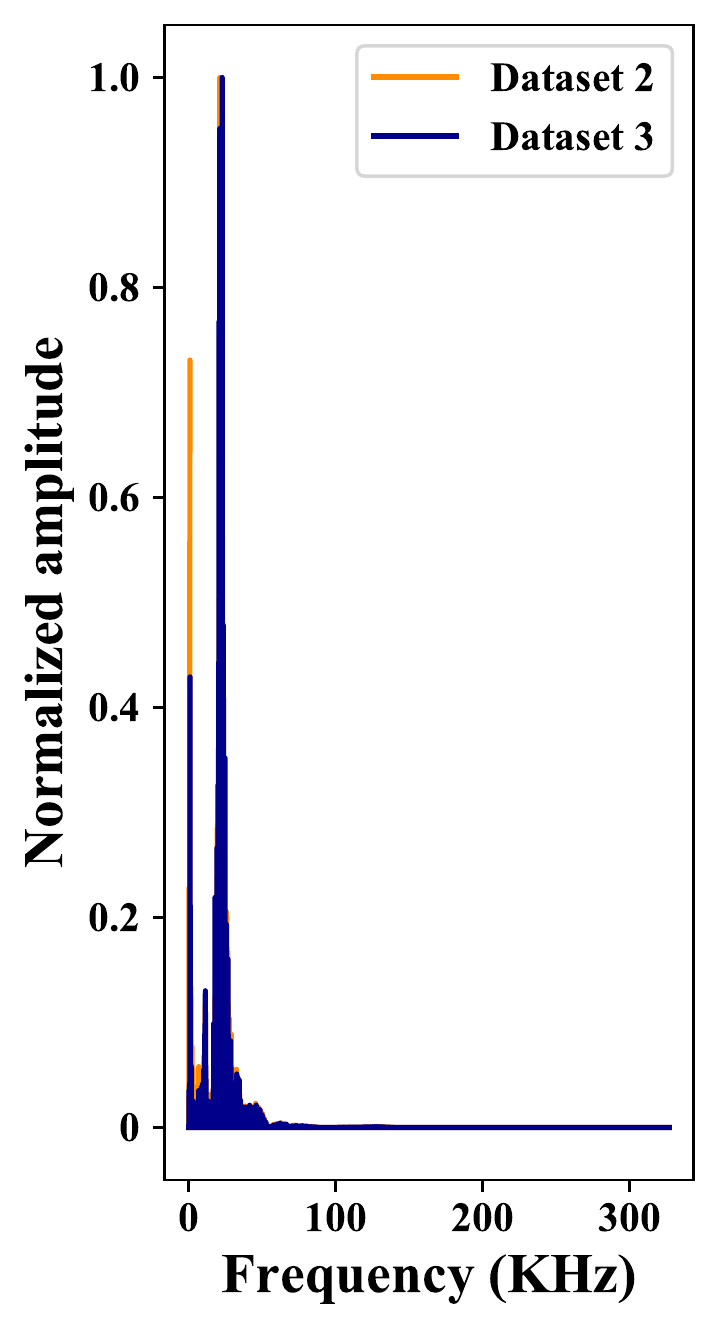}
\end{minipage}%
}%
\subfigure[Constant Cavitation]{
\begin{minipage}[t]{0.5\linewidth}
\centering
\includegraphics[width=0.7\textwidth,height=65mm]{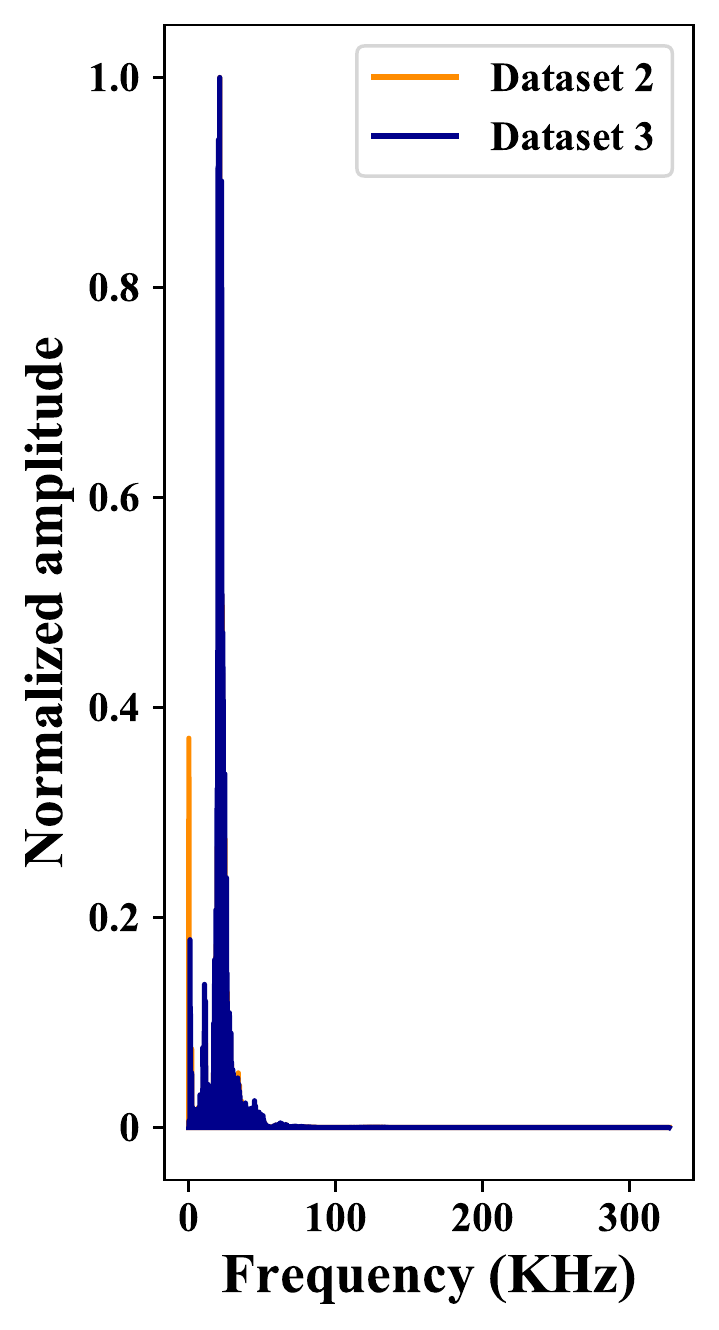}
\end{minipage}%
}%

\subfigure[Incipient Cavitation]{
\begin{minipage}[t]{0.5\linewidth}
\centering
\includegraphics[width=0.7\textwidth,height=65mm]{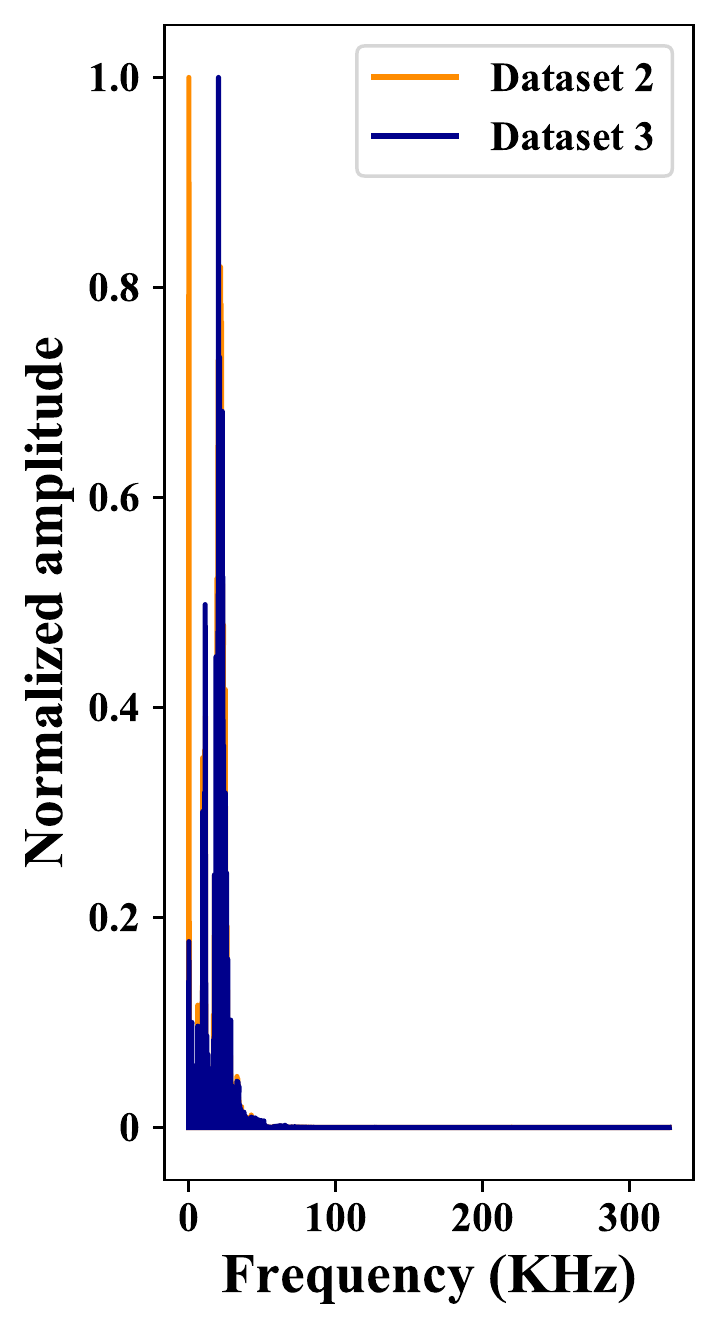}
\end{minipage}%
}%
\subfigure[Non cavitation]{
\begin{minipage}[t]{0.5\linewidth}
\centering
\includegraphics[width=0.7\textwidth,height=65mm]{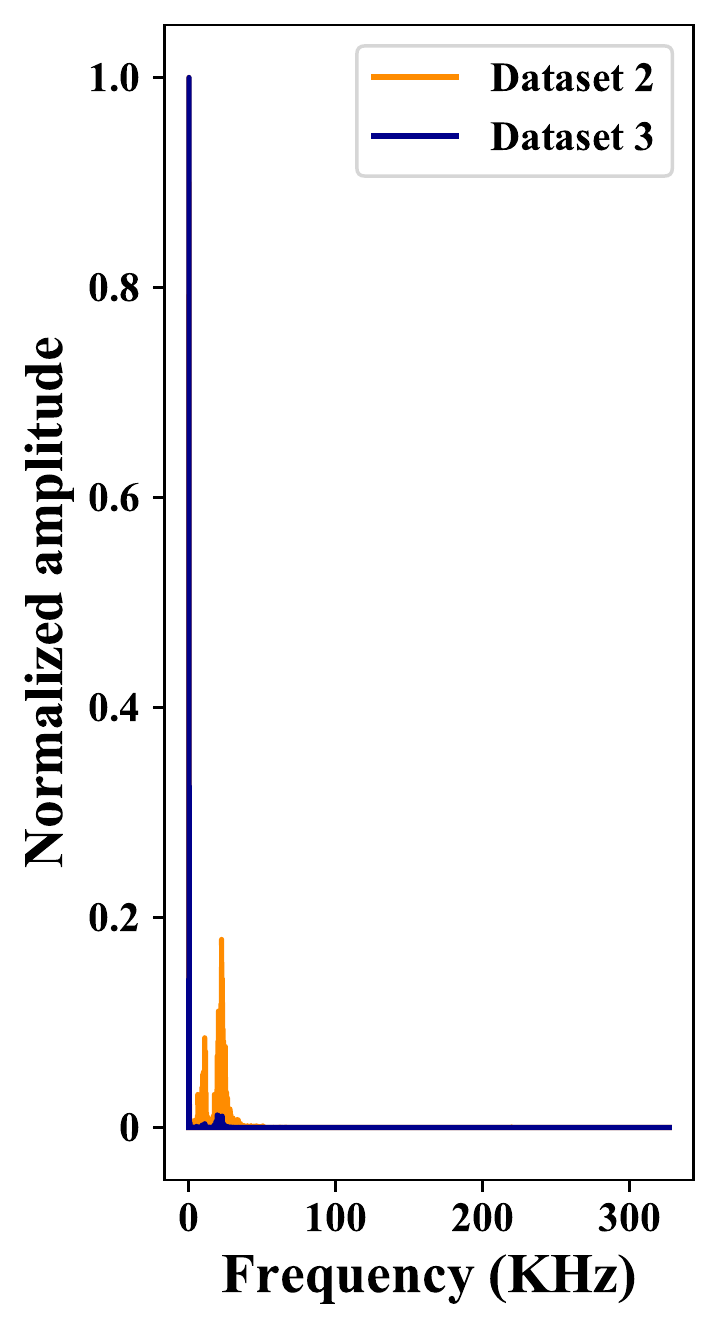}
\end{minipage}%
}%
\centering
\caption{Examples of cavitation choked flow (a), constant cavitation (b), incipient cavitation (c) and non-cavitation (d) for \emph{\textbf{Dataset 2}} and \emph{\textbf{Dataset 3}} converted by Swin-FFT.}
\label{fig: analysis-Swin-FFT}
\end{figure}

In order to verify that the Swin-FFT can filter out most of the noise, we select the data from \emph{\textbf{Dataset 2}} which was obtained under the same operation as \emph{\textbf{Dataset 3}}. Then, the data are processed using the Swin-FFT. Finally, we compare the shape of each cavitation state from \emph{\textbf{Dataset 2}} without noise and \emph{\textbf{Dataset 3}} with noise, as shown in Figure \ref{fig: analysis-Swin-FFT}.
From Figure \ref{fig: analysis-Swin-FFT}, it can be seen that the spectral structures of \emph{\textbf{Dataset 2}} without noise and \emph{\textbf{Dataset 3}} with noise are basically similar when transformed into the frequency domain by Swin-FFT. This indicates that Swin-FFT can filter most of the noise.

\subsection{Analysis of Data Acquisition Operations}
\label{sec analysis of data acquisition operations}
\begin{figure}
    \centering
    \includegraphics[width=0.45\textwidth,height=65mm]{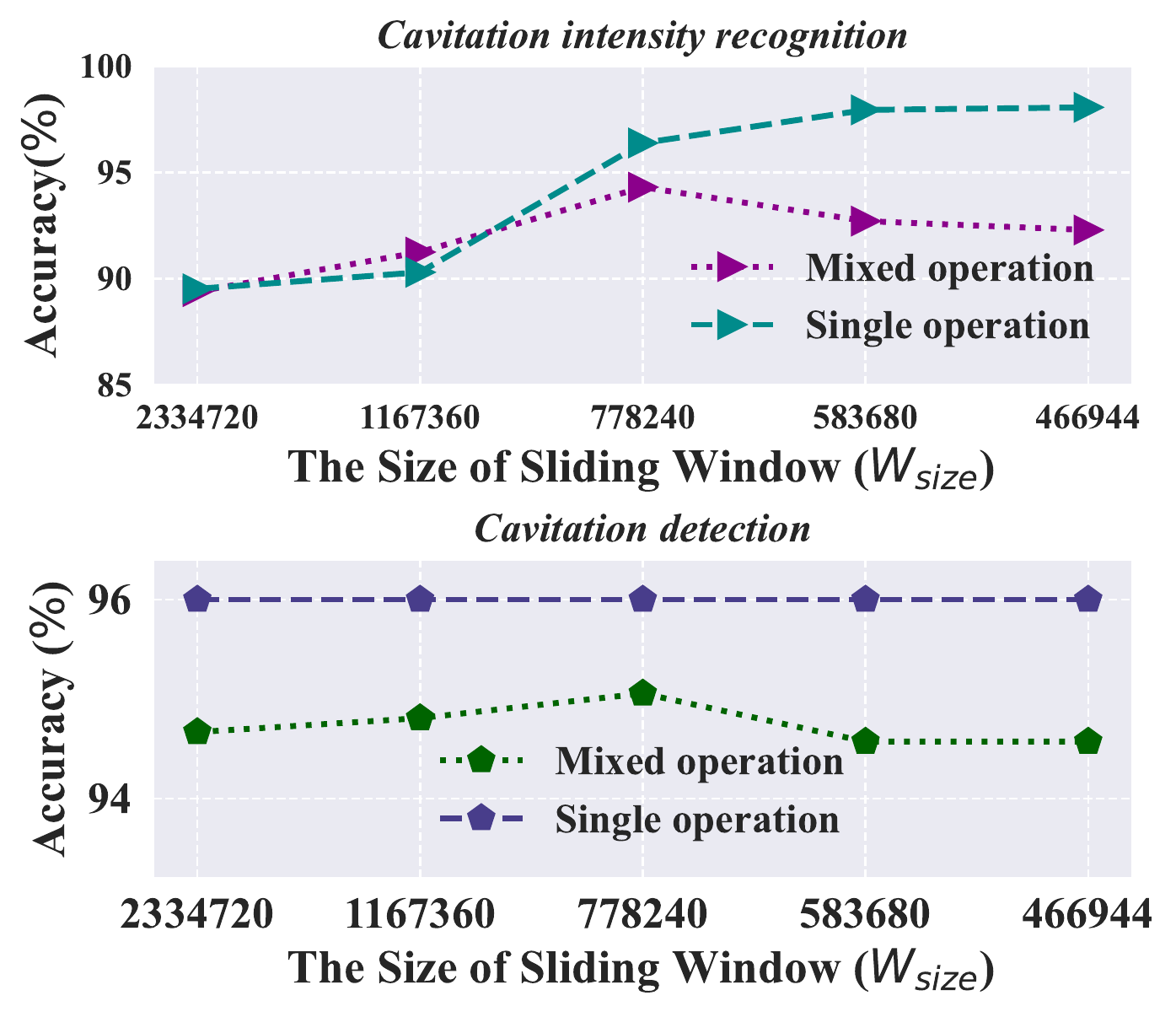}
    \caption{The effect of data diversity, with data obtained from single valve stroke and upstream pressure and mixed valve strokes and upstream pressures, on the results.}
    \label{fig: analysis data acquisition operation}
\end{figure}

In order to determine whether the data obtained by the mixed operation and the single operation have an impact on the results, a total of 46 data are selected from \emph{\textbf{Dataset 2}} for valve stroke equals to 15 $mm$ and upstream pressure equals to 10 $bar(a)$, where the numbers of cavitation choked flow, constant cavitation, incipient cavitation and non cavitation are 15, 24, 3 and 4, respectively. Then, the selected data are processed by Swin-FFT and fed into the 1-D DHRN. Finally, the results of cavitation detection and cavitation intensity recognition are obtained and shown in Figure \ref{fig: analysis data acquisition operation}.

As can be seen from Figure \ref{fig: analysis data acquisition operation}, the accuracies of cavitation intensity recognition and cavitation detection are higher with single operation data compared to mixed operation data. Therefore, we conclude that for data obtained from a single valve stroke and upstream pressure, the performance of cavitation intensity recognition and cavitation detection is better. We noted that the accuracy of cavitation intensity recognition is not $100\%$ in this test, it can be thus inferred that the balance of data for each cavitation state is important.

\subsection{Analysis of Downsampling}
\label{sec: downsampling analysis}
\begin{figure}
    \centering
    \includegraphics[width=0.45\textwidth,height=110mm]{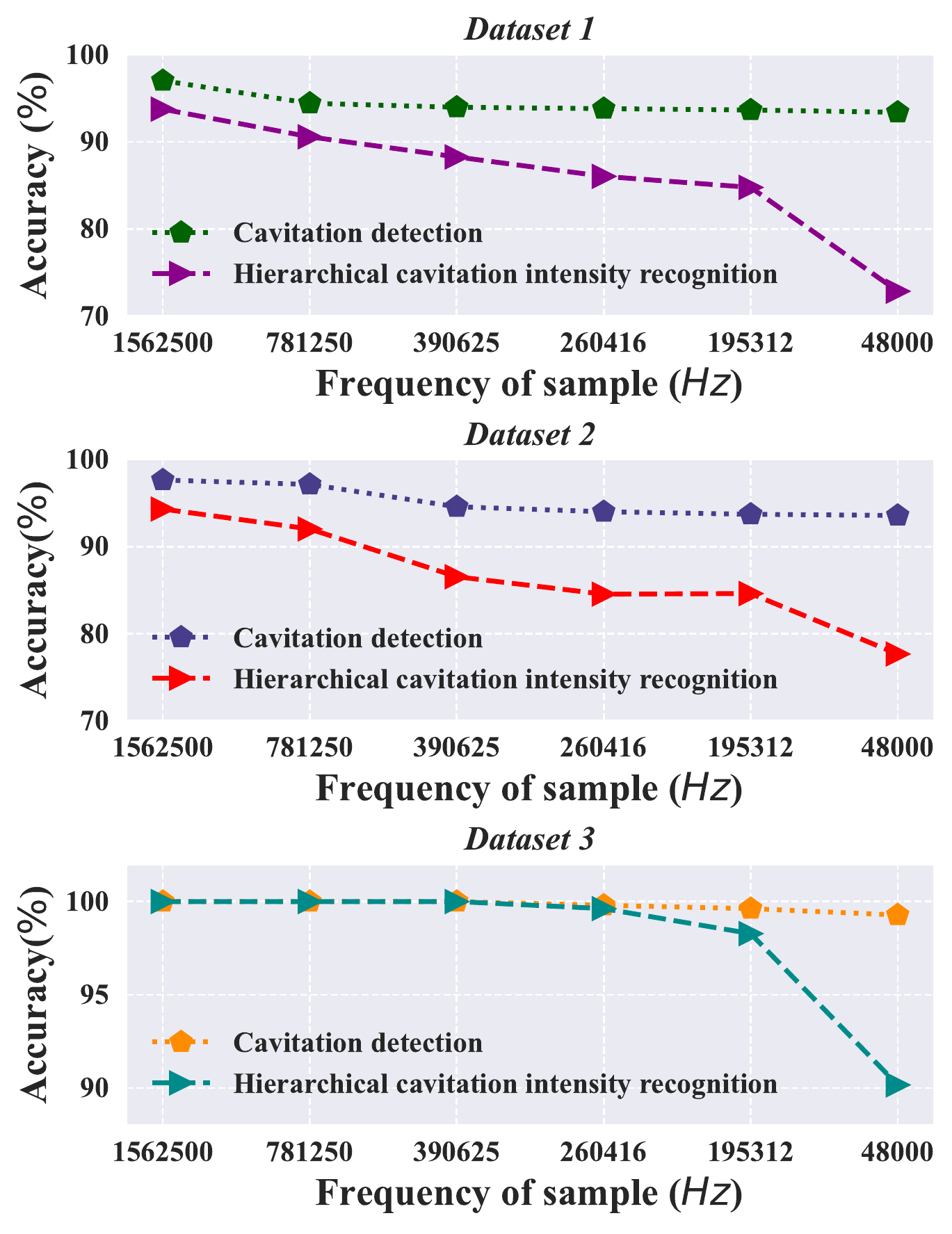}
    \caption{The results of cavitation intensity recognition and cavitation detection on different frequencies of samples for \emph{\textbf{Dataset 1}}, \emph{\textbf{Dataset 2}} and \emph{\textbf{Dataset 3}}.}
    \label{fig:Downsampling}
\end{figure}
In practical applications, the resolution of sensors represent the quality of the data obtained. The ability to recognize different levels of cavitation intensity with the data obtained from low level sensors becomes very significant and challenging. In this research, 1-D DHRN is evaluated for cavitation intensity recognition and cavitation detection under the original frequency of samples ($Fs$ = 1562500 Hz), one-half, one-quarter, one-sixth, one-eighth of the original frequency (sampling rate) of samples (781250 Hz, 390625 Hz, 260416 Hz and 195312 Hz) and that of the mobile phone can mostly accommodate (48000 Hz $\approx Fs/32$), respectively. The cavitation intensity recognition and cavitation detection accuracy of the 1-D DHRN model has been compared again in \emph{\textbf{Dataset 1}}, \emph{\textbf{Dataset 2}} and \emph{\textbf{Dataset 3}} with different frequencies of samples, as shown in Figure \ref{fig:Downsampling}. 

$\bullet$ \textbf{Dataset 1 and Dataset 2}. For cavitation intensity recognition, although the 1-D DHRN model begins to suffer from gradual reduction in accuracy as the frequency of samples decreases, the accuracy always remains above $84\%$ when 1562500 Hz $\geq  Fs\geq$ 195312 Hz. And even for mobile phone setup with the frequency of samples equals to 48000 Hz (one-32nd of the original signal frequency of samples), our 1-D DHRN has achieved $72.83\%$ and $77.65\%$ accuracy on the two datasets. For cavitation detection, the accuracy of the 1-D DHRN model progressively decreases as the frequency of samples decreases, but the accuracy of cavitation detection remains above $93\%$ at each all values of sampling frequency. When the frequency of samples equals to 48000 Hz, our method gives $93.29\%$ and $93.57\%$ accuracies, respectively. And when the frequency of samples is reduced to one-32nd of the original signal frequency of samples, the accuracy of 1-D DHRN is only decreased by $3.73\%$ and $4.07\%$.

$\bullet$ \textbf{Dataset 3}. For cavitation intensity recognition, the accuracy of our method begins to reduce when the frequency of sample is 260416 Hz (260416 Hz $\approx Fs/6$). The accuracy of our method always remains above $98\%$ when 260416 Hz $\geq  Fs\geq$ 195312 Hz. And our method has achieved $90.17\%$ accuracy when the frequency of sample is equal to $48000 Hz$. For cavitation detection, the accuracy of our method always remains above $99\%$ at each particular value of $Fs$. When the frequency of samples is equal to 48000 Hz, our method obtains $99.28\%$, which is reduced by only $0.72\%$ of the accuracy when compared to the original frequency of samples.

\subsection{Learning and Optimization}
In our paper, cavitation detection and cavitation intensity recognition are multi-class classification problem in data-driven supervised learning. The essence of supervised learning is to minimize the cumulative classification error, i.e. $\arg {\min _\theta }\sum\nolimits_i {e(f({x_i};\theta ),y_i^*)}$, where ${y^*}$ denotes reference or "ground truth" and $f(x;\theta )$ is the output of the neural network. We typically optimize, i.e. train, with a stochastic gradient descent (SGD) optimizer. And we rely on auto-diff to compute the gradient w.r.t. weights $\theta $, ${{\partial f} \mathord{\left/{\vphantom {{\partial f} {\partial \theta }}} \right.\kern-\nulldelimiterspace} {\partial \theta }}$. In recent years, there has been a lot of research on optimizers \cite{bock2018improvement,duchi2011adaptive,zamfirache2022policy,pu2013fractional}, and the Adam optimizer is used in our research to optimize errors. In addition, the choice of error function is important for learning and optimization of supervised learning. For classification problems, the cross-entropy function is usually employed as the error function, since it can avoid the problem of learning rate decreasing due to gradient dispersion.

\subsection{Extending Applications}
Our proposed 1-D DHRB is motivated by the residual block, and the residual block is the core idea of the ResNet \cite{he2016deep}. As ResNet has been proven to be a very effective convolutional neural network structure, which is widely employed in fields such as image classification \cite{lu2007survey}, medical image recognition \cite{ayyachamy2019medical} and object recognition. In addition, 1D CNN have been successfully applied to vibration signals \cite{rashid2020sustainable}. In addition, 1D CNN have been successfully applied to vibration signals \cite{zhang2018deep}, ECG \cite{bouaziz2019automatic,albu2019results} and EEG \cite{upadhyay2020wavelet,borlea2021unified}. Therefore, our proposed 1D DHRN can also be employed for vibration signals, ECG, EEG and other signals.

\section{Conclusion and Outlook}
\label{sec:6-Conclusion}
In this work we present a novel multi-task learning framework for cavitation intensity recognition ("cavitation choked flow", "constant cavitation", "incipient cavitation" and "non cavitation") and cavitation detection ("cavitation" and "non cavitation") using 1-D double hierarchical residual networks (1-D DHRN). To the best of our knowledge, besides to be the first application of 1-D convolutional neural networks based on 1-D residual blocks to perform cavitation intensity recognition and cavitation detection, other main contributions to this paper are summarized as the following four points. Firstly, a sliding window with fast fourier transform (Swin-FFT) data augmentation method is introduced to mitigate the few-short learning problem. Secondly, the 1-D double hierarchical residual blocks (1-D DHRB) with large kernels is proposed as a feature extractor to capture sensitive features of valves acoustic signals. Thirdly, a new structure of 1-D DHRN has been constructed through stacking 1-D DHRB for cavitation intensity recognition, which has in total 18 layers for the whole structure. With deeper network and better feature extraction layers, the proposed 1-D DHRN could improve the final prediction performance on cavitation intensity recognition and cavitation detection. Finally, 1-D DHRN has been tested on three dataset of valve acoustic signals and achieved significant results for the two tasks simultaneously. Moreover, the structure of 1-D DHRN has also been evaluated on different frequencies of samples and showed excellent results for even the frequency of samples which the mobile phones can hold. These results have validated the good performance of our proposed 1-D DHRN both in the field of cavitation intensity recognition and cavitation detection.

Although the proposed method has improved recognition of the incipient cavitation recognition, which is still relatively low. In future work, convolutional neural networks (CNN) have achieved remarkable results in mechanical device health management due to their stronger capability of representation learning. Therefore, we directly extract deeper and more expressive valve cavitation features using the 1D CNN to enhance the performance of valve cavitation detection and cavitation intensity recognition.

\section*{Acknowledgements}
\label{sec:acknowledgements}
This research is supported by Xidian - FIAS International Joint Research Center (Y. S.), by the AI grant at FIAS through SAMSON AG (J. F., K. Z.), by the BMBF funding through the ErUM - Data project (K. Z.), by SAMSON AG (D. V., T. S., A. W.), by the Walter GreinerGesellschaft zur F\"orderung der physikalischen Grundla - genforschung e.V. through the Judah M. Eisenberg Laureatus Chair at Goethe Universit\"at Frankfurt am Main (H. S.), by the NVIDIA GPU grant through NVIDIA Corporation (K. Z.).

\bibliographystyle{unsrt}
\bibliography{refss}
\clearpage
\appendix
\setcounter{table}{0}
\setcounter{figure}{0}
\setcounter{equation}{0}
\renewcommand{\thetable}{A\arabic{table}}
\renewcommand{\thefigure}{A\arabic{figure}}
\renewcommand{\theequation}

\subsection*{Examples of cavitation states}
\begin{figure*}
\centering
\subfigure[Cavitation Choked Flow]{
\begin{minipage}[t]{0.45\linewidth}
\centering
\includegraphics[width=\textwidth,height=35mm]{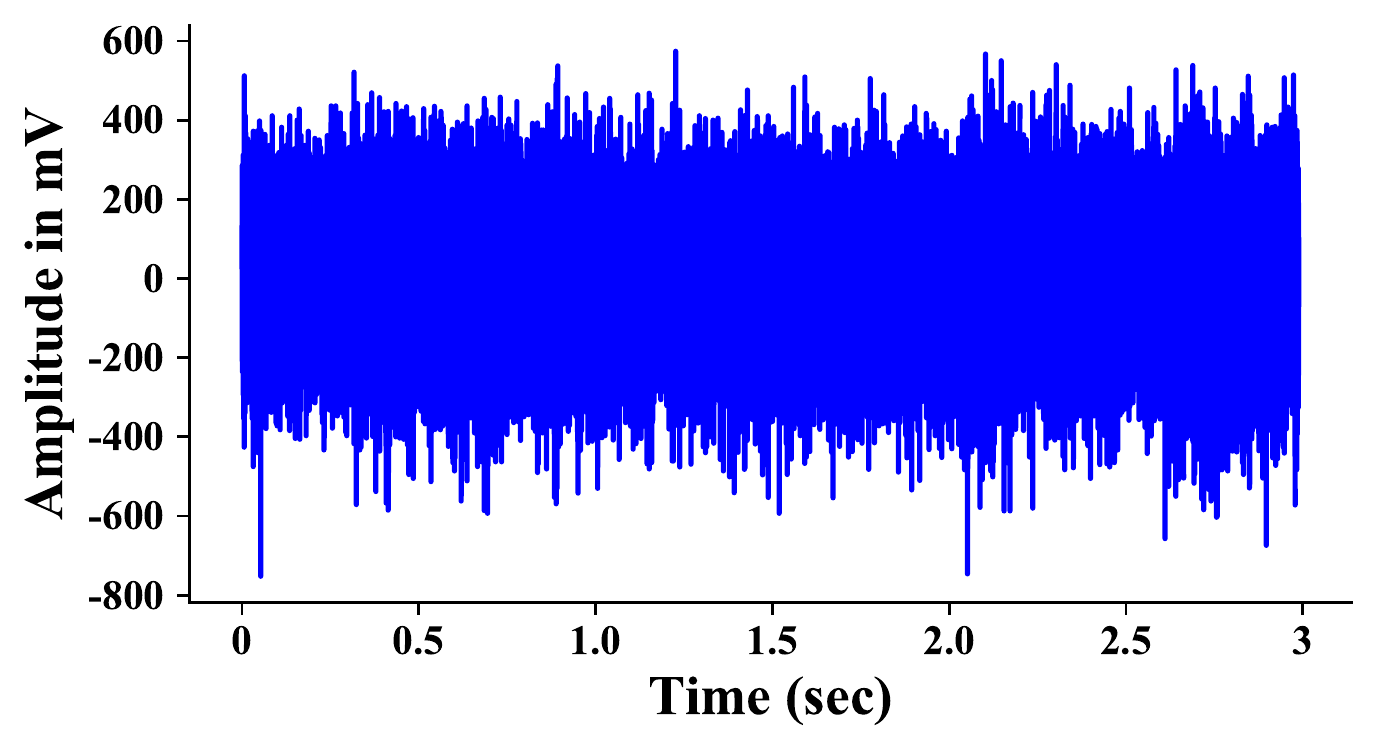}
\end{minipage}%
}%
\subfigure[Constant Cavitation]{
\begin{minipage}[t]{0.45\linewidth}
\centering
\includegraphics[width=\textwidth,height=35mm]{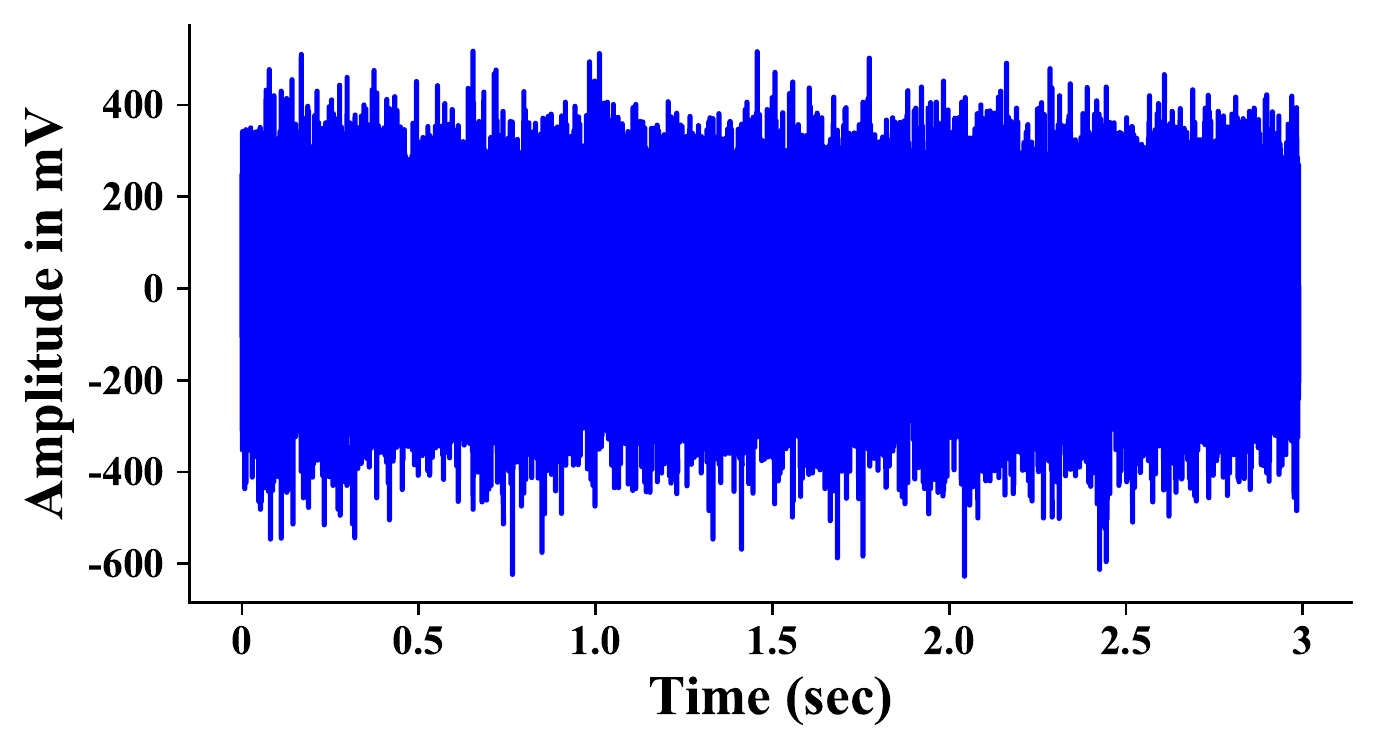}
\end{minipage}%
}%

\subfigure[Incipient Cavitation]{
\begin{minipage}[t]{0.45\linewidth}
\centering
\includegraphics[width=\textwidth,height=35mm]{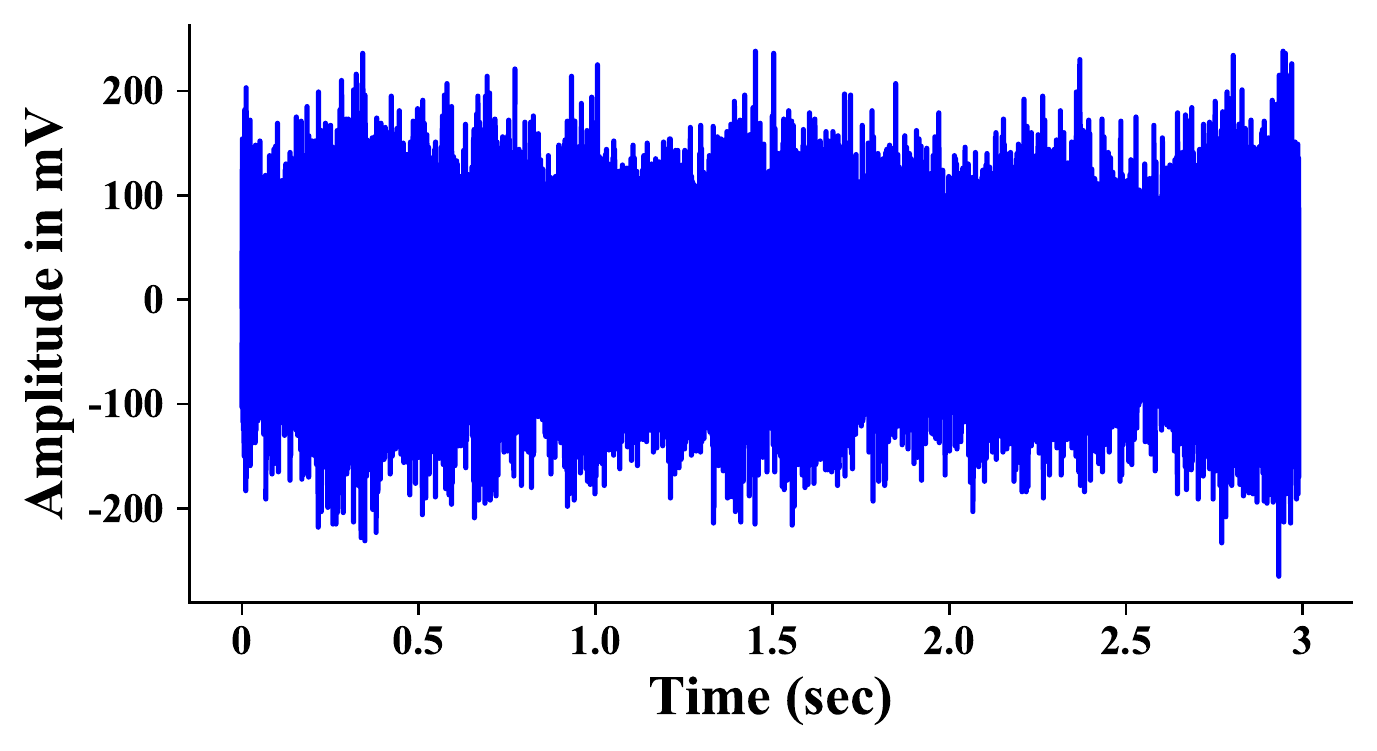}
\end{minipage}%
}%
\subfigure[Non cavitation]{
\begin{minipage}[t]{0.45\linewidth}
\centering
\includegraphics[width=\textwidth,height=35mm]{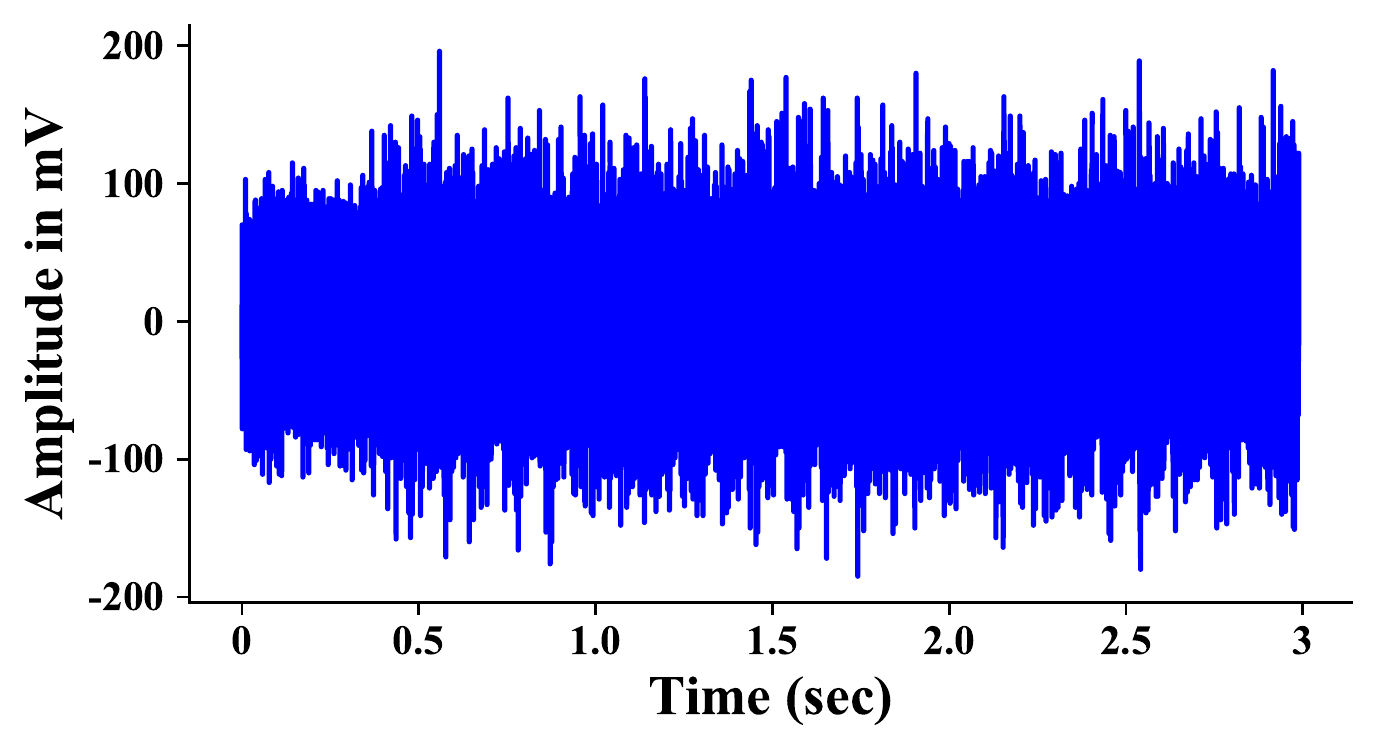}
\end{minipage}%
}%
\centering
\caption{Examples of (a)-(d) for cavitation choked flow, constant cavitation, incipient cavitation and non-cavitation in \emph{\textbf{Dataset 1}}.}
\label{fig: state example dataset1}
\end{figure*}

\begin{figure*}
\centering
\subfigure[Cavitation Choked Flow]{
\begin{minipage}[t]{0.45\linewidth}
\centering
\includegraphics[width=\textwidth,height=35mm]{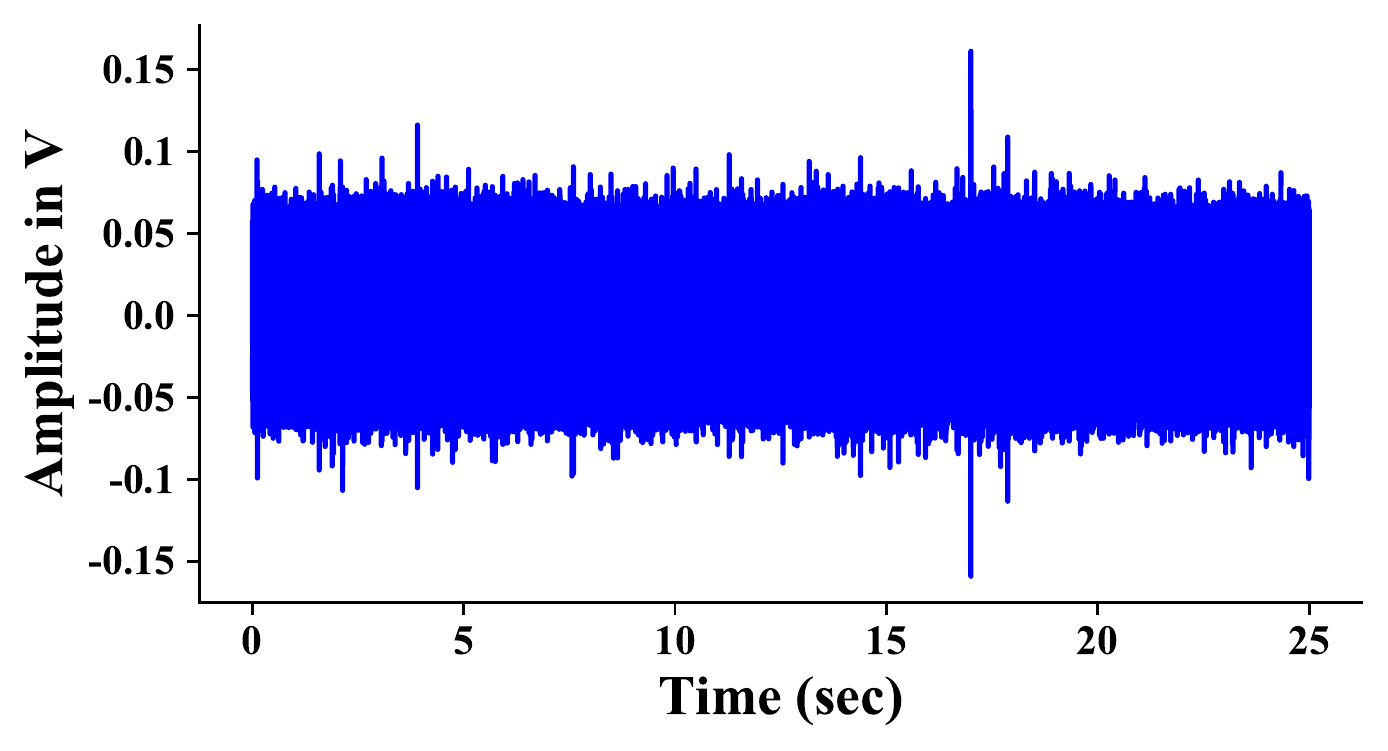}
\end{minipage}%
}%
\subfigure[Constant Cavitation]{
\begin{minipage}[t]{0.45\linewidth}
\centering
\includegraphics[width=\textwidth,height=35mm]{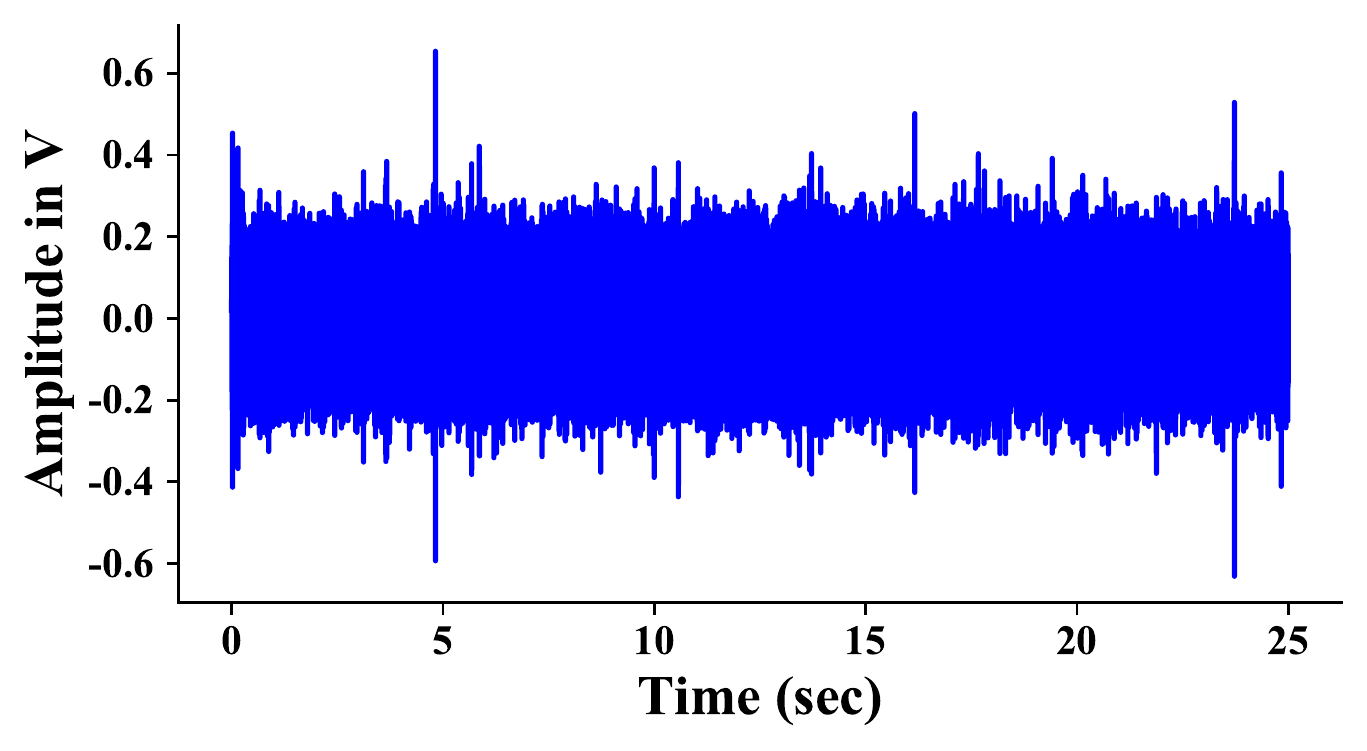}
\end{minipage}%
}%

\subfigure[Incipient Cavitation]{
\begin{minipage}[t]{0.45\linewidth}
\centering
\includegraphics[width=\textwidth,height=35mm]{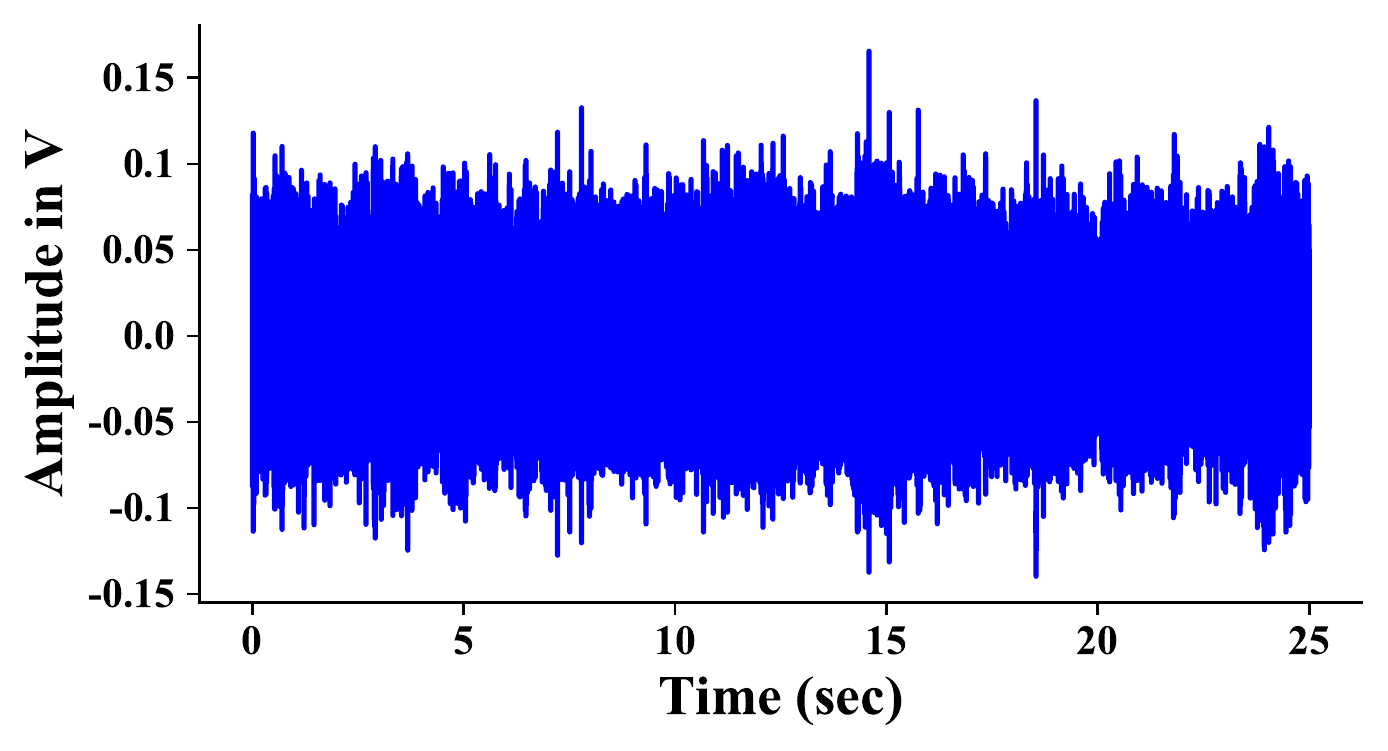}
\end{minipage}%
}%
\subfigure[Non cavitation]{
\begin{minipage}[t]{0.45\linewidth}
\centering
\includegraphics[width=\textwidth,height=35mm]{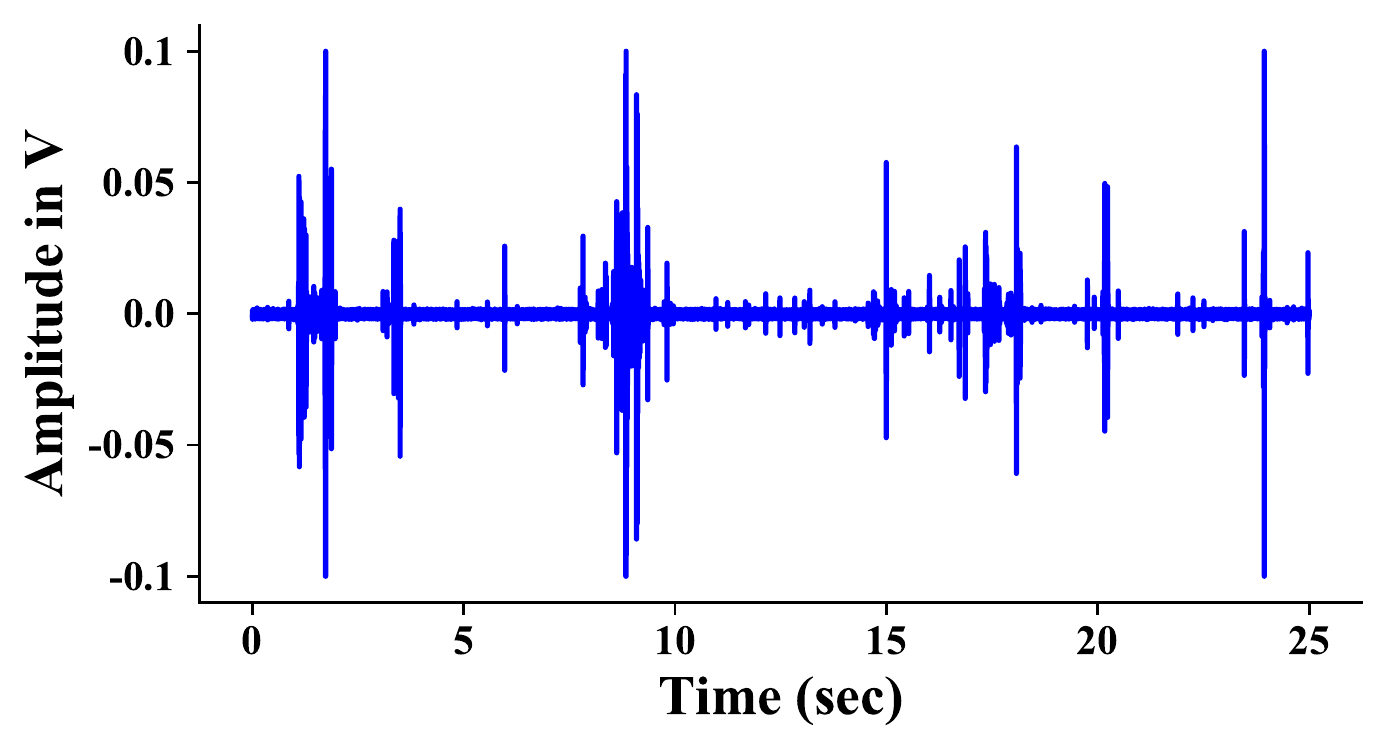}
\end{minipage}%
}%
\centering
\caption{Examples of (a)-(d) for cavitation choked flow, constant cavitation, incipient cavitation and non-cavitation in \emph{\textbf{Dataset 2}}.}
\label{fig: state example dataset2}
\end{figure*}

\begin{figure*}
\centering
\subfigure[Cavitation Choked Flow]{
\begin{minipage}[t]{0.45\linewidth}
\centering
\includegraphics[width=\textwidth,height=35mm]{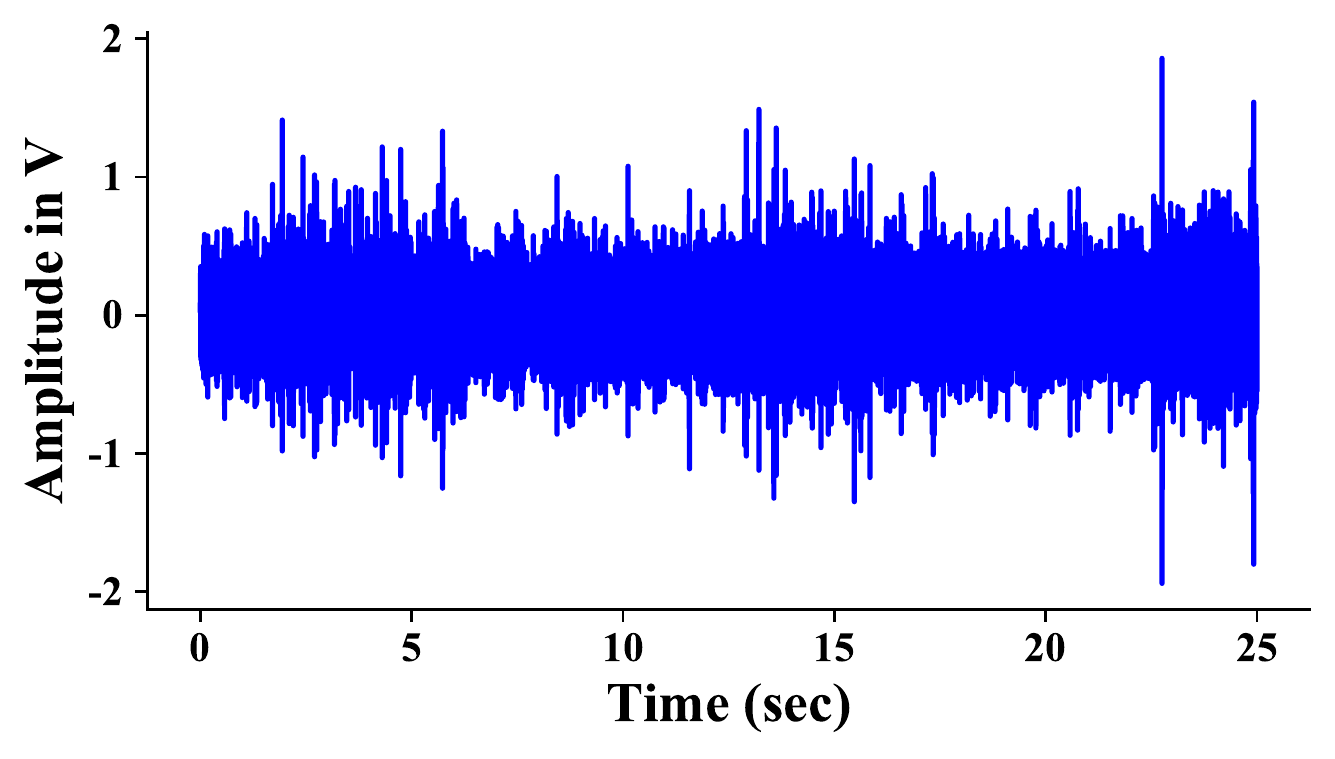}
\end{minipage}%
}%
\subfigure[Constant Cavitation]{
\begin{minipage}[t]{0.45\linewidth}
\centering
\includegraphics[width=\textwidth,height=35mm]{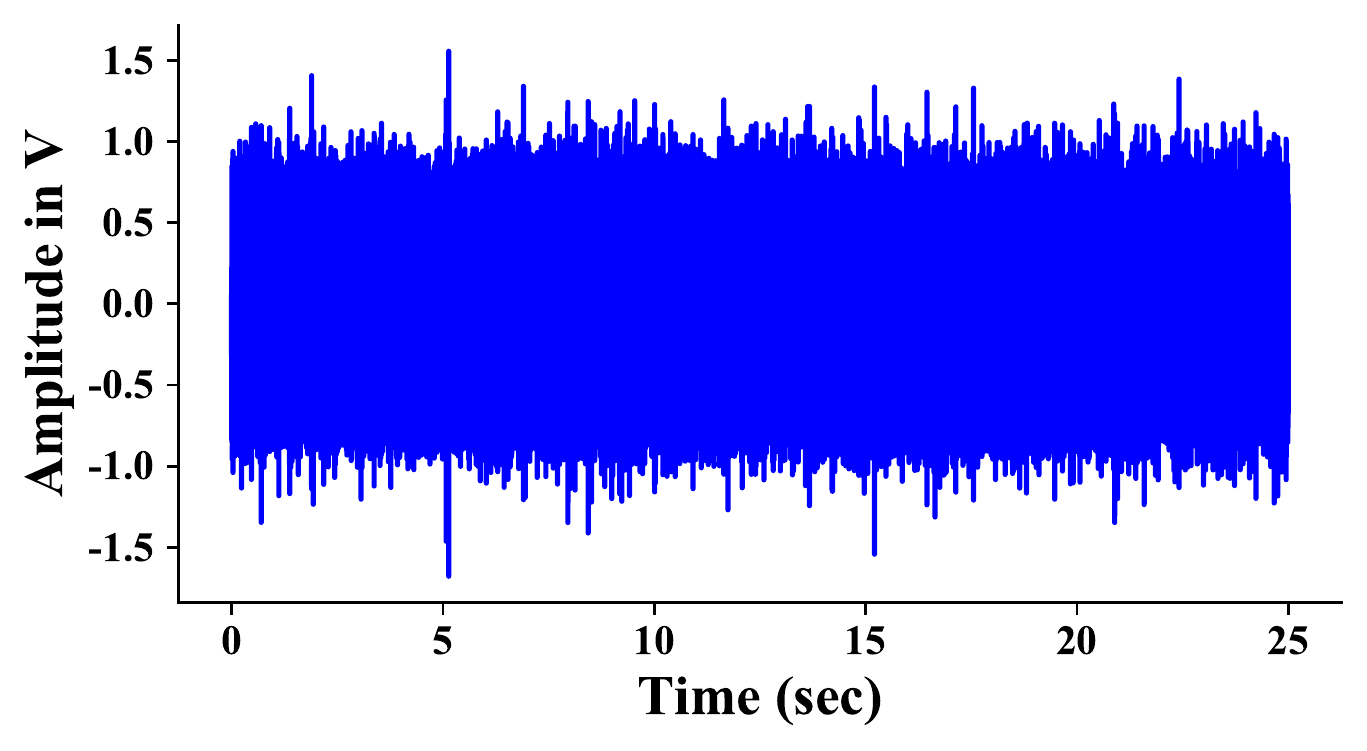}
\end{minipage}%
}%

\subfigure[Incipient Cavitation]{
\begin{minipage}[t]{0.45\linewidth}
\centering
\includegraphics[width=\textwidth,height=35mm]{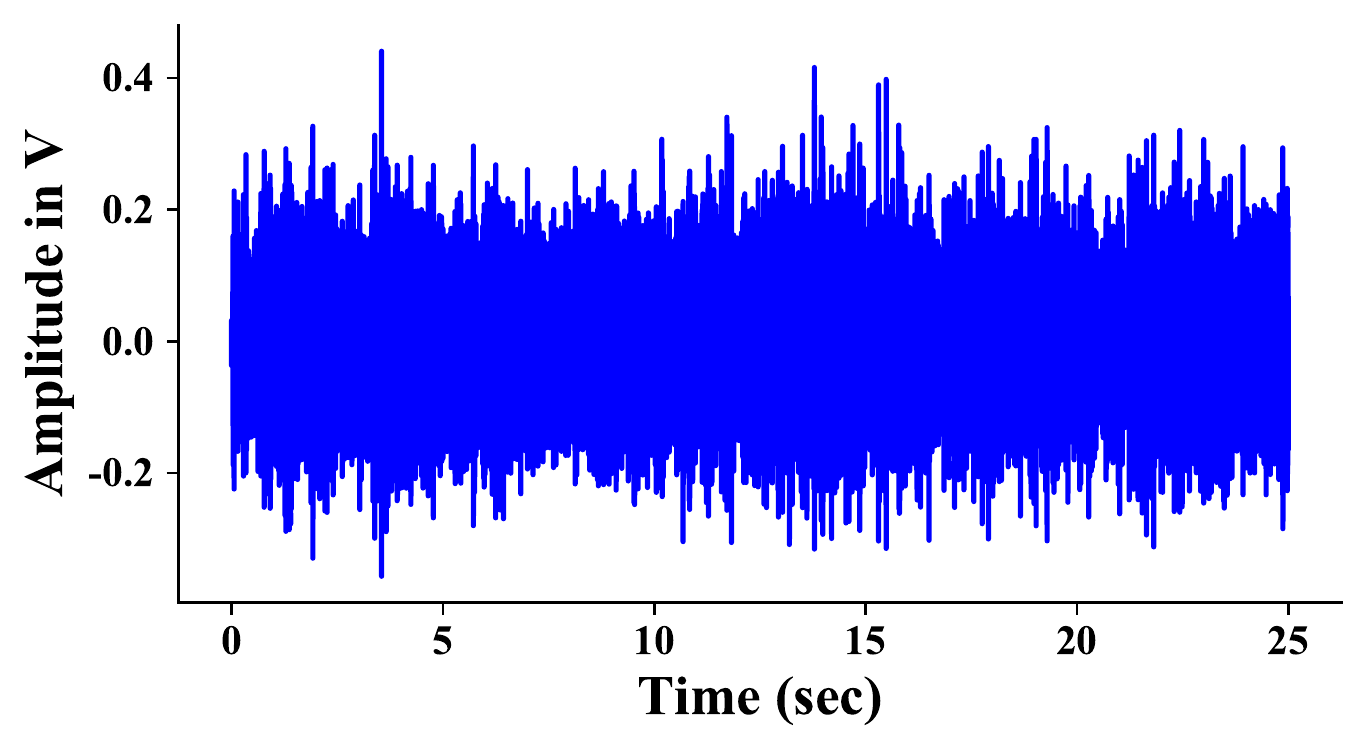}
\end{minipage}%
}%
\subfigure[Non cavitation]{
\begin{minipage}[t]{0.45\linewidth}
\centering
\includegraphics[width=\textwidth,height=35mm]{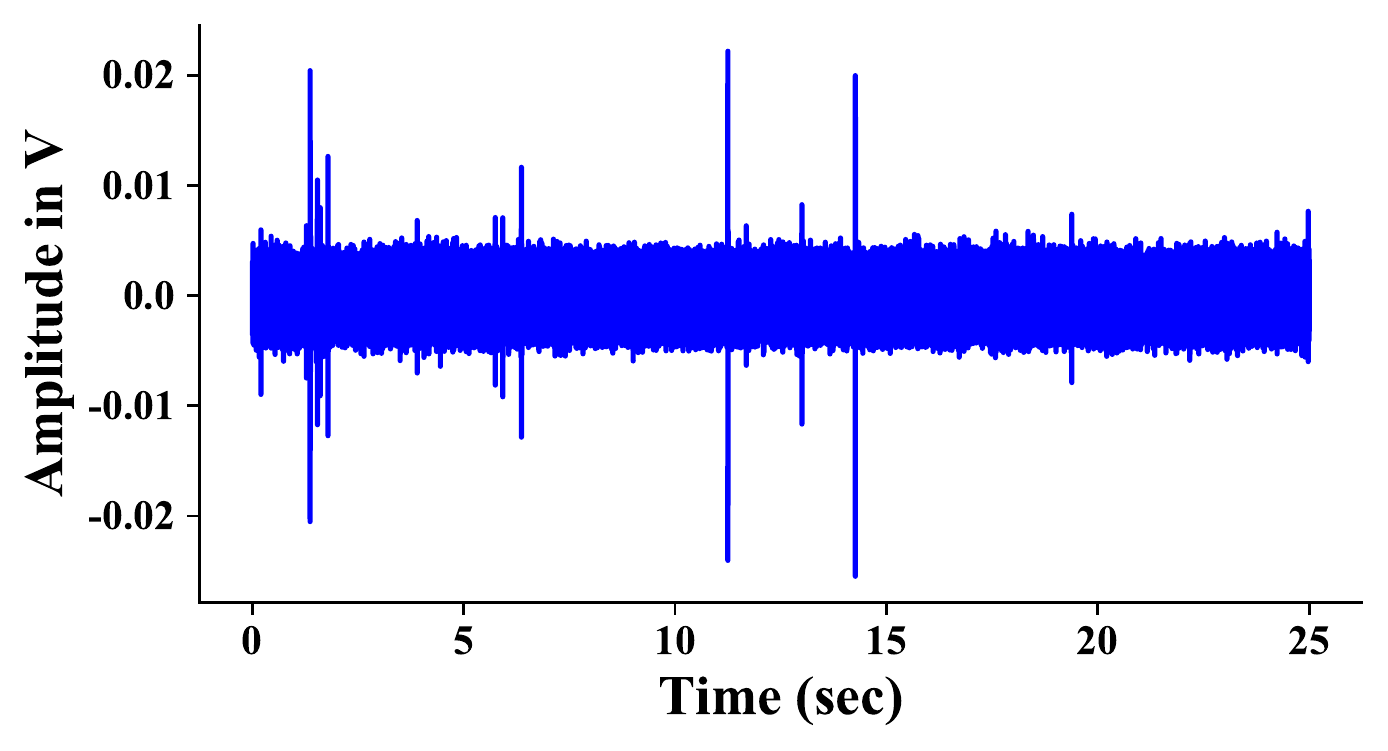}
\end{minipage}%
}%
\centering
\caption{Examples of (a)-(d) for cavitation choked flow, constant cavitation, incipient cavitation and non-cavitation in \emph{\textbf{Dataset 3}}.}
\label{fig: state example dataset3}
\end{figure*}

Figures \ref{fig: state example dataset1}, \ref{fig: state example dataset2} and \ref{fig: state example dataset3} show examples of different cavitation states for \emph{\textbf{Dataset 1}}, \emph{\textbf{Dataset 2}} and \emph{\textbf{Dataset 3}}, respectively.

\subsection*{Training and validation sets loss curves}
\begin{figure*}
\centering
\subfigure[\emph{\textbf{Dataset 1}}]{
\begin{minipage}[t]{0.3\linewidth}
\centering
\includegraphics[width=\textwidth,height=40mm]{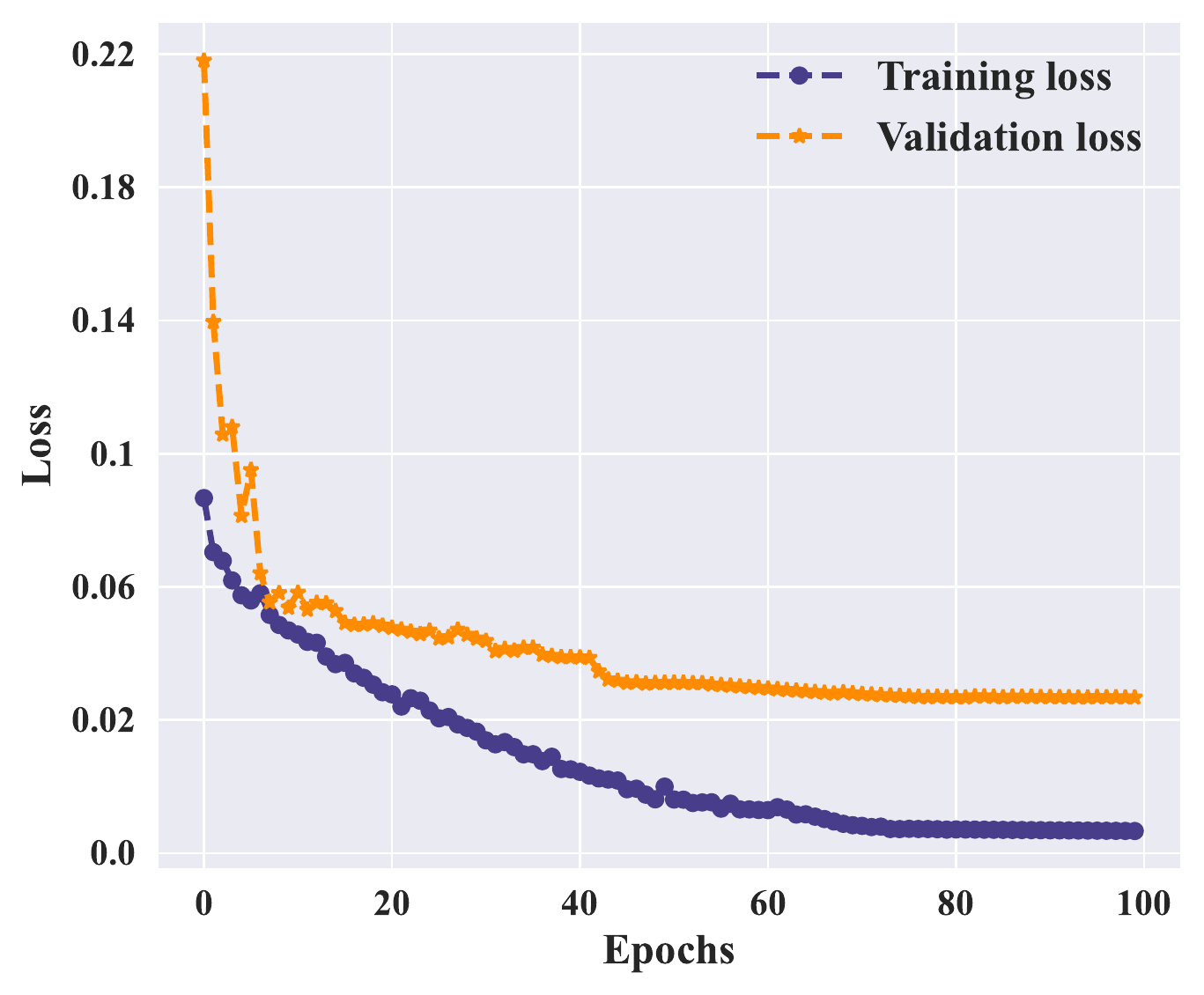}
\end{minipage}%
}%
\subfigure[\emph{\textbf{Dataset 2}}]{
\begin{minipage}[t]{0.3\linewidth}
\centering
\includegraphics[width=\textwidth,height=40mm]{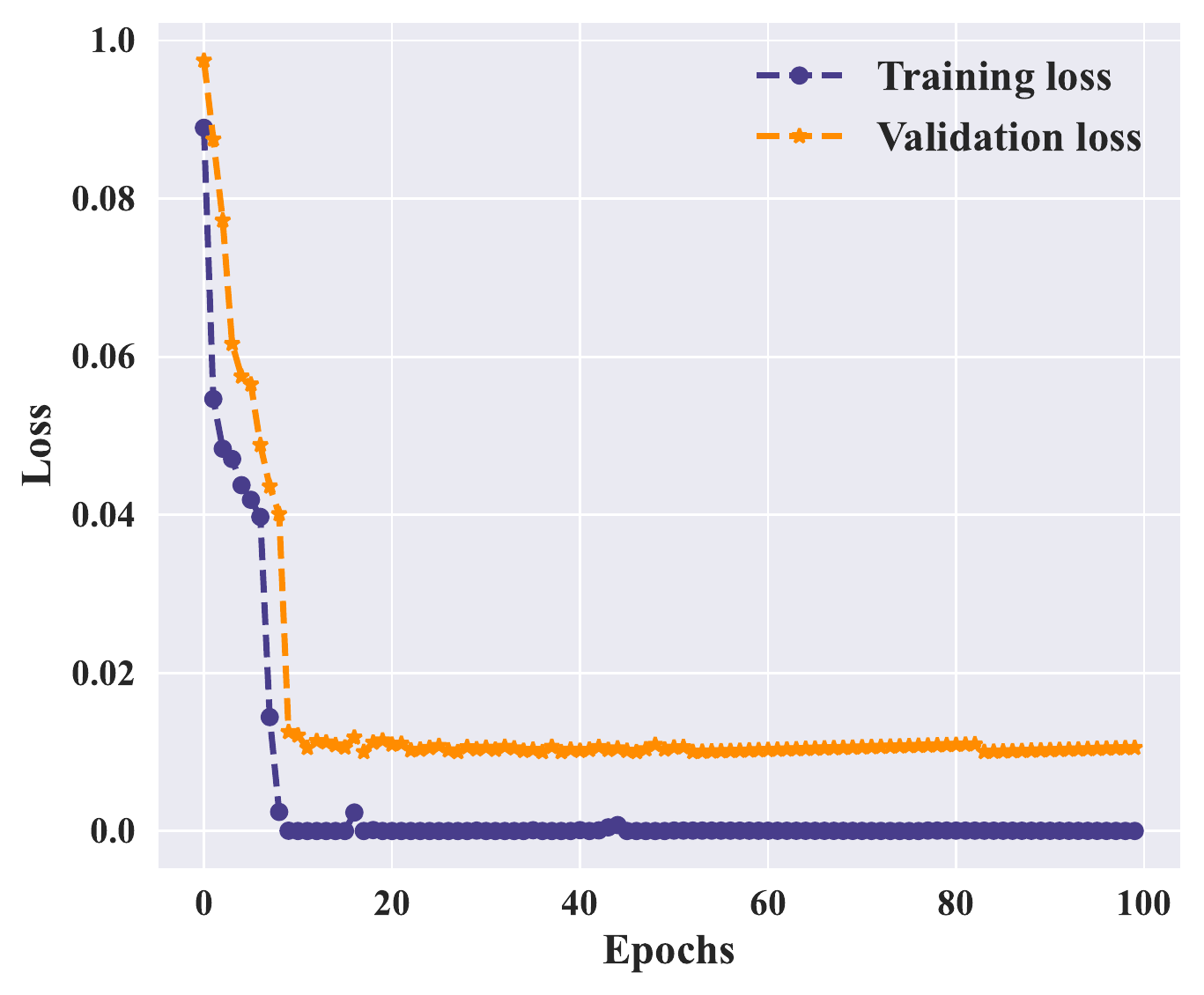}
\end{minipage}%
}%
\subfigure[\emph{\textbf{Dataset 3}}]{
\begin{minipage}[t]{0.3\linewidth}
\centering
\includegraphics[width=\textwidth,height=40mm]{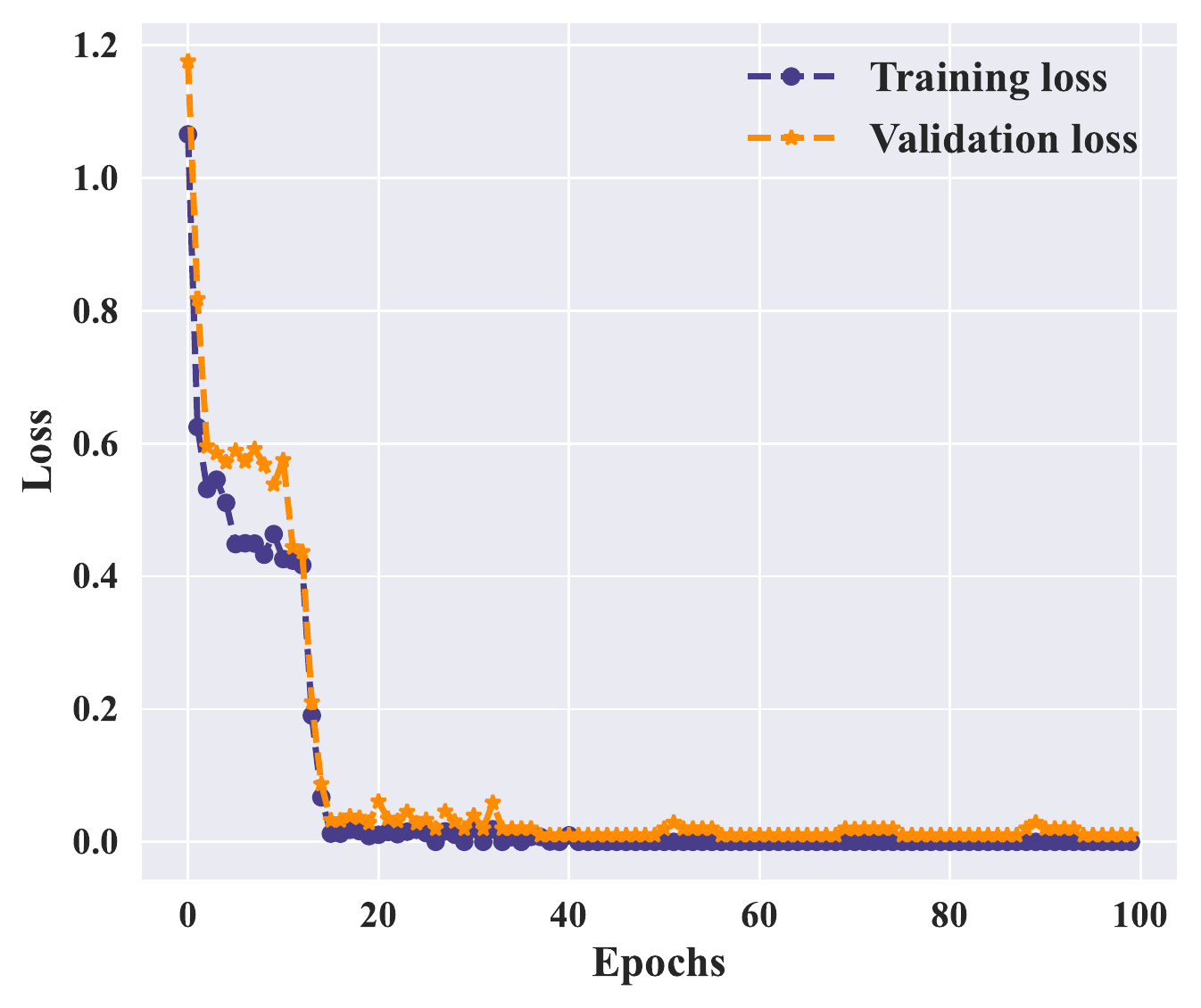}
\end{minipage}%
}%
\centering
\caption{The loss curves of our method for cavitation intensity recognition in three different real-world cavitation datasets (\emph{\textbf{Dataset 1}}, \emph{\textbf{Dataset 2}} and \emph{\textbf{Dataset 3}}).}
\label{fig: loss cavitation recognition}
\end{figure*}

\begin{figure*}
\centering
\subfigure[\emph{\textbf{Dataset 1}}]{
\begin{minipage}[t]{0.3\linewidth}
\centering
\includegraphics[width=\textwidth,height=40mm]{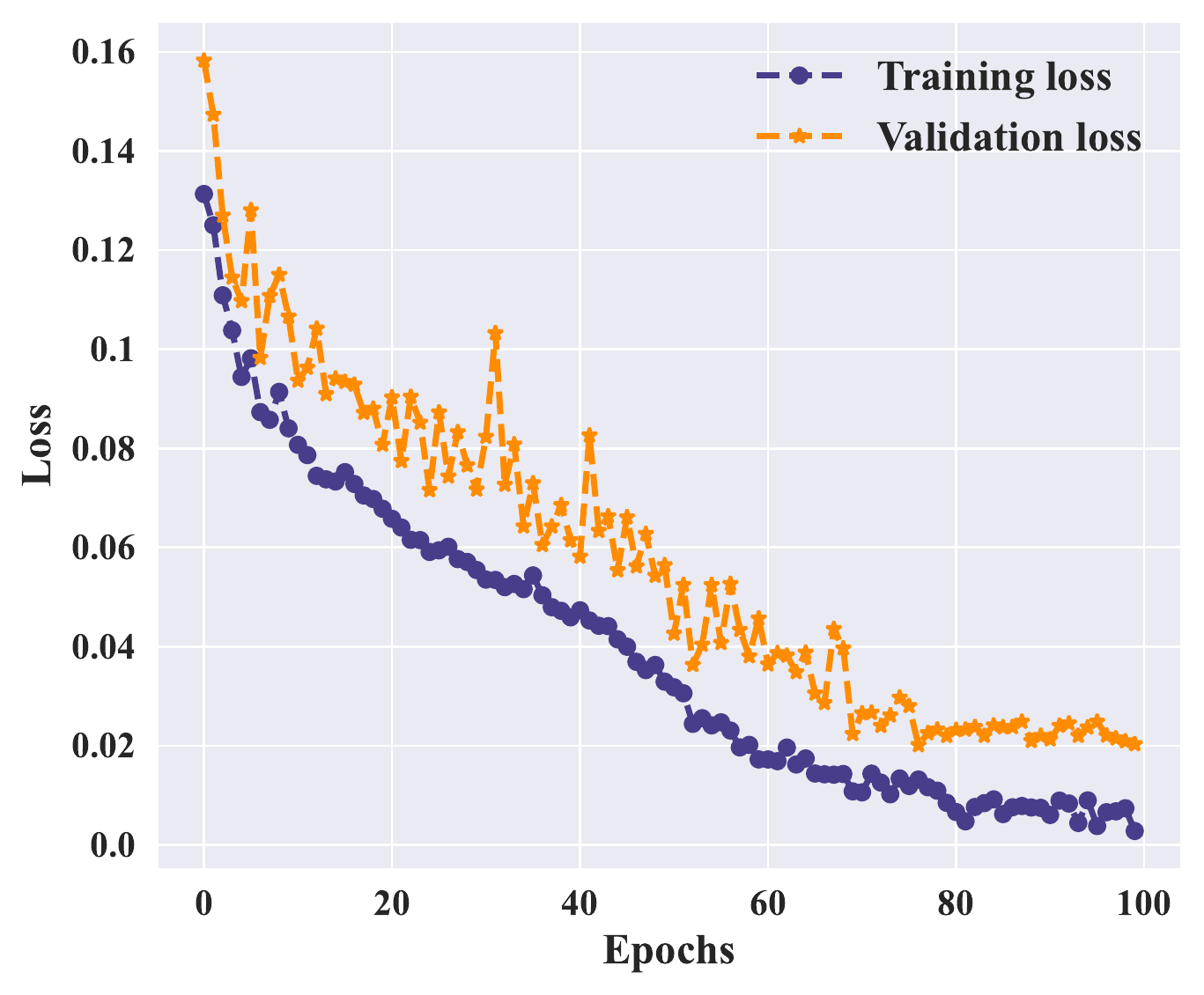}
\end{minipage}%
}%
\subfigure[\emph{\textbf{Dataset 2}}]{
\begin{minipage}[t]{0.3\linewidth}
\centering
\includegraphics[width=\textwidth,height=40mm]{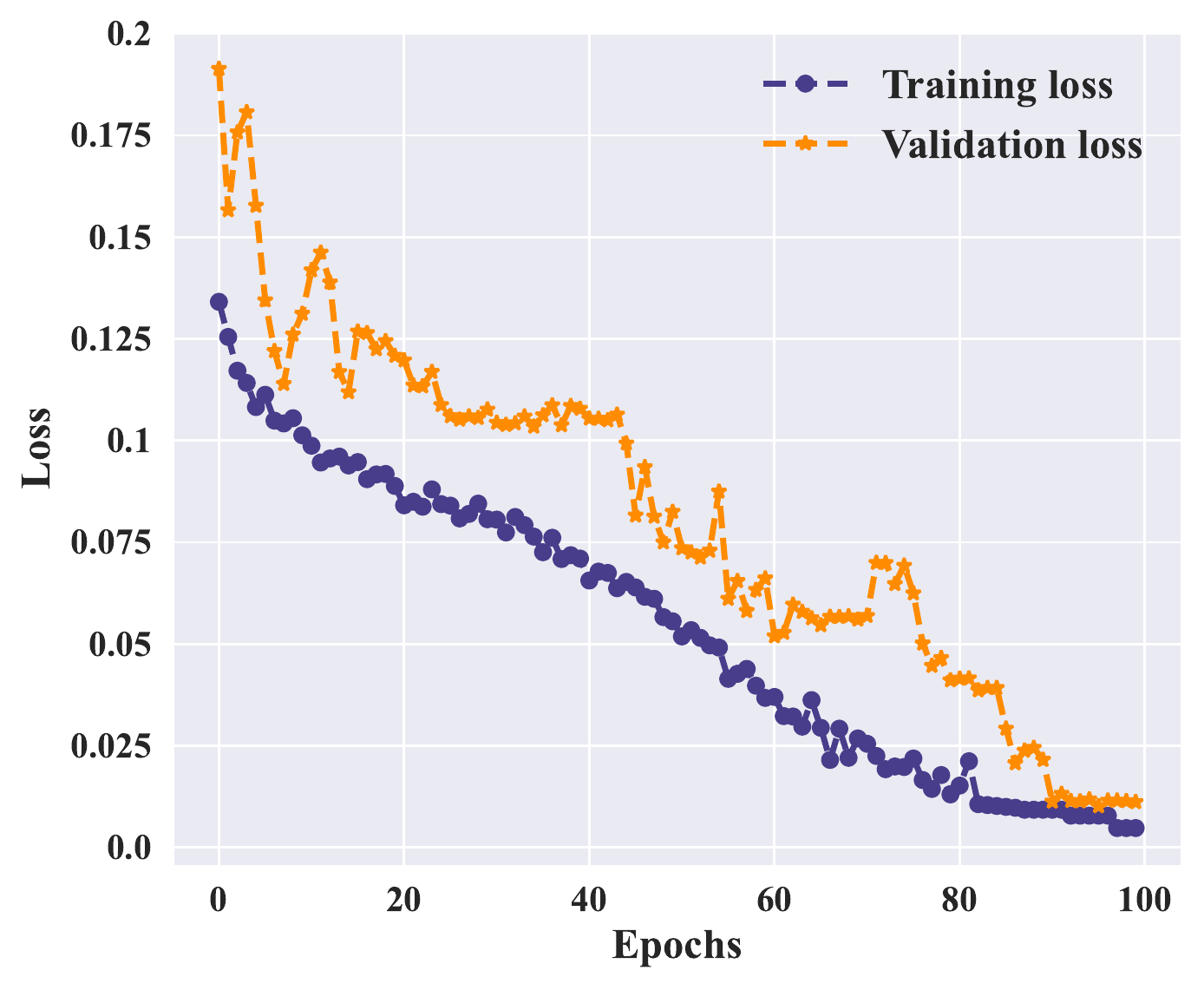}
\end{minipage}%
}%
\subfigure[\emph{\textbf{Dataset 3}}]{
\begin{minipage}[t]{0.3\linewidth}
\centering
\includegraphics[width=\textwidth,height=40mm]{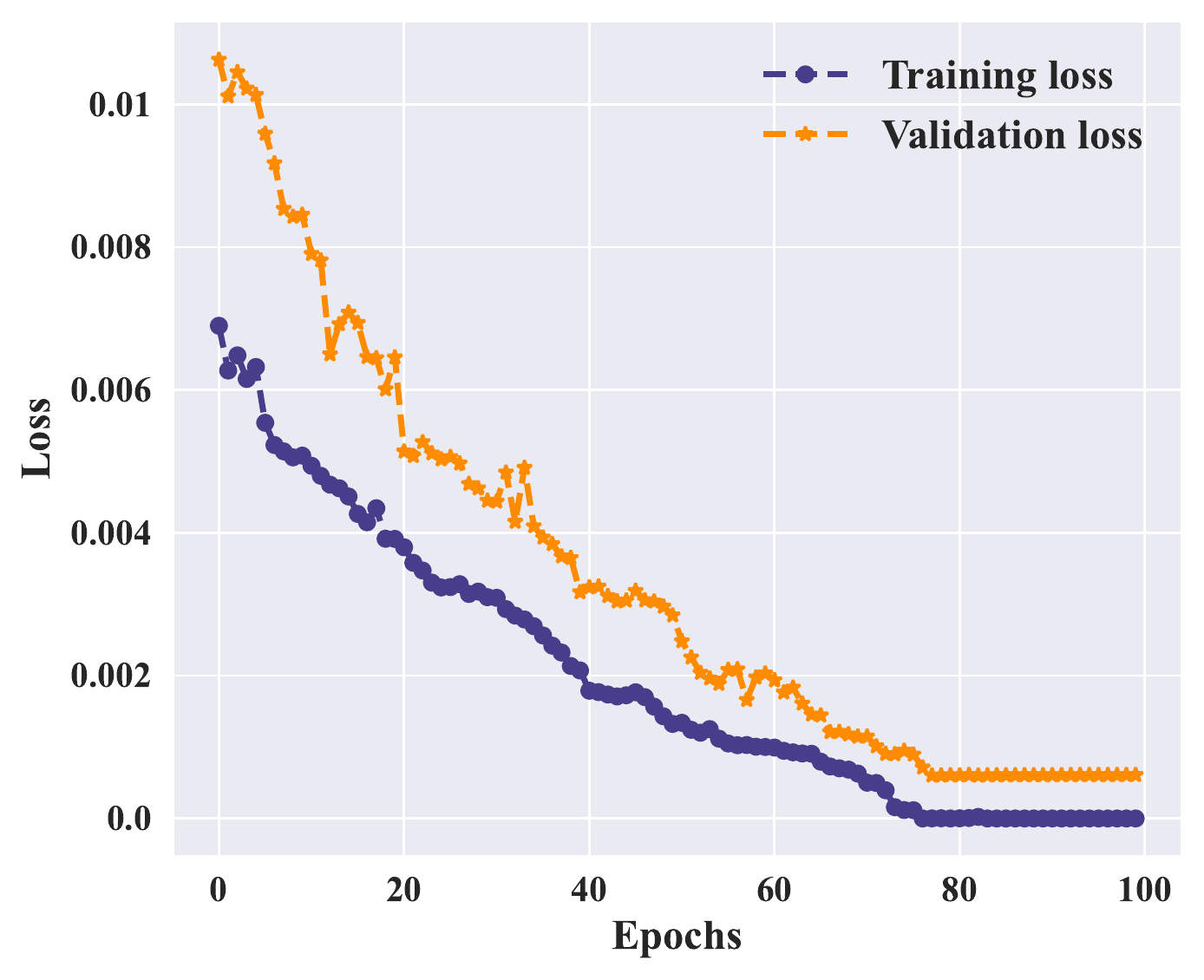}
\end{minipage}%
}%
\centering
\caption{The loss curves of our method for cavitation detection in three different real-world cavitation datasets (\emph{\textbf{Dataset 1}}, \emph{\textbf{Dataset 2}} and \emph{\textbf{Dataset 3}}).}
\label{fig: loss cavitation detection}
\end{figure*}

In order to quantitatively show that our good results are not due to overfitting of our model. We show the loss curves of our method on three different real-world cavitation datasets (\emph{\textbf{Dataset 1}}, \emph{\textbf{Dataset 2}} and \emph{\textbf{Dataset 3}}), as shown in Figures \ref{fig: loss cavitation recognition} and \ref{fig: loss cavitation detection}. As can be seen in Figures \ref{fig: loss cavitation recognition} and \ref{fig: loss cavitation detection}, our method does not overfitting on three different real-world cavitation datasets (\emph{\textbf{Dataset 1}}, \emph{\textbf{Dataset 2}} and \emph{\textbf{Dataset 3}}). By monitoring the loss of validationin in learning curve, we observe saturation which indicates that our model has converged.

\subsection*{Cavitation Intensity Recognition}
The precision, recall and F1-score of 1-D DHRN (our method) and the compared methods in \emph{\textbf{Dataset 1}}, \emph{\textbf{Dataset 2}} and \emph{\textbf{Dataset 3}} are shown in Table \ref{tab: Precision, Recall and F1-scor intensity recognition dataset1}, Table \ref{tab: Precision, Recall and F1-scor intensity recognition dataset2} and \ref{tab: Precision, Recall and F1-scor intensity recognition dataset3}.
\begin{table*}
	\caption{Precision, Recall and F1-score results of different ${W_{size}}$ of the 1-D DHRN and comparison methods for cavitation intensity recognition in \emph{\textbf{Dataset 1}} ($\%$).}
	\label{tab: Precision, Recall and F1-scor intensity recognition dataset1}
	\centering
	\setlength{\tabcolsep}{1.2mm}{
	\begin{tabular}{llcccccccccc}
	\toprule
	\multirow{2}{*}{Metric}&\multirow{2}{*}{Method} &\multicolumn{10}{c}{Window Size (${W_{size}}$) } \\
	\cmidrule(r){3-12}
	&& 2334720 & 1167360 &778240 &583680 &466944 &389120 &333531 &291840 & 259413 &233472\\
	\midrule
	 \multirow{6}{*}{Precision} &SVM \cite{yang2005cavitation}                 & 40.93 & 31.90 & 31.77 & 31.59 & 49.15 & 31.09 & 31.78 & 31.71 & 31.71 & 31.84 \\
	 &Decision Tree \cite{sakthivel2010vibration}   & 46.51 & 43.84 & 58.12 & 44.47 & 48.81 & 48.33 & 48.25 & 44.79 & 52.32 & 45.78  \\
	 &1-D CNN \cite{shervani2018cavitation}         & 64.90 & 61.34 & 73.86 & 84.33 & 90.42 & 88.13 & 81.41 & 86.22 & 84.03 & 82.55  \\
	 &XGBoost + ASFE \cite{SHA2022110897}                             & 89.21 & 83.28 & 85.07 & 85.59 & 86.99 & 80.29 & 84.03 & 78.46 & 83.33 & 83.38   \\
	 &1-D Resnet-18$^{*}$                          & 59.02 & 56.41 & 78.24 & 56.05 & 74.43 & 58.93 & 87.82 & 77.60 & 88.27 & 69.89 \\
	 &1-D DHRN                                      & 59.20 & 80.96 & 91.28 & 91.61 & 95.72 & 90.93 & 92.60 & 91.09 & 91.18 & 93.08 \\
	\midrule
	\midrule
	 \multirow{6}{*}{Recall} &SVM \cite{yang2005cavitation}                 & 48.65  & 46.81  & 46.53 & 45.99 & 50.38 & 45.28 & 46.15 & 45.92 & 46.16 & 46.71  \\
	 &Decision Tree \cite{sakthivel2010vibration}   & 43.86  & 38.78  & 46.32 & 47.10 & 51.38 & 45.53 & 47.66 & 45.66 & 49.26 & 44.31  \\
	 &1-D CNN \cite{shervani2018cavitation}         & 65.98  & 67.83  & 72.71 & 72.77 & 74.20 & 75.72 & 77.60 & 75.55 & 73.81 & 73.94   \\
	 &XGBoost + ASFE \cite{SHA2022110897}                             & 89.58  & 83.33  & 85.19 & 85.59 & 87.22 & 81.83 & 85.62 & 84.20 & 85.49 & 85.42   \\
	 &1-D Resnet-18$^{*}$                          & 62.70  & 65.49  & 70.16 & 59.77 & 73.84 & 68.09 & 80.43 & 75.33 & 72.85 & 67.45 \\
	 &1-D DHRN                                      & 65.38  & 72.40  & 83.27 & 84.24 & 88.00 & 84.12 & 79.74 & 82.30 & 84.56 & 83.09 \\
	\midrule
	\midrule
	 \multirow{6}{*}{F1-score} &SVM \cite{yang2005cavitation}                 & 41.01  & 37.93  & 37.75 & 37.39 & 42.35 & 36.87 & 37.64 & 37.51 & 37.59 & 37.28  \\
	 &Decision Tree \cite{sakthivel2010vibration}   & 40.96  & 37.19  & 43.66 & 42.03 & 43.25 & 41.60 & 42.86 & 41.87 & 47.34 & 41.47  \\
	 &1-D CNN \cite{shervani2018cavitation}         & 63.52  & 64.23  & 71.69 & 72.81 & 71.61 & 74.64 & 77.37 & 75.18 & 73.82 & 73.28   \\
	 &XGBoost + ASFE \cite{SHA2022110897}                             & 89.46  & 82.57  & 85.06 & 85.39 & 87.09 & 80.52 & 84.32 & 81.20 & 83.72 & 83.41   \\
	 &1-D Resnet-18$^{*}$                          & 60.37  & 60.06  & 72.16 & 56.58 & 70.92 & 62.98 & 83.14 & 75.70 & 77.36 & 66.81 \\
	 &1-D DHRN                                      & 62.05  & 74.05  & 85.40 & 86.32 & 90.53 & 85.70 & 80.25 & 83.87 & 86.24 & 85.46 \\
	\bottomrule
	\end{tabular}
	}
\end{table*}
\begin{table}
	\caption{Precision, Recall and F1-score results of different ${W_{size}}$ of the 1-D DHRN and comparison methods for cavitation intensity recognition in \emph{\textbf{Dataset 2}} ($\%$).}
	\label{tab: Precision, Recall and F1-scor intensity recognition dataset2}
	\centering
	\scriptsize
	\setlength{\tabcolsep}{0.4mm}{
	\begin{tabular}{llccccc}
	\toprule
    \multirow{2}{*}{Metric}    &\multirow{2}{*}{Method} &\multicolumn{5}{c}{Window Size (${W_{size}}$)} \\
	\cmidrule(r){3-7}
	                                                                         & & 2334720 & 1167360 &778240 &583680 &466944 \\
	 \midrule
	 \multirow{6}{*}{Precision} & SVM \cite{yang2005cavitation}                & 52.37 & 51.78 & 49.50 & 49.76 & 51.62   \\
	 &Decision Tree \cite{sakthivel2010vibration}                               & 66.52 & 63.48 & 63.20 & 63.20 & 60.94   \\
	 &1-D CNN \cite{shervani2018cavitation}                                     & 85.54 & 86.51 & 87.25 & 86.76 & 86.56    \\
	 &XGBoost + ASFE \cite{SHA2022110897}                                       & 85.44 & 88.82 & 89.65 & 89.21 & 87.66  \\
	 &1-D Resnet-18$^{*}$                                                       & 88.26 & 89.93 & 91.93 & 93.53 & 89.43  \\
	 &1-D DHRN                                                                  & 89.41 & 91.19 & 95.14 & 92.17 & 92.03  \\
	 \midrule	 
	 \midrule
	 \multirow{6}{*}{Recall} &SVM \cite{yang2005cavitation}                 & 50.66 & 49.89 & 48.07 & 49.06 & 50.40   \\
	 &Decision Tree \cite{sakthivel2010vibration}   & 64.84 & 62.63 & 62.06 & 62.17 & 60.04   \\
	 &1-D CNN \cite{shervani2018cavitation}         & 85.10 & 85.13 & 86.31 & 85.42 & 85.31   \\
	 &XGBoost + ASFE \cite{SHA2022110897}                             & 84.89 & 88.04 & 89.18 & 88.67 & 87.01 \\
	 &1-D Resnet-18$^{*}$                          & 87.38 & 89.47 & 86.87 & 84.06 & 84.57 \\
	 &1-D DHRN                                      & 89.38 & 82.83 & 94.02 & 89.20 & 87.43  \\
	 \midrule	 
	 \midrule
	 \multirow{6}{*}{F1-score} &SVM \cite{yang2005cavitation}                 & 51.02 & 50.28 & 48.24 & 49.41 & 50.63   \\
	 &Decision Tree \cite{sakthivel2010vibration}   & 64.96 & 62.47 & 62.31 & 62.40 & 60.28  \\
	 &1-D CNN \cite{shervani2018cavitation}         & 84.84 & 85.14 & 86.78 & 86.08 & 85.93    \\
	 &XGBoost + ASFE \cite{SHA2022110897}                             & 84.75 & 87.82 & 88.87 & 88.38 & 86.82  \\
	 &1-D Resnet-18$^{*}$                          & 87.22 & 89.70 & 88.85 & 86.94 & 86.11  \\
	 &1-D DHRN                                      & 89.39 & 86.81 & 94.58 & 90.18 & 89.24  \\
	\bottomrule
	\end{tabular}
	}
\end{table}

\begin{table}
	\caption{Precision, Recall and F1-score results of different ${W_{size}}$ of the 1-D DHRN and comparison methods for cavitation intensity recognition in \emph{\textbf{Dataset 3}} ($\%$).}
	\label{tab: Precision, Recall and F1-scor intensity recognition dataset3}
	\centering
	\scriptsize
	\setlength{\tabcolsep}{0.4mm}{
	\begin{tabular}{llccccc}
	\toprule
     \multirow{2}{*}{Metric}    &\multirow{2}{*}{Method} &\multicolumn{5}{c}{Window Size (${W_{size}}$)} \\
	\cmidrule(r){3-7}
	 && 2334720 & 1167360 &778240 &583680 &466944 \\
	 \midrule
	 \multirow{6}{*}{Precision} &SVM \cite{yang2005cavitation}                 & 52.37 & 51.78 & 49.50 & 49.76 & 51.62   \\
	 &Decision Tree \cite{sakthivel2010vibration}   & 66.52 & 63.48 & 63.20 & 63.20 & 60.94   \\
	 &1-D CNN \cite{shervani2018cavitation}         & 85.54 & 86.51 & 87.25 & 86.76 & 86.56    \\
	 &XGBoost + ASFE \cite{SHA2022110897}                             & 85.44 & 88.82 & 89.65 & 89.21 & 87.66  \\
	 &1-D Resnet-18$^{*}$                          & 88.26 & 89.93 & 91.93 & 93.53 & 89.43  \\
	 &1-D DHRN                                      & 89.41 & 91.19 & 95.14 & 92.17 & 92.03  \\
	 \midrule	 
	 \midrule
	 \multirow{6}{*}{Recall} &SVM \cite{yang2005cavitation}                 & 50.66 & 49.89 & 48.07 & 49.06 & 50.40   \\
	 &Decision Tree \cite{sakthivel2010vibration}   & 64.84 & 62.63 & 62.06 & 62.17 & 60.04   \\
	 &1-D CNN \cite{shervani2018cavitation}         & 85.10 & 85.13 & 86.31 & 85.42 & 85.31   \\
	 &XGBoost + ASFE \cite{SHA2022110897}                             & 84.89 & 88.04 & 89.18 & 88.67 & 87.01 \\
	 &1-D Resnet-18$^{*}$                          & 87.38 & 89.47 & 86.87 & 84.06 & 84.57 \\
	 &1-D DHRN                                      & 89.38 & 82.83 & 94.02 & 89.20 & 87.43  \\
	 \midrule	 
	 \midrule
	 \multirow{6}{*}{F1-score} &SVM \cite{yang2005cavitation}                 & 51.02 & 50.28 & 48.24 & 49.41 & 50.63   \\
	 &Decision Tree \cite{sakthivel2010vibration}   & 64.96 & 62.47 & 62.31 & 62.40 & 60.28  \\
	 &1-D CNN \cite{shervani2018cavitation}         & 84.84 & 85.14 & 86.78 & 86.08 & 85.93    \\
	 &XGBoost + ASFE \cite{SHA2022110897}                             & 84.75 & 87.82 & 88.87 & 88.38 & 86.82  \\
	 &1-D Resnet-18$^{*}$                          & 87.22 & 89.70 & 88.85 & 86.94 & 86.11  \\
	 &1-D DHRN                                      & 89.39 & 86.81 & 94.58 & 90.18 & 89.24  \\
	\bottomrule
	\end{tabular}
	}
\end{table}

\subsection*{Cavitation Detection}
The precision, recall and F1-score of 1-D DHRN (our method) and the compared methods in \emph{\textbf{Dataset 1}}, \emph{\textbf{Dataset 2}} and \emph{\textbf{Dataset 3}} are shown in Table \ref{tab: Precision, Recall and F1-scor cavitation detection dataset1}, Table \ref{tab: Precision, Recall and F1-scor cavitation detection dataset2} and \ref{tab: Precision, Recall and F1-scor cavitation detection dataset3}.
\begin{table*}
	\caption{Precision, Recall and F1-score results of different ${W_{size}}$ of the 1-D DHRN and comparison methods for cavitation detection in \emph{\textbf{Dataset 1}} ($\%$).}
	\label{tab: Precision, Recall and F1-scor cavitation detection dataset1}
	\centering
	\setlength{\tabcolsep}{1.2mm}{
	\begin{tabular}{llcccccccccc}
	\toprule
	\multirow{2}{*}{Metric} &\multirow{2}{*}{Method} &\multicolumn{10}{c}{Window Size (${W_{size}}$)} \\
	\cmidrule(r){3-12}
	&& 2334720 & 1167360 &778240 &583680 &466944 &389120 &333531 &291840 & 259413 &233472\\
	\midrule
	 \multirow{6}{*}{Precision} &SVM \cite{yang2005cavitation}                   & 73.61  & 73.37  & 71.95 & 70,82 & 73.40 & 69.87 & 69.62 & 70.85 & 72.38 & 79.15  \\
	 &Decision Tree \cite{sakthivel2010vibration}     & 92.59  & 88.41  & 88.30 & 90.51 & 99.29 & 89.00 & 98.56 & 84.42 & 97.31 & 92.61  \\
	 &1-D CNN \cite{shervani2018cavitation}           & 94.64  & 94.17  & 93.95 & 95.14 & 94.66 & 95.35 & 94.96 & 95.63 & 95.53 & 95.84   \\
	 &XGBoost + ASFE \cite{SHA2022110897}                               & 93.14  & 92.61  & 92.48 & 91.63 & 91.57 & 91.83 & 89.26 & 90.28 & 89.83 & 90.29  \\
	 &1-D Resnet-18$^{*}$                            & 94.72  & 95.63  & 95.45 & 95.26 & 95.59 & 95.91 & 95.85 & 95.63 & 95.29 & 95.37 \\
	 &1-D DHRN                                        & 96.15  & 96.55  & 96.45 & 96.16 & 96.19 & 96.75 & 96.58 & 96.02 & 96.05 & 96.09 \\
	 \midrule	 
	 \midrule
	 \multirow{6}{*}{Recall} &SVM \cite{yang2005cavitation}                   & 88.33 & 94.17 & 88.33 & 90.00 & 95.67 & 90.83 & 91.67 & 90.63 & 88.33 & 85.83  \\
	 &Decision Tree \cite{sakthivel2010vibration}     & 41.67 & 50.83 & 46.11 & 51.67 & 46.67 & 51.67 & 49.05 & 54.17 & 53.52 & 62.67  \\
	 &1-D CNN \cite{shervani2018cavitation}           & 93.75 & 94.17 & 93.89 & 93.75 & 93.50 & 94.93 & 94.61 & 94.48 & 95.76 & 62.67   \\
	 &XGBoost + ASFE \cite{SHA2022110897}                               & 93.56 & 92.36 & 92.36 & 91.32 & 91.25 & 91.09 & 88.99 & 89.93 & 89.51 & 90.00  \\
	 &1-D Resnet-18$^{*}$                            & 95.42 & 95.63 & 96.25 & 95.99 & 95.08 & 96.63 & 96.58 & 96.12 & 95.51 & 95.90  \\
	 &1-D DHRN                                        & 96.88 & 96.25 & 97.08 & 96.77 & 96.46 & 97.40 & 96.07 & 96.30 & 96.25 & 95.90 \\
	 \midrule	 
	 \midrule
	 \multirow{6}{*}{F1-score} &SVM \cite{yang2005cavitation}                   & 80.30 & 82.48 & 79.30 & 79.27 & 83.06 & 78.99 & 79.14 & 79.52 & 79.57 & 77.91  \\
	 &Decision Tree \cite{sakthivel2010vibration}     & 57.47 & 64.55 & 60.58 & 65.78 & 63.49 & 65.38 & 65.51 & 65.99 & 69.06 & 74.75  \\
	 &1-D CNN \cite{shervani2018cavitation}           & 94.11 & 94.17 & 93.92 & 94.27 & 93.95 & 95.12 & 94.77 & 94.93 & 95.64 & 95.05   \\
	 &XGBoost + ASFE \cite{SHA2022110897}                               & 93.08 & 92.39 & 92.39 & 91.38 & 91.31 & 91.19 & 89.06 & 90.00 & 89.58 & 90.07  \\
	 &1-D Resnet-18$^{*}$                            & 94.94 & 95.63 & 95.67 & 95.49 & 95.31 & 96.15 & 96.08 & 95.83 & 95.39 & 95.56 \\
	 &1-D DHRN                                        & 96.51 & 96.40 & 96.76 & 96.46 & 96.32 & 97.07 & 96.32 & 96.16 & 96.15 & 95.99 \\
	\bottomrule
	\end{tabular}
	}
\end{table*}
\begin{table}
	\caption{Precision, Recall and F1-score results of different ${W_{size}}$ of the 1-D DHRN and comparison methods for cavitation detection in \emph{\textbf{Dataset 2}} ($\%$).}
	\label{tab: Precision, Recall and F1-scor cavitation detection dataset2}
	\centering
	\scriptsize
	\setlength{\tabcolsep}{0.4mm}{
	\begin{tabular}{llccccc}
	\toprule
	\multirow{2}{*}{Metric} &\multirow{2}{*}{Method} &\multicolumn{5}{c}{Window Size (${W_{size}}$)} \\
	\cmidrule(r){3-7}
	&& 2334720 & 1167360 &778240 &583680 &466944 \\
	\midrule
	 \multirow{6}{*}{Precision} &SVM \cite{yang2005cavitation}                 & 87.88 & 86.98 & 85.60 & 86.19 & 84.70   \\
	 &Decision Tree \cite{sakthivel2010vibration}   & 88.20 & 82.64 & 86.21 & 86.14 & 82.42  \\
	 &1-D CNN \cite{shervani2018cavitation}         & 94.64 & 95.35 & 93.95 & 95.14 & 94.66    \\
	 &XGBoost + ASFE \cite{SHA2022110897}                             & 90.54 & 90.75 & 91.53 & 91.24 & 90.37  \\
	 &1-D Resnet-18$^{*}$                          & 95.85 & 95.91 & 95.45 & 95.26 & 95.59  \\
	 &1-D DHRN                                      & 96.45 & 98.32 & 98.47 & 96.16 & 96.19  \\
	 \midrule	 
	 \midrule
	 \multirow{6}{*}{Recall} &SVM \cite{yang2005cavitation}                 & 69.71 & 76.66 & 68.62 & 72.00 & 70.81   \\
	 &Decision Tree \cite{sakthivel2010vibration}   & 68.27 & 69.93 & 72.15 & 71.69 & 72.10   \\
	 &1-D CNN \cite{shervani2018cavitation}         & 93.75 & 94.93 & 93.89 & 93.75 & 93.50    \\
	 &XGBoost + ASFE \cite{SHA2022110897}                             & 90.56 & 90.73 & 91.33 & 91.02 & 90.43  \\
	 &1-D Resnet-18$^{*}$                          & 95.48 & 96.63 & 96.25 & 95.99 & 95.08  \\
	 &1-D DHRN                                      & 94.08 & 95.40 & 95.38 & 96.77 & 96.46 \\
	 \midrule	 
	 \midrule
	 \multirow{6}{*}{F1-score} &SVM \cite{yang2005cavitation}                 & 77.75 & 89.04 & 76.17 & 78.46 & 77.13  \\
	 &Decision Tree \cite{sakthivel2010vibration}   & 76.96 & 75.76 & 78.56 & 78.25 & 76.92   \\
	 &1-D CNN \cite{shervani2018cavitation}         & 94.11 & 95.12 & 93.92 & 94.27 & 93.95    \\
	 &XGBoost + ASFE \cite{SHA2022110897}                             & 90.16 & 90.32 & 90.90 & 90.54 & 90.04  \\
	 &1-D Resnet-18$^{*}$                          & 95.47 & 96.15 & 95.67 & 95.49 & 95.31  \\
	 &1-D DHRN                                      & 96.76 & 96.83 & 96.90 & 96.46 & 96.32  \\
	\bottomrule
	\end{tabular}
	}
\end{table}
\begin{table}
	\caption{Precision, Recall and F1-score results of different ${W_{size}}$ of the 1-D DHRN and comparison methods for cavitation detection in \emph{\textbf{Dataset 3}} ($\%$).}
	\label{tab: Precision, Recall and F1-scor cavitation detection dataset3}
	\centering
	\scriptsize
	\setlength{\tabcolsep}{0.4mm}{
	\begin{tabular}{llccccc}
	\toprule
	\multirow{2}{*}{Metric} &\multirow{2}{*}{Method} &\multicolumn{5}{c}{Window Size (${W_{size}}$)} \\
	\cmidrule(r){3-7}
	&& 2334720 & 1167360 &778240 &583680 &466944 \\
	\midrule
	 \multirow{6}{*}{Precision} &SVM \cite{yang2005cavitation}                 & 100.00  & 100.00  & 100.00 & 100.00 & 100.00    \\
	 &Decision Tree \cite{sakthivel2010vibration}   & 100.00  & 100.00  & 100.00 & 100.00 & 100.00   \\
	 &1-D CNN \cite{shervani2018cavitation}         & 100.00  & 100.00  & 100.00 & 100.00 & 100.00     \\
	 &XGBoost + ASFE \cite{SHA2022110897}                             & 100.00  & 100.00  & 100.00 & 100.00 & 100.00   \\
	 &1-D Resnet-18$^{*}$                          & 100.00  & 100.00  & 100.00 & 100.00 & 100.00  \\
	 &1-D DHRN                                      & 100.00  & 100.00  & 100.00 & 100.00 & 100.00   \\
	 \midrule	 
	 \midrule
	 \multirow{6}{*}{Recall} &SVM \cite{yang2005cavitation}                 & 100.00  & 100.00  & 100.00 & 100.00 & 100.00    \\
	 &Decision Tree \cite{sakthivel2010vibration}   & 100.00  & 100.00  & 100.00 & 100.00 & 100.00    \\
	 &1-D CNN \cite{shervani2018cavitation}         & 100.00  & 100.00  & 100.00 & 100.00 & 100.00     \\
	 &XGBoost + ASFE \cite{SHA2022110897}                             & 100.00  & 100.00  & 100.00 & 100.00 & 100.00   \\
	 &1-D Resnet-18$^{*}$                          & 100.00  & 100.00  & 100.00 & 100.00 & 100.00   \\
	 &1-D DHRN                                      & 100.00  & 100.00  & 100.00 & 100.00 & 100.00  \\
	 \midrule	 
	 \midrule
	 \multirow{6}{*}{F1-score} &SVM \cite{yang2005cavitation}                 & 100.00  & 100.00  & 100.00 & 100.00 & 100.00   \\
	 &Decision Tree \cite{sakthivel2010vibration}   & 100.00  & 100.00  & 100.00 & 100.00 & 100.00   \\
	 &1-D CNN \cite{shervani2018cavitation}         & 100.00  & 100.00  & 100.00 & 100.00 & 100.00     \\
	 &XGBoost + ASFE \cite{SHA2022110897}                            & 100.00  & 100.00  & 100.00 & 100.00 & 100.00   \\
	 &1-D Resnet-18$^{*}$                          & 100.00  & 100.00  & 100.00 & 100.00 & 100.00   \\
	 &1-D DHRN                                      & 100.00  & 100.00  & 100.00 & 100.00 & 100.00   \\
	\bottomrule
	\end{tabular}
	}
\end{table}

\begin{figure}
    \centering
    \includegraphics[width=0.5\textwidth,height=60mm]{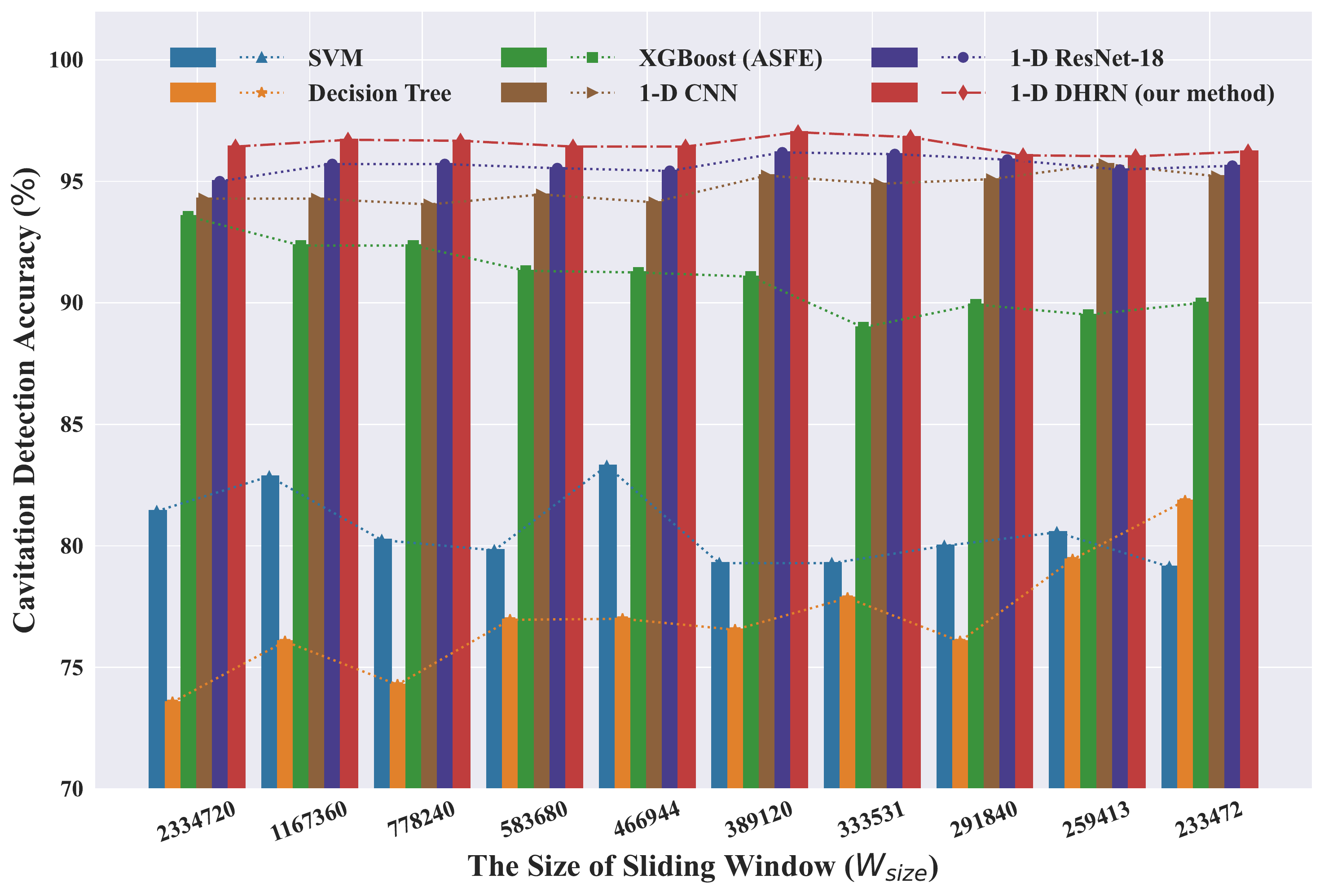}
    \caption{The effect of ${W_{size}}$ on cavitation detection accuracy under different methods in \emph{\textbf{Dataset 1}}.}
    \label{fig: Accuracy cavitation detection Dataset1}
\end{figure}

\begin{figure}
    \centering
    \includegraphics[width=0.5\textwidth,height=60mm]{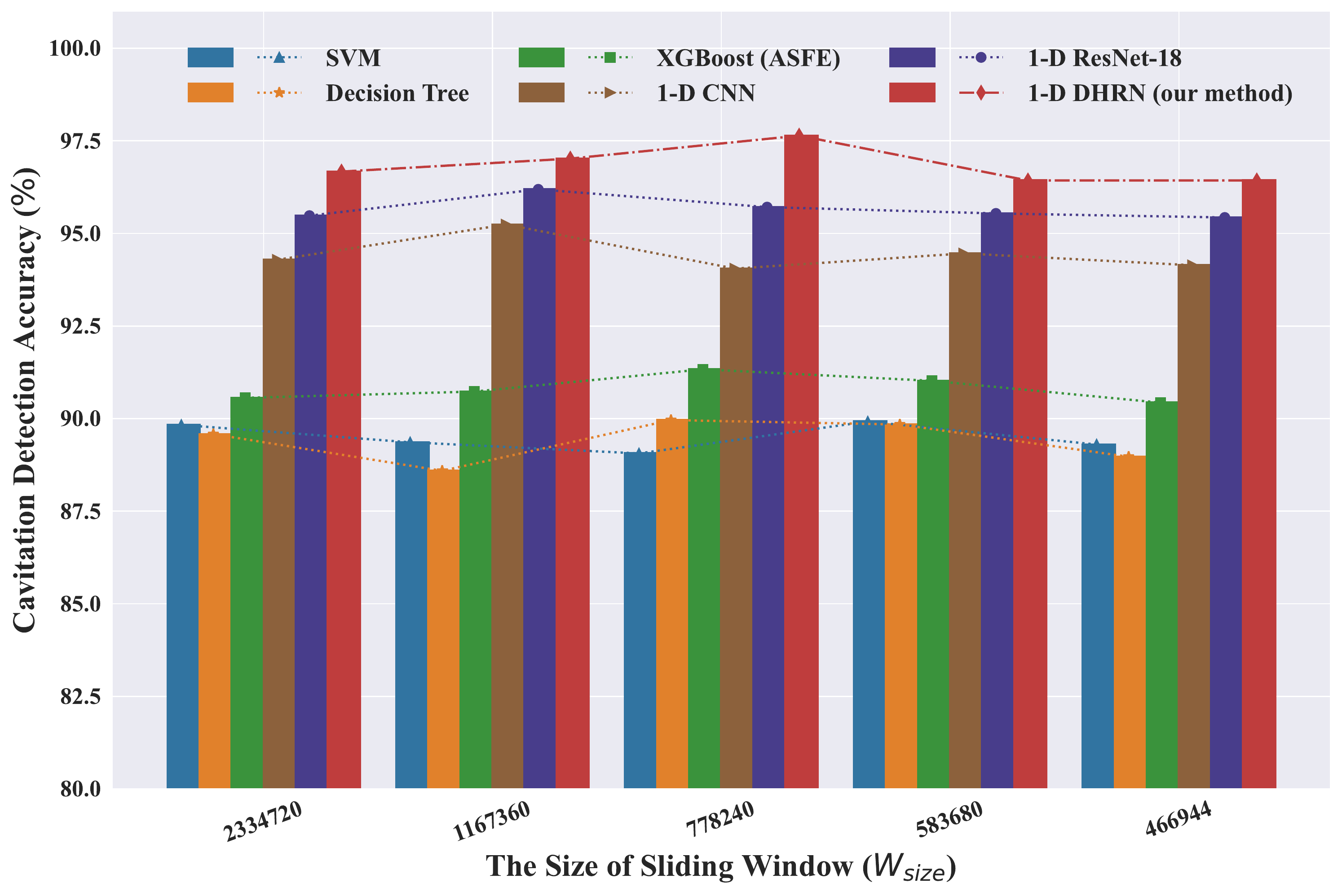}
    \caption{The effect of ${W_{size}}$ on cavitation detection accuracy under different methods in \emph{\textbf{Dataset 2}}.}
    \label{fig: Accuracy cavitation detection Dataset2}
\end{figure}

\begin{figure}
    \centering
    \includegraphics[width=0.5\textwidth,height=60mm]{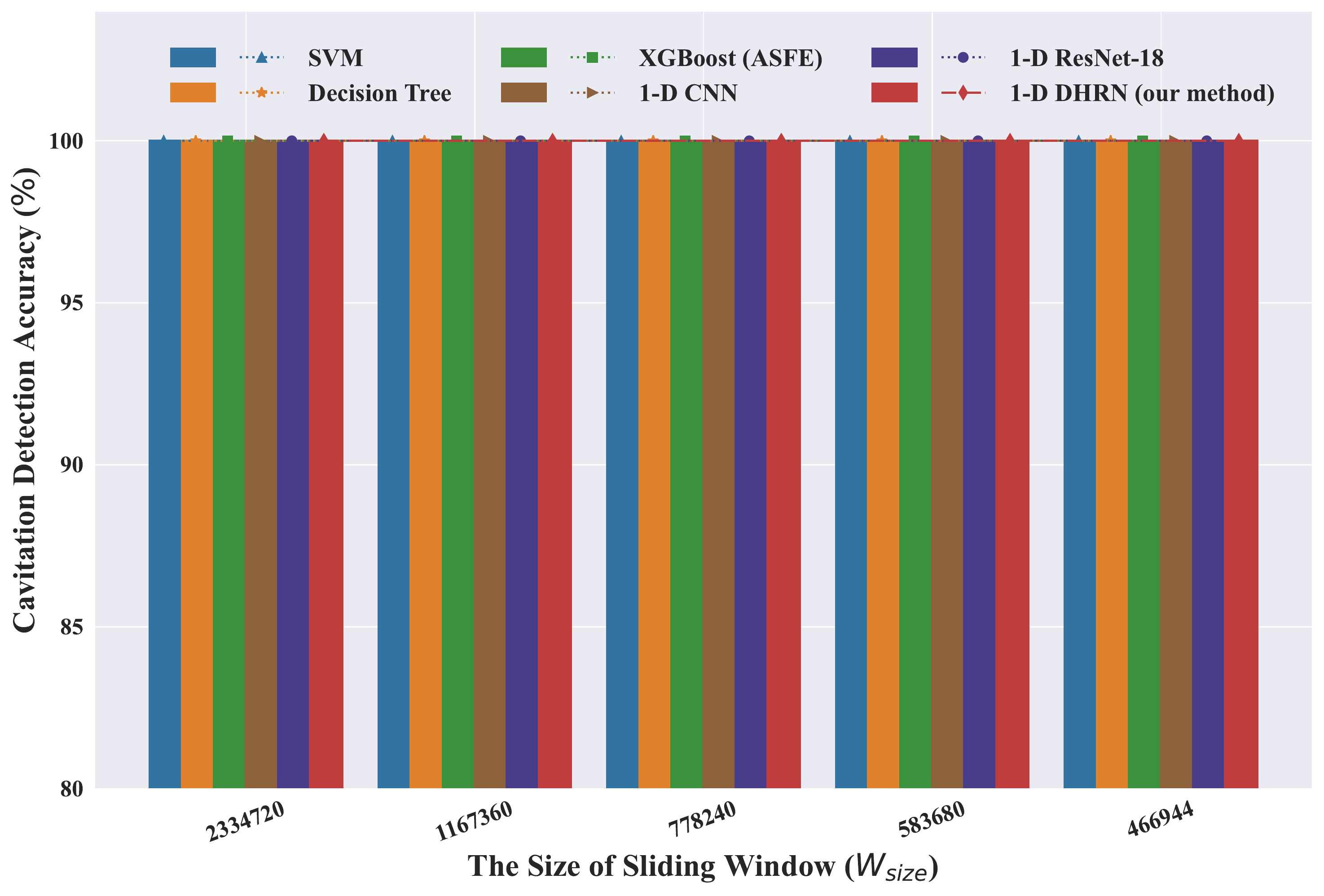}
    \caption{The effect of ${W_{size}}$ on cavitation detection accuracy under different methods in \emph{\textbf{Dataset 3}}.}
    \label{fig: Accuracy cavitation detection Dataset3}
\end{figure}

\begin{figure}
    \centering
    \includegraphics[width=0.45\textwidth,height=62mm]{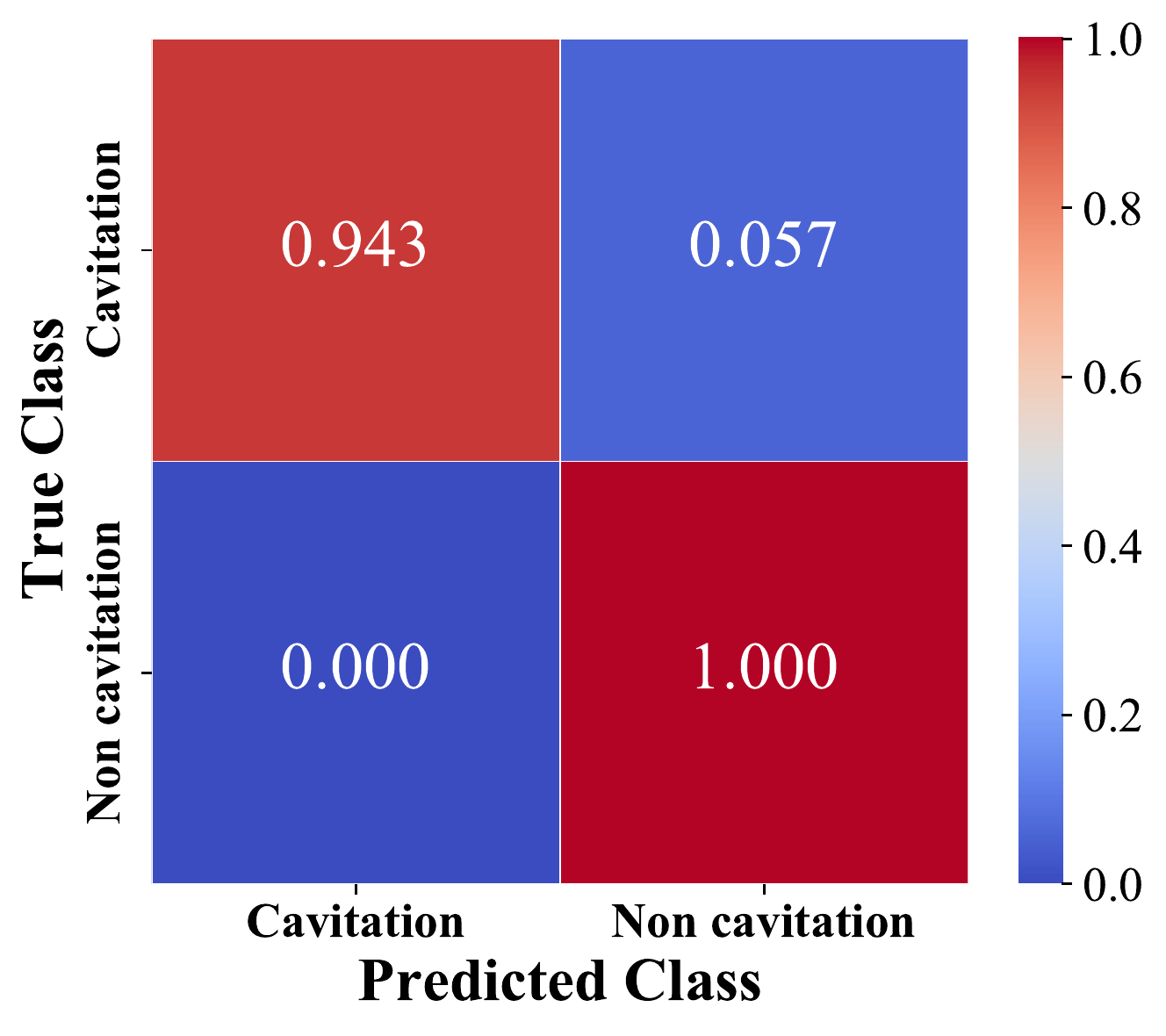}
    \caption{The confusion matrix for the best cavitation detection accuracy of the 1-D DHRN in \emph{\textbf{Dataset 1}}.}
    \label{fig: Confusion matrix TwoDataset1}
\end{figure}

\begin{figure}
    \centering
    \includegraphics[width=0.45\textwidth,height=62mm]{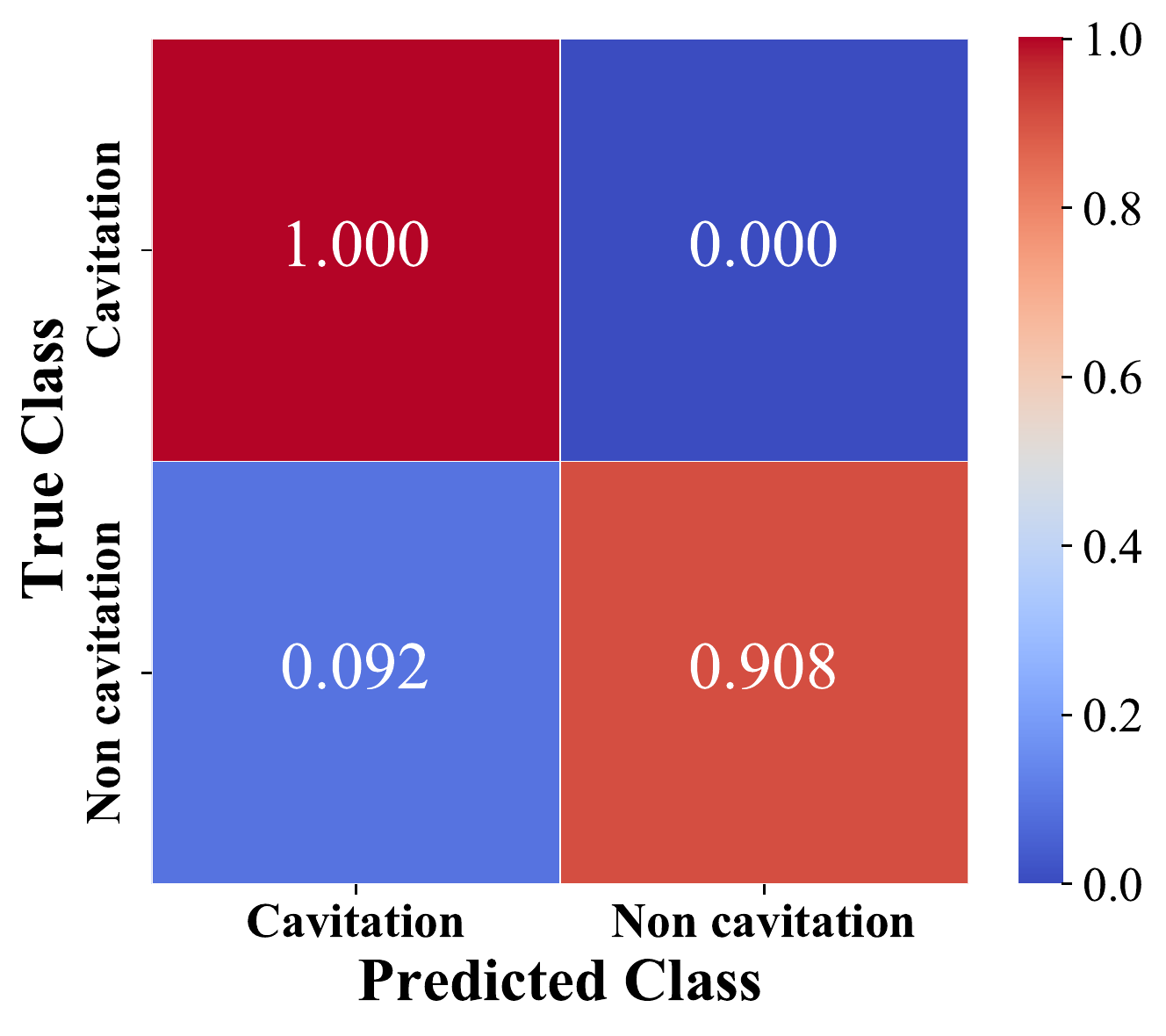}
    \caption{The confusion matrix for the best cavitation detection accuracy of the 1-D DHRN in \emph{\textbf{Dataset 2}}.}
    \label{fig: Confusion matrix TwoDataset2}
\end{figure}

\begin{figure}
    \centering
    \includegraphics[width=0.45\textwidth,height=62mm]{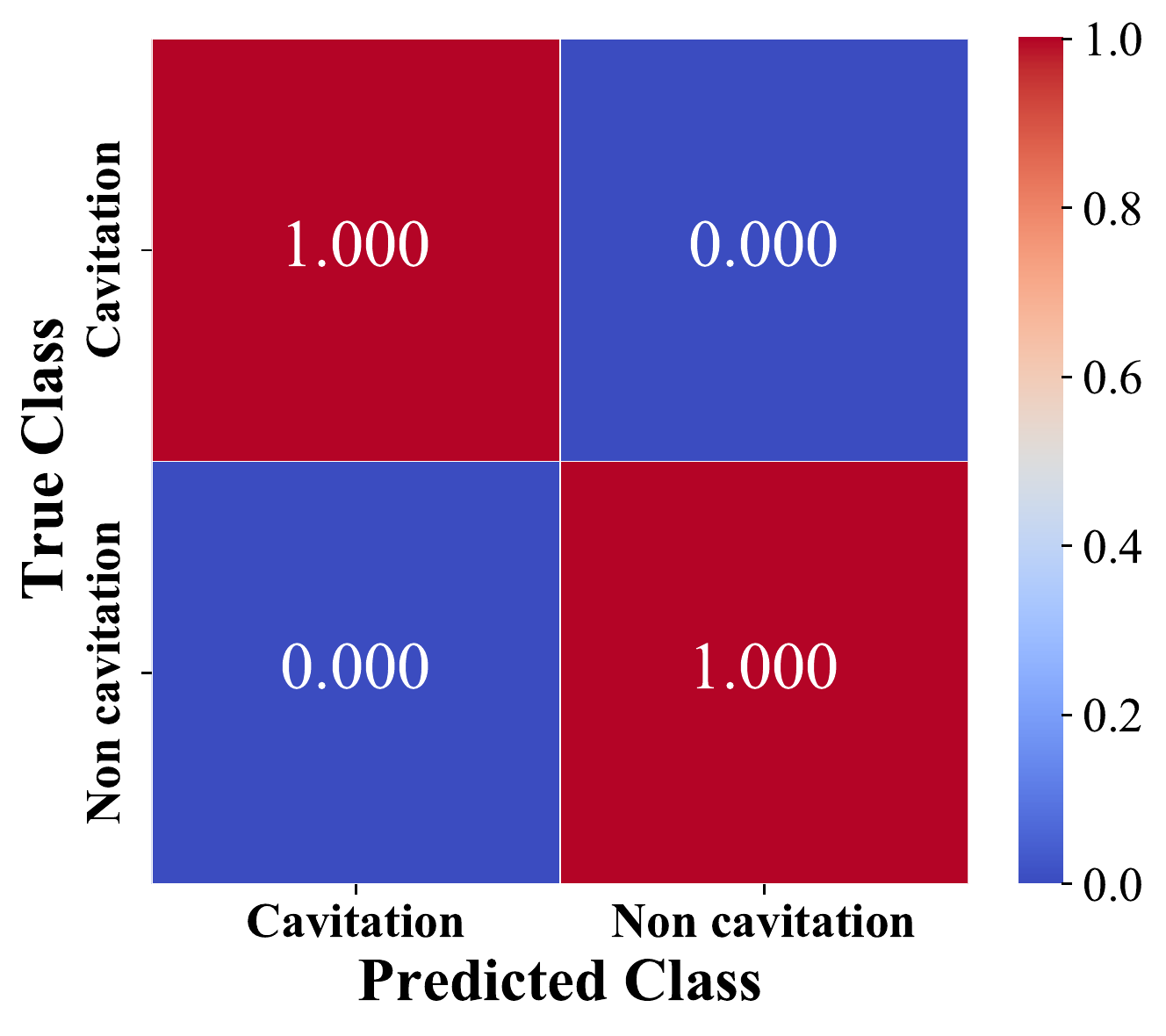}
    \caption{The confusion matrix for the best cavitation detection accuracy of the 1-D DHRN in \emph{\textbf{Dataset 3}}.}
    \label{fig: Confusion matrix TwoDataset3}
\end{figure}

The accuracy of the 1-D DHRN (our method) and comparison methods during different ${W_{size}}$ in \emph{\textbf{Dataset 1}}, \emph{\textbf{Dataset 2}} and \emph{\textbf{Dataset 3}} are shown in Figures \ref{fig: Accuracy cavitation detection Dataset1}, \ref{fig: Accuracy cavitation detection Dataset2} and \ref{fig: Accuracy cavitation detection Dataset3}. The confusion probability matrix of our method for best accuracy in \emph{\textbf{Dataset 1}}, \emph{\textbf{Dataset 2}} and \emph{\textbf{Dataset 3}} are shown in Figures \ref{fig: Confusion matrix TwoDataset1}, \ref{fig: Confusion matrix TwoDataset2} and \ref{fig: Confusion matrix TwoDataset3}.


\subsection*{Downsampling analysis}
The accuracy, precision, recall and F1-scores of 1-D DHRN for cavitation detection during different frequencies of samples in \emph{\textbf{Dataset 1}}, \emph{\textbf{Dataset 2}} and \emph{\textbf{Dataset 3}} are shown in Table \ref{tab:  Accuracy, Precision, Recall and F1-scor cavitation detection dataset123}.

The accuracy, precision, recall and F1-scores of 1-D DHRN for cavitation intensity recognition during different frequencies of samples in \emph{\textbf{Dataset 1}}, \emph{\textbf{Dataset 2}} and \emph{\textbf{Dataset 3}} are shown in Table \ref{tab:  Accuracy, Precision, Recall and F1-scor cavitation intensity recognition dataset123}.
\begin{table}
\caption{ Accuracy, Precision, Recall and F1-score results of different frequencies of samples $FS$ of the 1-D DHRN for cavitation detection in \emph{\textbf{Dataset 1}}, \emph{\textbf{Dataset 2}} and \emph{\textbf{Dataset 3}} ($\%$).}
\label{tab:  Accuracy, Precision, Recall and F1-scor cavitation detection dataset123}
\centering
\setlength{\tabcolsep}{0.8mm}{
\begin{tabular}{llcccc}
\toprule
\multicolumn{1}{c}{\multirow{2}{*}{Data}} & \multicolumn{1}{c}{\multirow{2}{*}{FS}} & \multicolumn{4}{c}{Metrics}  \\ 
\cmidrule{3-6} 
\multicolumn{1}{c}{}& \multicolumn{1}{c}{} & \multicolumn{1}{l}{Accuracy} & \multicolumn{1}{l}{Precision} & \multicolumn{1}{l}{Recall} & \multicolumn{1}{l}{F1-score} \\ 
\midrule
\multirow{6}{*}{Dataset1} & 1562500 & 97.02     & 96.75.  & 97.40  & 97.07      \\
                          & 781250  & 94.43     & 94.25   & 95.12  & 94.39      \\
                          & 390625  & 93.97     & 93.71   & 94.49  & 93.91       \\
                          & 260416  & 93.81     & 93.52   & 94.12  & 93.73   \\
                          & 195312  & 93.65     & 93.55   & 94.44  & 93.61  \\
                          & 48000   & 93.39     & 93.29   & 94.18  & 93.35  \\ 

\midrule
\midrule
\multirow{6}{*}{Dataset2} & 1562500 & 97.64     & 98.47  & 95.38  & 96.90       \\
                          & 781250  & 97.14     & 98.36  & 95.29  & 96.80       \\
                          & 390625  & 94.57     & 94.40  & 94.54  & 94.47       \\
                          & 260416  & 94.00     & 93.78  & 94.63  & 93.95   \\
                          & 195312  & 93.71     & 93.45  & 94.21  & 93.65  \\
                          & 48000   & 93.57     & 93.45  & 94.33  & 93.53  \\ 
\midrule
\midrule
\multirow{6}{*}{Dataset3} & 1562500 & 100.00    & 100.00 & 100.00 & 100.00 \\
                          & 781250  & 100.00    & 100.00 & 100.00 & 100.00      \\
                          & 390625  & 100.00    & 100.00 & 100.00 & 100.00       \\
                          & 260416  & 99.80     & 99.61  & 99.87  & 99.74   \\
                          & 195312  & 99.62     & 99.26  & 99.75  & 99.50  \\
                          & 48000   & 99.28     & 98.61  & 99.52  & 99.06  \\ 
\bottomrule
\end{tabular}

}

\end{table}
\begin{table}
\caption{ Accuracy, Precision, Recall and F1-score results of different frequencies of samples $FS$ of the 1-D DHRN for cavitation intensity recognition in \emph{\textbf{Dataset 1}}, \emph{\textbf{Dataset 2}} and \emph{\textbf{Dataset 3}} ($\%$).}
\label{tab:  Accuracy, Precision, Recall and F1-scor cavitation intensity recognition dataset123}
\centering
\setlength{\tabcolsep}{0.8mm}{
\begin{tabular}{llcccc}
\toprule
\multicolumn{1}{c}{\multirow{2}{*}{Data}} & \multicolumn{1}{c}{\multirow{2}{*}{FS}} & \multicolumn{4}{c}{Metrics}  \\ 
\cmidrule{3-6} 
\multicolumn{1}{c}{}& \multicolumn{1}{c}{} & \multicolumn{1}{l}{Accuracy} & \multicolumn{1}{l}{Precision} & \multicolumn{1}{l}{Recall} & \multicolumn{1}{l}{F1-score} \\ 
\midrule
\multirow{6}{*}{Dataset1} & 1562500 & 93.75    & 95.72  & 88.00     & 90.53       \\
                          & 781250  & 90.57    & 91.49  & 84.10     & 86.18       \\
                          & 390625  & 88.26    & 89.47  & 81.63     & 83.03       \\
                          & 260416  & 86.03    & 89.38  & 78.25     & 81.19   \\
                          & 195312  & 84.76    & 88.04  & 77.43     & 80.76  \\
                          & 48000   & 72.83    & 73.26  & 69.92     & 65.59  \\ 
                          
\midrule
\midrule
\multirow{6}{*}{Dataset2} & 1562500 & 94.31    & 95.14  & 94.02      & 94.58       \\
                          & 781250  & 92.02    & 92.17  & 91.03      & 91.60       \\
                          & 390625  & 86.53    & 87.25  & 86.56      & 86.39      \\
                          & 260416  & 84.53    & 85.57  & 84.52      & 84.50   \\
                          & 195312  & 84.60    & 85.35  & 84.52      & 84.43  \\
                          & 48000   & 77.65    & 80.37  & 77.66      & 78.09  \\ 
\midrule
\midrule
\multirow{6}{*}{Dataset3} & 1562500 & 100.00   & 100.00 & 100.00      & 100.00       \\
                          & 781250  & 100.00   & 100.00 & 100.00      & 100.00       \\
                          & 390625  & 100.00   & 100.00 & 100.00      & 100.00       \\
                          & 260416  & 99.62    & 99.26  & 99.75       & 99.50   \\
                          & 195312  & 98.28    & 98.39  & 98.28       & 98.28  \\
                          & 48000   & 90.17    & 92.95  & 90.17       & 89.78  \\ 
\bottomrule
\end{tabular}

}

\end{table}

\end{document}